\documentclass[10pt,twocolumn,letterpaper]{article}

\usepackage[pagenumbers]{wacv}  
\usepackage[accsupp]{axessibility}  

\usepackage{amsfonts}
\usepackage{amsmath}
\usepackage{booktabs}
\usepackage{multirow}
\usepackage{rotating} 
\usepackage{algorithm}
\usepackage{algpseudocode}
\usepackage{tikz}

\newcommand{\dscell}[1]{%
  \raisebox{-0.5\height}{\includegraphics[width=0.225\linewidth]{media/dataset/#1}}%
}

\definecolor{wacvblue}{rgb}{0.21,0.49,0.74}
\usepackage[breaklinks,colorlinks,allcolors=wacvblue]{hyperref}

\def\wacvPaperID{1639}
\def\confName{WACV}
\def\confYear{2027}

\title{SSA-3DGS: Unsupervised Removal of Screen-Space Artifacts for 3D Gaussian Splatting}

\author{Kristof Overdulve \quad Lode Jorissen \quad Nick Michiels\\
Digital Future Lab, Flanders Make, Hasselt University\\
{\tt\small \{kristof.overdulve, lode.jorissen, nick.michiels\}@uhasselt.be}
}

\begin{document}
\makeatletter
\g@addto@macro\@maketitle{%
  \vspace{12pt}%
  \begin{center}
  \setlength{\tabcolsep}{1.2pt}%
  \renewcommand{\arraystretch}{0.5}%
  \newcommand{\tincg}[1]{\includegraphics[width=0.161\linewidth]{media/teaser2/#1}}%
  \footnotesize
  \begin{tabular}{@{}cccccc@{}}
    \multicolumn{3}{c}{\textbf{Input:} posed images corrupted by screen-space artifacts}
      & Vanilla 3DGS
      & \textbf{Ours} (reconstruction)
      & \textbf{Ours} (extracted artifact) \\[2pt]
    \tincg{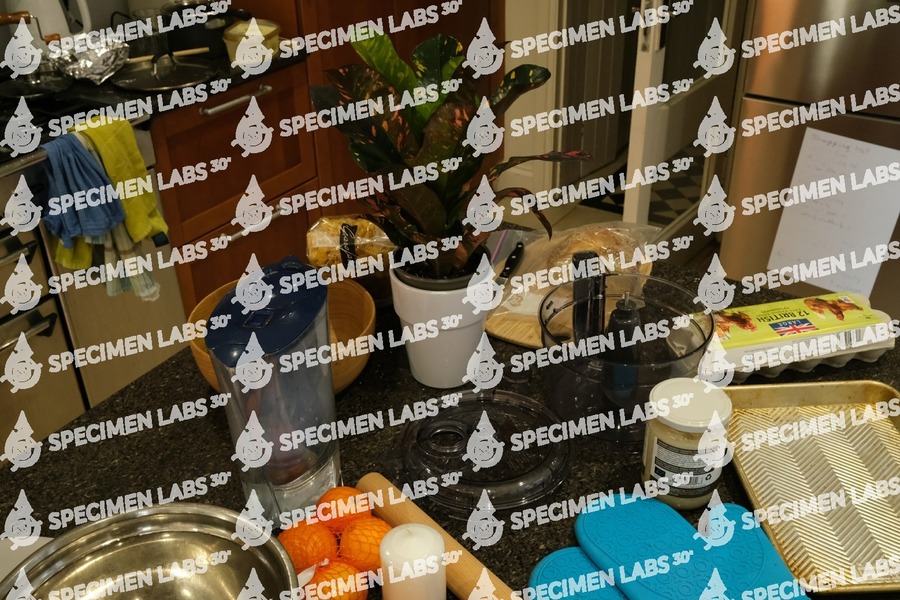} &
    \tincg{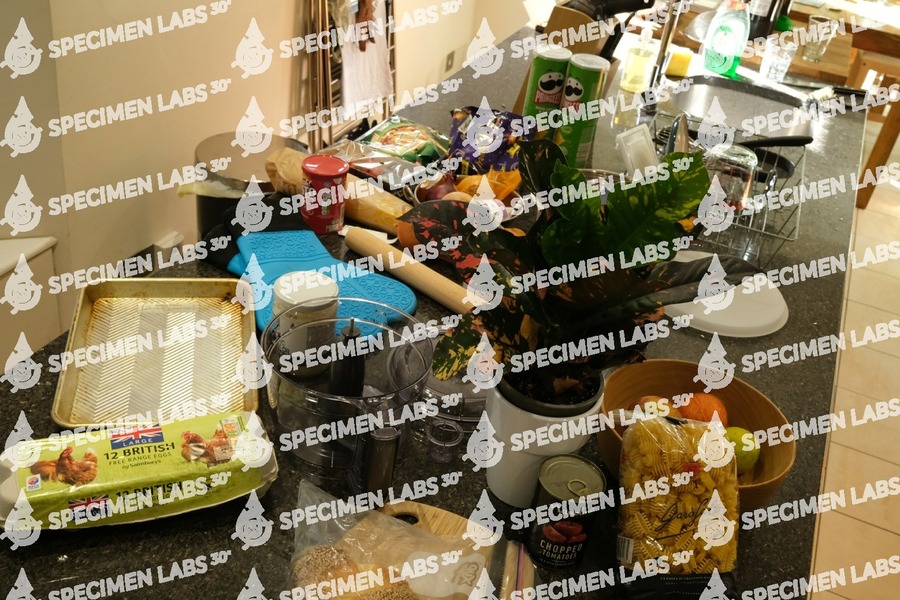} &
    \tincg{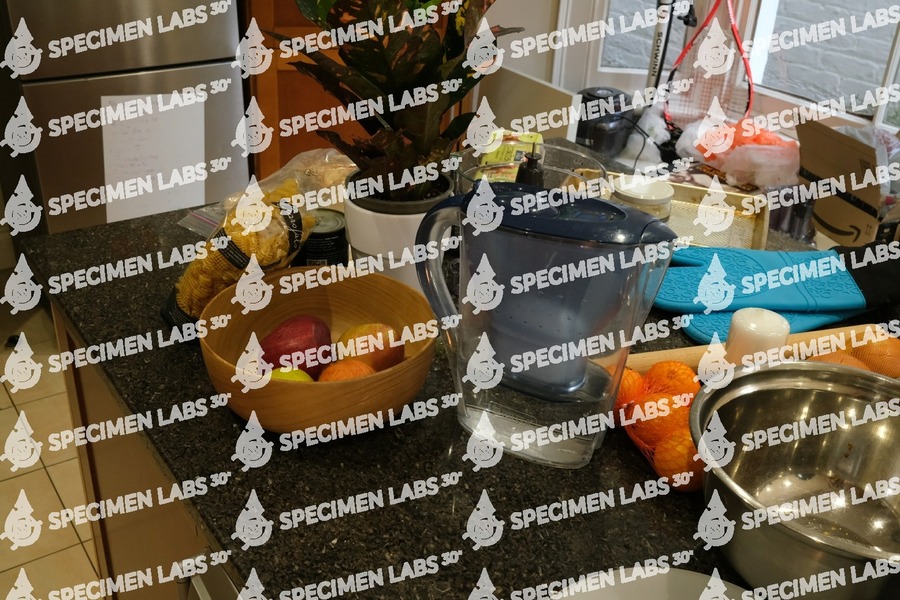} &
    \tincg{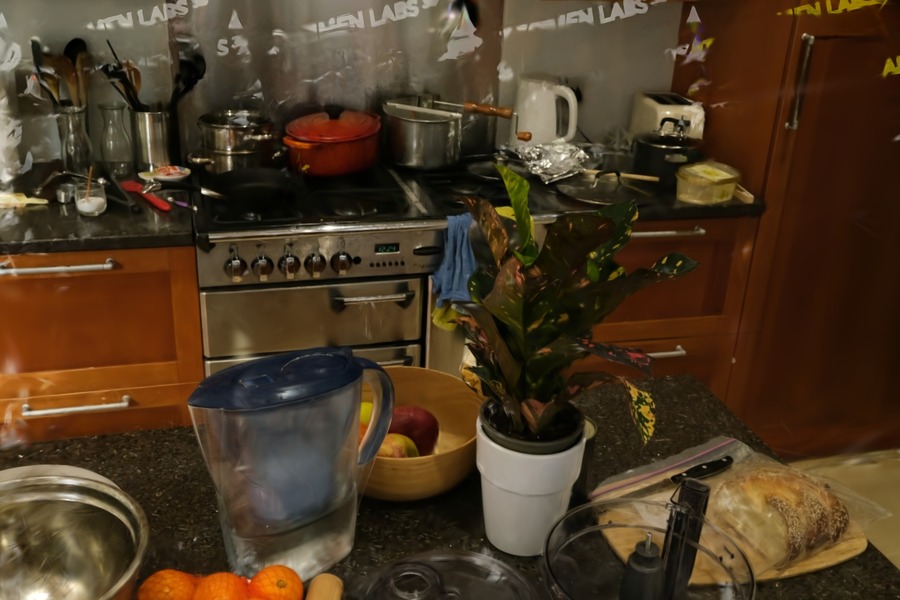} &
    \tincg{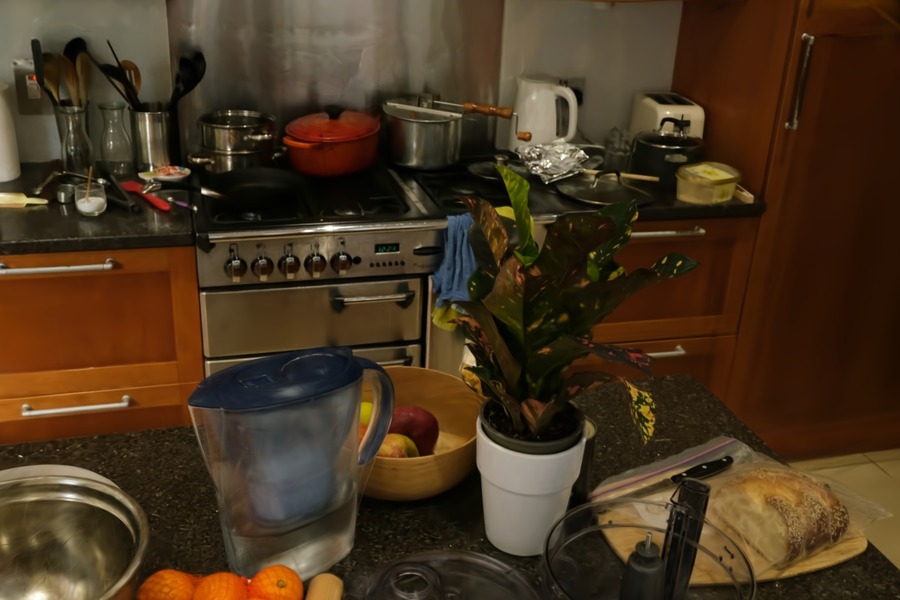} &
    \tincg{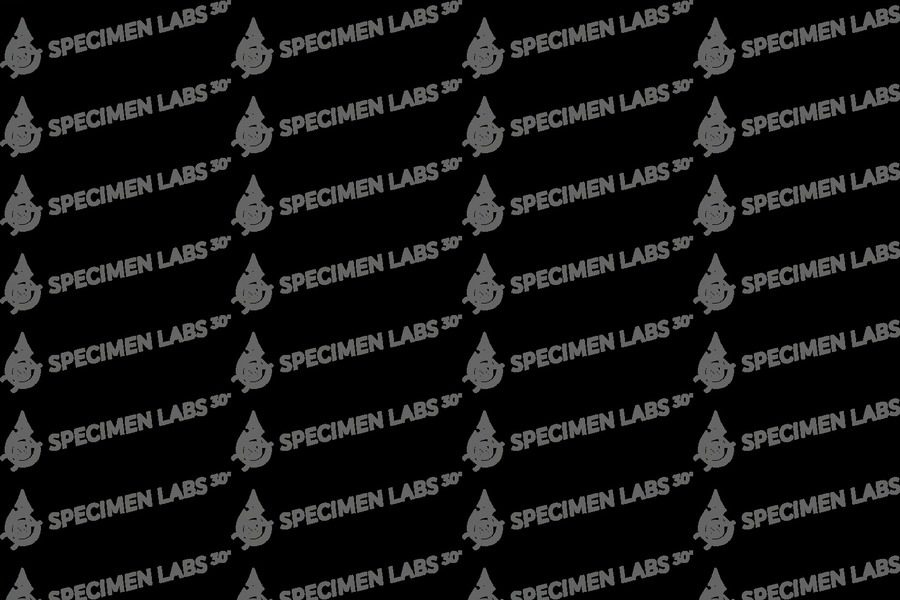} \\
    \tincg{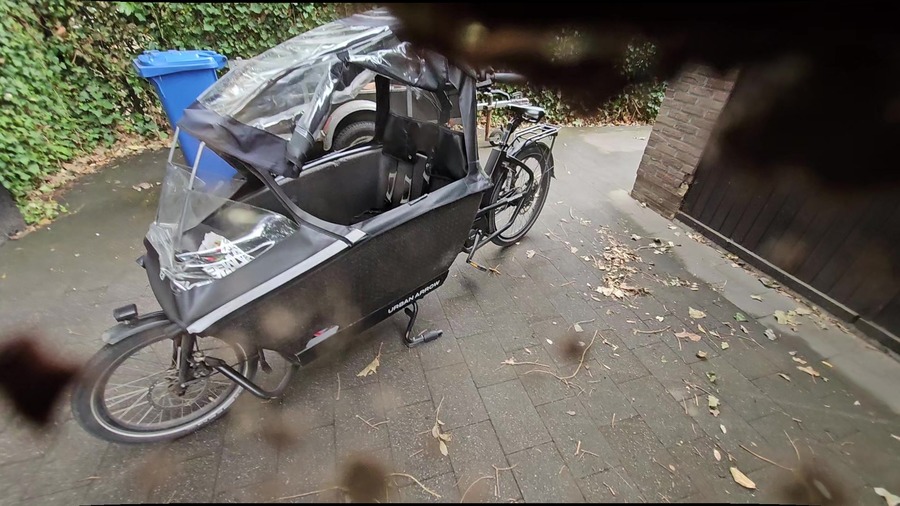} &
    \tincg{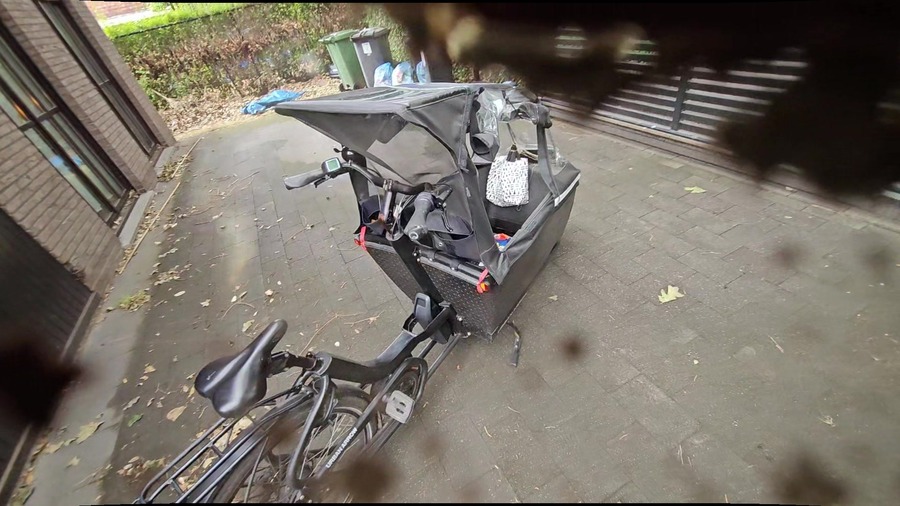} &
    \tincg{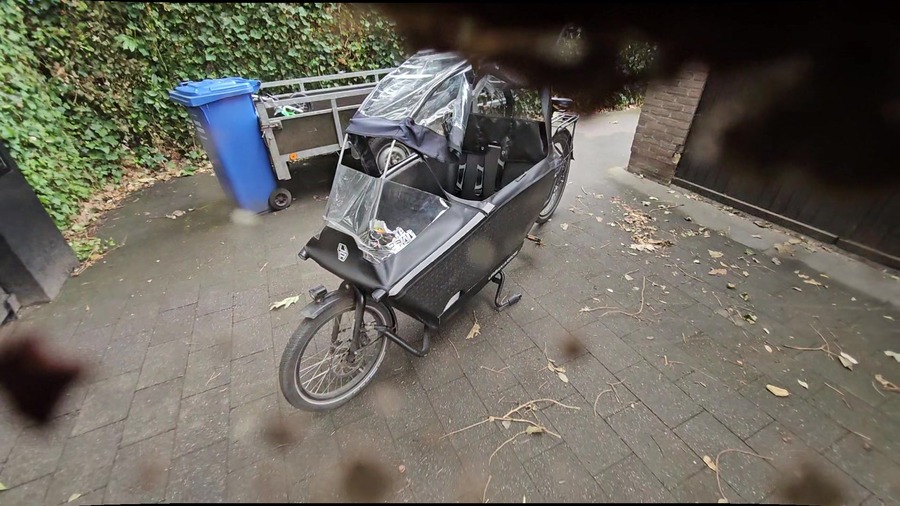} &
    \tincg{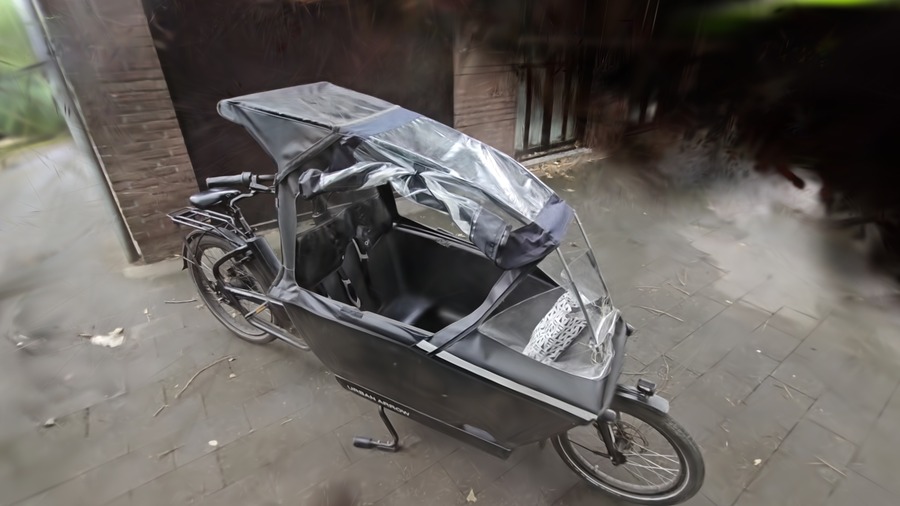} &
    \tincg{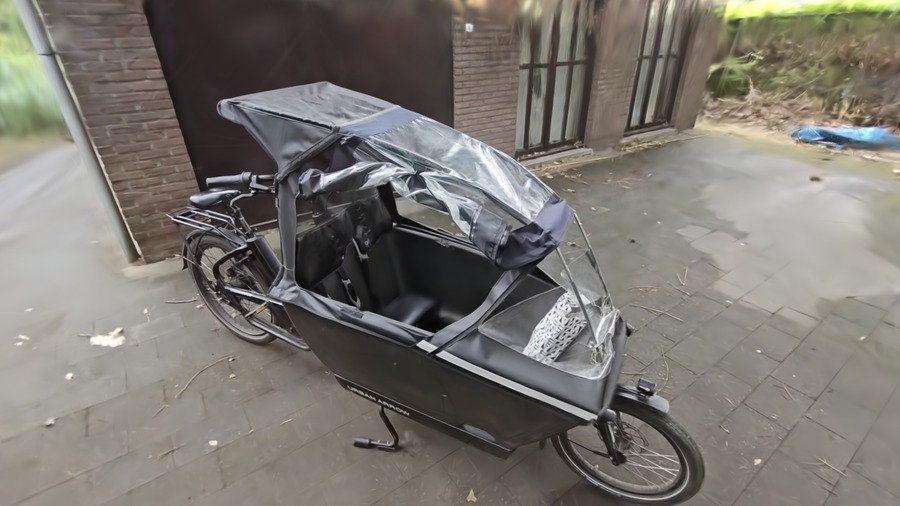} &
    \includegraphics[width=0.161\linewidth]{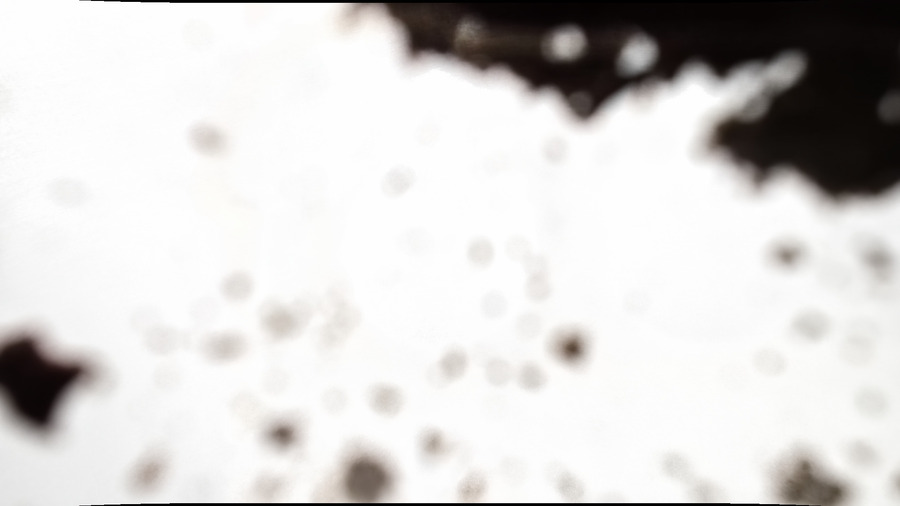} \\
  \end{tabular}
  \captionof{figure}{\textbf{Unsupervised removal of screen-space artifacts.} Given posed images corrupted by an artifact fixed in the camera frame (left: three inputs), vanilla 3DGS bakes the artifact into the scene, whereas \textit{SSA-3DGS} recovers the clean scene \emph{and} the artifact, without masks or supervision. Top: synthetic watermark; bottom: real mud capture. Extracted artifacts are composited over a black or white background, whichever is more legible.}
  \label{fig:teaser}
  \end{center}%
}
\makeatother

\maketitle
\begin{abstract}
Novel View Synthesis (NVS) methods, such as 3D Gaussian Splatting (3DGS), rely on the assumption of clean, multi-view consistent, posed input images. Real-world captures can violate this assumption due to \textbf{screen-space artifacts}---static occlusions fixed to the 2D image plane rather than to the 3D world. Common examples include physical sensor defects, environmental obstructions (such as rain or mud on the lens enclosure), capture obstructions (such as a thumb over the camera sensor or a dashboard visible in dashcam footage), and digital overlays (such as watermarks or UI elements). When present, they are erroneously baked into the 3D geometry as ``floaters'' or near-camera artifacts, degrading the quality of novel-view rendering. In this work, we propose \textit{SSA-3DGS}, an unsupervised framework that jointly optimizes a 3D scene and a learnable 2D overlay to recover a clean 3D scene and the corrupting artifacts. By exploiting geometric consensus across views, our method effectively disentangles static artifacts from the 3D scene geometry without supervision or manual input. Across diverse synthetic corruptions and a self-captured real-world dataset, SSA-3DGS improves reconstruction fidelity by up to ${\sim}8$~dB PSNR over 3DGS trained on the same corrupted inputs, while also recovering the artifact itself.
\end{abstract}

\section{Introduction}

The goal of Novel View Synthesis (NVS) is to capture the visual essence of a real-world scene from multiple viewpoints and render it from arbitrary novel angles. This field has witnessed a paradigm shift with the advent of Neural Radiance Fields (NeRF) \cite{mildenhall2021nerf} and, more recently, 3D Gaussian Splatting (3DGS) \cite{kerbl20233d}. 3DGS enables high-fidelity, real-time rendering by representing scenes as a collection of explicit 3D Gaussians rather than implicit neural weights.

\begin{figure}[tbhp]
  \centering
  \includegraphics[width=0.24\columnwidth]{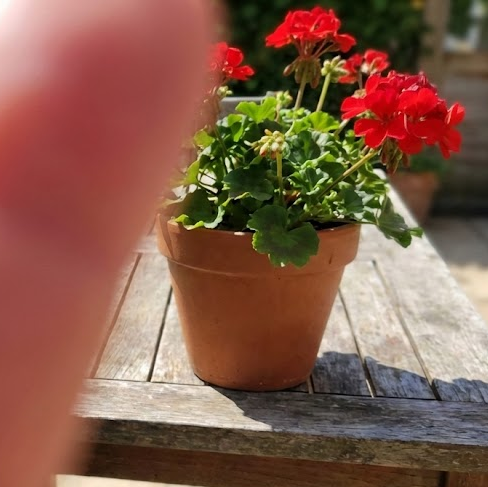}
  \includegraphics[width=0.24\columnwidth]{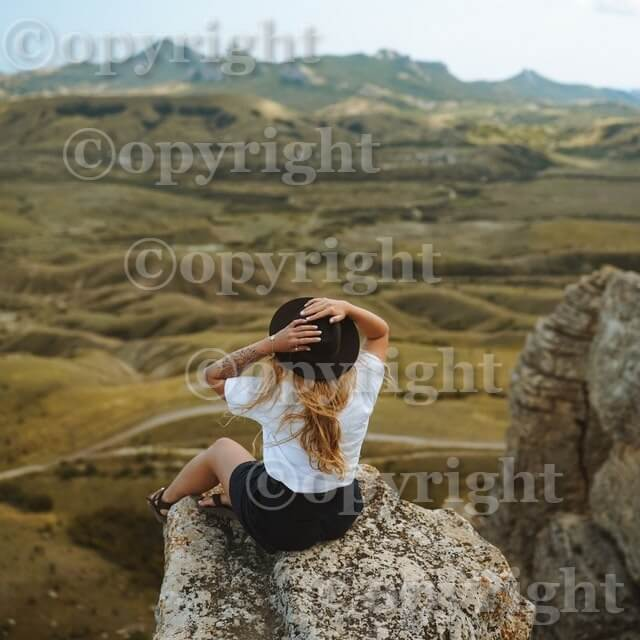}
  \includegraphics[width=0.24\columnwidth]{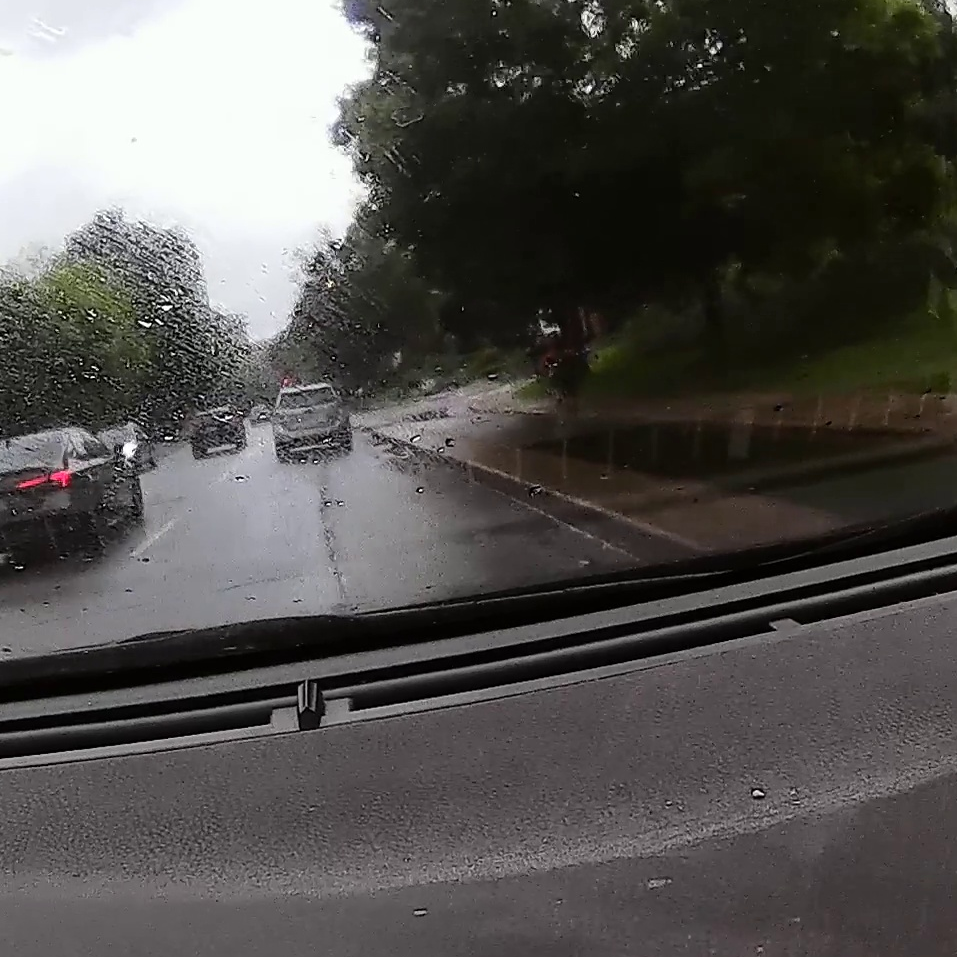}
  \includegraphics[width=0.24\columnwidth]{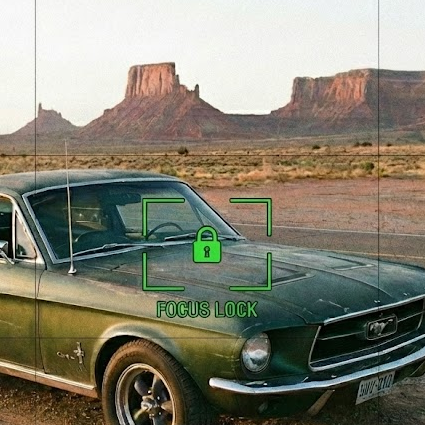}
  \caption{Example screen-space artifacts. Left to right: a thumb over the sensor, watermark labels, the dashboard in dashcam footage, and UI elements.}
  \label{fig:intro}
\end{figure}

The reconstruction quality of novel-view synthesis algorithms relies heavily on a fundamental assumption in Multi-View Stereo (MVS): that the input images provide a multi-view-consistent representation of the underlying scene geometry. While recent research has effectively modeled \textit{world-space} inconsistencies such as illumination changes \cite{qiao2025restorgs} or transient moving elements like pedestrians \cite{xu2024wild, martin2021nerf, sabour2025spotlesssplats, wang2025desplat}, these approaches struggle to handle \textit{screen-space artifacts} (see \cref{fig:intro} for example screen-space artifacts). Distinct from world-space objects, these artifacts are static relative to the camera sensor, manifesting as sensor defects (dead pixels or dust), environmental effects (rain or mud spatters on the lens enclosure), capture obstructions (visible fingers, or car dashboards in dashcam footage), or digital overlays (watermarks or UI elements). Because these elements are fixed to the image plane, they effectively ``move'' relative to the world geometry as the camera moves. Consequently, standard novel view synthesis and approaches that focus on world-space inconsistencies often fail to reject them, thereby erroneously baking the artifacts into the 3D scene as near-camera ``floaters'' or view-dependent effects.

Addressing this issue is non-trivial. Single-view image restoration and inpainting techniques \cite{suvorov2022resolution, wan2020reflection, su2025deep, cui2025bioir, zamfir2025moceir, flux2025kontext} operate on a per-image basis, lack knowledge of the 3D scene structure, and hallucinate frame-to-frame texture variations. These lead to temporal flickering and geometric inconsistencies when used as a multi-view pre-processing step. Inpainting, moreover, requires manual annotation, which is not scalable when processing hundreds of images.

In this work, we propose \textit{SSA-3DGS}, an unsupervised framework built upon 3DGS that disentangles 3D scene geometry from 2D screen-space artifacts. Our method achieves this by formulating a simultaneous \emph{dual objective}: (1) accurately separating and segmenting the 2D artifacts from the scene, and (2) reconstructing a high-fidelity, artifact-free 3D asset. The key to making this joint optimization work is leveraging motion parallax. As the camera moves through the world, screen-space artifacts remain static relative to the image plane, allowing the two signals to be decoupled without supervision. We model screen-space artifacts as learnable, semi-transparent overlays that are shared across all training views and jointly optimized with the 3DGS parameters. During training, the gradients of the reconstruction loss are distributed between the 3D model and the 2D overlay. To prevent the overlay from absorbing valid scene details or altering the scene's exposure, we introduce a sparsity regularization scheme.

We evaluate SSA-3DGS on diverse synthetic corruptions of real-world images and on a self-captured real-world dataset containing genuine, physically induced artifacts. These experiments demonstrate that our method learns clean, artifact-free 3D scenes from heavily corrupted inputs (\cref{fig:teaser}). In summary, our contributions are:

\begin{enumerate}
    \item An unsupervised method to, given a multi-view recording corrupted by screen-space artifacts, learn a clean 3DGS scene and the screen-space artifact via a shared learnable 2D overlay composited with the 3DGS render.
    \item A sparsity regularization scheme that biases the optimization toward explaining pixels through 3D geometry, only using the artifact layer if needed.
    \item A real-world, self-captured dataset~\cite{overdulve2026} on which we benchmark our proposed algorithm.
\end{enumerate}
\section{Related work}
\label{sec:related}

\subsection{NeRF \& 3D Gaussian Splatting}
NeRF \cite{mildenhall2021nerf}, which represents scenes as continuous volumetric functions optimized via differentiable ray marching, has revolutionized the field of Novel View Synthesis (NVS). While NeRF achieves photorealistic quality, its implicit nature leads to slow training and rendering; subsequent approaches accelerate training and rendering using voxel grids \cite{fridovich2022plenoxels} or multi-resolution hash encodings \cite{muller2022instant}. Most notably, Kerbl et al. introduced 3D Gaussian Splatting (3DGS) \cite{kerbl20233d}, representing the scene as a set of anisotropic 3D Gaussians rasterized onto the image plane. 3DGS combines state-of-the-art quality with real-time performance. All these methods rely heavily on multi-view consistent inputs to avoid floating artifacts and to optimize towards the correct ground-truth 3D geometry using differentiable rendering.

\subsection{Image and multi-frame restoration}
Removing unwanted occlusions is a long-standing problem in image processing. Significant research has focused on removing rain, dirt, and reflections from \emph{single} images using deep learning: Eigen et al.\ \cite{eigen2013restoring} train CNNs to remove dirt and rain, others utilize attention mechanisms \cite{luo2020weakly} or synthetic physics-based datasets \cite{hao2019learning} to hallucinate clean backgrounds behind raindrops, and single-image reflection removal \cite{wan2020reflection, zhu2024revisiting} separates background and reflection layers. State-of-the-art inpainting methods, such as LaMa \cite{suvorov2022resolution} and supervised watermark-removal models \cite{hertz2019blind, su2025deep}, hallucinate missing textures but require training a separate model on large, paired datasets. While powerful, these 2D methods lack knowledge of 3D scene structure, so applying them independently to NVS input frames as a pre-processing step results in temporal flickering and geometric inconsistencies that break multi-view stereo assumptions. \emph{All-in-one} restoration models such as BioIR~\cite{cui2025bioir} and MoCE-IR~\cite{zamfir2025moceir} invert a fixed, closed set of \emph{natural} degradations (noise, haze, rain, blur, low light) \emph{blindly}; screen-space artifacts are not covered by these models, so these networks leave corrupting artifacts in place. Instruction-tuned image editors, such as FLUX.1-Kontext~\cite{flux2025kontext}, can use text prompts to instruct models on what content to remove from images. They work well for thin, semi-transparent overlays but hallucinate content incorrectly for large, near-opaque occluders. Additionally, they require explicit, well-designed text prompts.

When multiple views are available, geometric constraints can facilitate separation. Early works utilized stereo disparity inconsistencies to detect and remove adherent water drops~\cite{tanaka2006removal} and reflections~\cite{guo2014robust}. Gu et al.\ \cite{gu2009removing} estimate a dirty lens's attenuation and scattering from statistics over many frames, assuming defocused, low-frequency dirt. None of these methods reconstructs the scene in 3D.

\subsection{Robust novel-view synthesis}
Pioneering work in NVS for handling data that violates multi-view consistency is NeRF-W \cite{martin2021nerf}, which introduced transient embeddings to handle photometric variations and moving objects (e.g., pedestrians) in unconstrained photo collections. In contrast, RobustNeRF \cite{sabour2023robustnerf} employs robust loss functions to ignore pixels that do not adhere to epipolar geometry. Recent works adapt these ideas to 3DGS: SpotlessSplats \cite{sabour2025spotlesssplats} and Wild-GS \cite{xu2024wild} use feature-based masking and robust losses to prune Gaussians associated with transient objects, DeSplat \cite{wang2025desplat} explicitly separates ``distractor'' Gaussians from scene Gaussians, and others combine training residuals with a pretrained segmentation network to mask distractors from the loss~\cite{ungermann2024robust}. A second line of work removes artifacts of the reconstruction itself: some approaches eliminate floaters by prioritizing low-frequency optimization \cite{wang2025low}, while \emph{generative} methods repair under-observed renders with diffusion priors, either with reference-conditioned video diffusion for sparse-view reconstructions (3DGS-Enhancer \cite{liu2024enhancer}, GSFixer \cite{yin2025gsfixer}) or with a single-step image-diffusion enhancer (Difix3D+ \cite{wu2025difix3d}). Multi-view inpainting within 3DGS relies on masks to annotate objects to be removed in the scene~\cite{huang20253d, zhou2025high}. Crucially, all of these methods treat the corruption as part of the 3D world, whether as a transient inconsistency across views or as a defect in the reconstructed geometry. Screen-space artifacts, in contrast, are static relative to the camera but effectively ``move'' relative to the world as the camera trajectory changes, so robust estimators fail to reject them, generative priors leave them in place, and they are baked into the scene as floaters.

Other methods target degradations of the capture itself: some model camera-specific effects, such as rolling shutter or motion blur \cite{seiskari2024gaussian}, and RestorGS \cite{qiao2025restorgs} extends 3DGS to degraded scenes (haze, underwater) by decoupling ``clear'' and ``degraded'' Gaussians. DerainNeRF \cite{li2024derainnerf} masks lens-adherent waterdrops found by a pretrained waterdrop detector out of the photometric loss, while DeRainGS \cite{liu2025deraings} and WeatherGS \cite{qian2025weathergs} first remove rain streaks or snow with pretrained restoration networks and then mask residual artifacts or lens drops. These weather-specific methods handle lens drops but rely on networks trained on rain data, and their masks exclude pixels even where a semi-transparent drop still reveals the scene behind it. Our method instead leverages motion parallax directly within the 3DGS optimization, using the photometric loss as a supervision signal to separate scene content from overlays without 2D pre-processing, ground-truth masks, or pre-trained models.
\section{Methodology}

Our goal is to reconstruct a high-fidelity 3D scene from a set of images $\mathcal{I} = \{I_1, \dots, I_N\}$ captured from known poses $P_1, \dots, P_N$. We assume that the input images are corrupted by screen-space artifacts that do not conform to the scene's epipolar geometry. Our approach jointly optimizes the explicit, clean scene geometry and a view-independent screen-space overlay, thereby separating the two.

\subsection{Preliminaries: 3D Gaussian Splatting}
We adopt 3D Gaussian Splatting (3DGS) \cite{kerbl20233d} as our scene representation. The scene is modeled as a set of 3D Gaussians, where each Gaussian $G_k$ is defined by a center position $\mu_k \in \mathbb{R}^3$, a covariance matrix $\Sigma_k$, an opacity $\alpha_k \in [0, 1]$, and view-dependent color coefficients $c_k$ represented via Spherical Harmonics. For a given viewpoint, these 3D Gaussians are projected onto the 2D image plane. The final color $\hat{C}_{3D}(\mathbf{u})$ of a pixel $\mathbf{u}$ is computed by sorting the $\mathcal{N}$ overlapping Gaussians by depth and applying $\alpha$-blending:
\begin{equation}
    \hat{C}_{3D}(\mathbf{u}) = \sum_{i \in \mathcal{N}} c_i \alpha_i \prod_{j=1}^{i-1} (1 - \alpha_j),
\end{equation}
where $\mathcal{N}$ is the set of depth-sorted Gaussians overlapping the pixel and $\alpha_i$ denotes the $i$-th Gaussian's opacity weighted by its projected 2D footprint. During optimization, the set of Gaussians is periodically grown and pruned to fit the scene, either via the heuristic Adaptive Density Control (ADC) in the original 3DGS or via the more recent Markov Chain Monte Carlo (MCMC) densification \cite{kheradmand20243d}. Our formulation is agnostic to this choice, and we evaluate with both strategies (\cref{sec:impl}).

\subsection{Modeling screen-space artifacts}
Vanilla 3DGS assumes that $\hat{C}_{3D}(\mathbf{u})$ should directly match the ground-truth image $I_{gt}(\mathbf{u})$. However, in the presence of screen-space artifacts, this assumption forces the optimization to generate ``floaters''---3D Gaussians placed arbitrarily close to the camera center to mimic the artifact---resulting in overfitting to the training images. To resolve this, we introduce a \textbf{screen-space overlay model} that defines the artifact as a learnable 2D tensor shared across all $N$ views in the dataset. This model consists of two components:
\begin{itemize}
    \item \textbf{Artifact Color Map ($W$):} an RGB image at the input resolution holding the colors of the artifact (e.g., the white of a text watermark).
    \item \textbf{Artifact Alpha Matte ($A$):} a single-channel map of the same resolution holding the per-pixel artifact opacity.
\end{itemize}
To guarantee that the alpha-blending below operates on valid intensities, $W$ and $A$ are stored as unconstrained real-valued tensors and mapped to $[0,1]$ via a logistic sigmoid $\operatorname{sigmoid}(\cdot)$; all equations in this section should be read with this implicit reparameterization (i.e.\ $W$ and $A$ denote $\operatorname{sigmoid}(\tilde W)$ and $\operatorname{sigmoid}(\tilde A)$ of the underlying learnable parameters $\tilde W, \tilde A$). Unlike Gaussian splats, which are projected from 3D space onto images differently per pose $P_i$, $W$ and $A$ are defined in pixel coordinates and remain constant for all $i$. We append a compositing stage to the standard rendering pipeline. For each training iteration, we first render the clean 3D scene to obtain $\hat{C}_{3D}$, and then compose the learned overlay on top of this render. The final predicted color $\hat{C}(\mathbf{u})$ for a pixel with 2D coordinates $\mathbf{u}$ is modeled as an alpha-blend between the artifact and the 3D scene:
\begin{equation}
    \hat{C}(\mathbf{u}) = A(\mathbf{u}) \cdot W(\mathbf{u}) + (1 - A(\mathbf{u})) \cdot \hat{C}_{3D}(\mathbf{u}).
    \label{eq:composition}
\end{equation}
This equation implies a competition between the 3D scene and the 2D overlay: for each pixel, the optimization decides whether it is better explained by consistent 3D geometry or the static 2D overlay. The clean render $\hat{C}_{3D}$---the 3D scene rendered \emph{without} the overlay---is the artifact-free output we ultimately care about: at test time, we discard the overlay and render $\hat{C}_{3D}$ from novel views.

\subsection{Optimization and regularization}
\label{sec:optimization}

Minimizing the reconstruction error between $\hat{C}$ and $I_{gt}$ is insufficient, because the decomposition in \cref{eq:composition} is under-constrained: the loss only sees the composited color $\hat{C}$, so many $(\hat{C}_{3D}, A, W)$ triples reproduce the same image. Two degenerate solutions are particularly harmful. First, pixels that are relatively static across views can be erroneously baked into the 2D overlay. Second, the overlay can collapse into a semi-transparent, spatially uniform \emph{exposure} layer: the 3D scene is reconstructed too bright, and a low-opacity dark overlay dims it back to the correct intensity. Such an overlay contains no genuine artifact. It merely rescales the exposure of an otherwise clean render so that it is invisible to the photometric loss. To suppress these solutions and ensure the overlay captures only true screen-space artifacts, we introduce a composite loss function that prioritizes explaining pixel data through the 3D scene structure over the 2D overlay. The loss is defined as:
\begin{equation}
    \mathcal{L} = \mathcal{L}_{rgb} + \lambda_{sparse} \mathcal{L}_{sparse},
\end{equation}
where $\lambda_{sparse}$ is a hyperparameter controlling the relative importance of the sparsity regularization. The loss comprises:

\noindent \textbf{Photometric Loss ($\mathcal{L}_{rgb}$)}: the standard L1 and SSIM loss used in 3DGS to measure the similarity between the composited prediction $\hat{C}$ and the ground truth $I_{gt}$:
\begin{equation}
    \mathcal{L}_{rgb} = (1 - \lambda_{ssim})\mathcal{L}_1(\hat{C}, I_{gt}) + \lambda_{ssim} \mathcal{L}_{ssim}(\hat{C}, I_{gt}).
\end{equation}

\noindent \textbf{Sparsity Regularization ($\mathcal{L}_{sparse}$)}: enforces the assumption that the density of artifacts is sparse relative to the scene content, encouraging the alpha matte $A$ to be zero wherever possible:
\begin{equation}
    \mathcal{L}_{sparse} = \underset{\mathbf{u}}{\operatorname{mean}}\, \bigl| A(\mathbf{u}) \bigr|,
\end{equation}
where the mean is taken over all pixels $\mathbf{u}$ of the alpha matte.

\noindent \textbf{Statistics-based initialization.} Rather than initializing the overlay from constants, we seed it from per-pixel training-set statistics. For each pixel $\mathbf{u}$ we compute the mean color $\mu(\mathbf{u})$ and the max-over-channels standard deviation $\sigma(\mathbf{u})$ across all training images, and set $W(\mathbf{u}) = \mu(\mathbf{u})$ and $A(\mathbf{u}) = \exp(-\sigma(\mathbf{u}) / s)$ with $s = 5 \times 10^{-2}$. Pixels whose intensity is stable across views (likely artifacts) therefore start nearly opaque, while pixels with strong multi-view variation (likely scene content) start near-transparent, giving the optimization a well-posed initial decomposition.
\section{Evaluation}

We evaluate \textit{SSA-3DGS} on synthetically corrupted images derived from standard real-world datasets and on a self-captured real-world dataset where captures are corrupted by physically induced screen-space artifacts. The aim is to answer three key questions: (1) \textbf{Robustness:} Can we consistently recover high-fidelity 3D geometry from multiple types of heavily corrupted inputs? (2) \textbf{Disentanglement:} Does the joint optimization successfully separate the clean 3D scene from the overlay? (3) \textbf{Regularization:} Are the proposed sparsity regularizer and statistics-based initialization necessary?

\subsection{Implementation details}
\label{sec:impl}
We implemented our method on top of \textbf{gsplat} \cite{ye2025gsplat}, using the default 3DGS hyperparameters and $30$k training iterations. Our method is agnostic to the densification strategy: we report results for both the default Adaptive Density Control (ADC) \cite{kerbl20233d} and Markov Chain Monte Carlo (MCMC) densification \cite{kheradmand20243d} where relevant. For MCMC, we use the same splat budget as the original dataset configurations. 

The overlay maps $W$ and $A$ are optimized using Adam with a fixed learning rate of $5 \times 10^{-3}$. We set $\lambda_{ssim} = 0.2$. The sparsity weight is $\lambda_{sparse} = 0.05$ for the synthetic experiments (all but mud) and $0.01$ for the real scenes and the synthetic mud artifacts, as these contain large areas of occluded content. The proposed algorithm is rather sensitive to the $\lambda_{sparse}$ value: a uniform $0.05$ costs $1.3$~dB on synthetic mud scenes and $0.9$~dB on the real scenes, a uniform $0.01$ costs $3.3$~dB on rain, while the other categories move by at most $0.3$~dB (Sec.~C of the supplementary material). All remaining settings are identical across all experiments on both the synthetic and the real-world dataset.

\subsection{Experimental setup}
\label{sec:setup}

We evaluate on two complementary datasets: a \emph{synthetic benchmark} built on Mip-NeRF~360, where we composite controlled artifacts onto otherwise clean, multi-view-consistent captures (\cref{sec:syntheticdata}), and a \emph{self-captured real-world dataset}, where the artifacts are physically induced (\cref{sec:realdataset}). The synthetic benchmark provides exact, clean ground truth for every corrupted view on an established benchmark dataset. It lets us isolate the effect of each artifact category, while the real-world data tests whether the method transfers to genuine, physically induced obstructions. Both datasets are publicly available~\cite{overdulve2026}. We report three standard reconstruction metrics: Peak Signal-to-Noise Ratio (\textbf{PSNR}), Structural Similarity Index (\textbf{SSIM}), and \textbf{FLIP} \cite{Andersson2020}. We prefer FLIP to the learned LPIPS metric \cite{zhang2018unreasonable} because it is explicitly designed to model the differences an observer perceives when alternating between a render and its reference, without depending on features learned from a particular image distribution. In all experiments, metrics are computed between the held-out clean test views and our clean render $\hat{C}_{3D}$ (\cref{eq:composition}), i.e., the 3D scene rendered without the learned overlay.

\subsection{Synthetic benchmark}
\label{sec:syntheticdata}
We build our synthetic benchmark on the seven publicly released scenes from the \textbf{Mip-NeRF~360} \cite{barron2022mip} dataset, an established benchmark in the NVS research community that provides high-quality, complex, unbounded indoor and outdoor 3D scenes with ground-truth camera poses. Following the established Mip-NeRF protocol, we downscale the images by a factor of $2$ for indoor scenes and $4$ for outdoor scenes and split them into train and test views, using 7/8 images for training and 1/8 for testing. Because these captures are clean and multi-view consistent, we composite synthetic artifacts onto the \emph{training} views, leaving the held-out test views untouched to serve as clean ground truth.

\noindent\textbf{Synthetic defect simulation.}
To strictly control the evaluation environment, we use the \texttt{albumentations} library to generate the synthetic artifacts. We simulate a range of screen-space artifacts:
\begin{enumerate}
    \item \textbf{Watermark:} a semi-transparent logo overlay ($\alpha=0.4$) typical of stock footage.
    \item \textbf{UI Elements:} synthetic buttons/HUD elements drawn on the frame periphery to mimic video-game or screen-recording captures.
    \item \textbf{Dashboard:} a fixed occlusion composited over the lower portion of every frame to mimic dashcam footage.
    \item \textbf{Rain spots / Mud:} environmental contamination of the lens enclosure generated with the \texttt{Spatter} transform in its rain and mud modes, respectively.
\end{enumerate}
To maintain temporal consistency of the artifacts relative to the sensor frame, we generate the augmentation parameters once for the first frame and apply the same artifact to all other training frames. We do not include sensor-level defects such as dead pixels, as they occur at such high signal frequencies that the multi-view averaging in vanilla 3DGS already suppresses them.

\begin{figure*}[htbp]
  \centering
  \setlength{\tabcolsep}{1pt}
  \renewcommand{\arraystretch}{0.5}
  \newcommand{\incgqual}[1]{\raisebox{-0.5\height}{\includegraphics[width=3cm]{media/qualitative2/#1}}}
  \newcommand{\rowlab}[1]{\rotatebox[origin=c]{90}{\scriptsize\emph{#1}}}
  \resizebox{\textwidth}{!}{%
  \begin{tabular}{@{}c cccccccc@{}}
    & Corrupted input & GT (clean) & 3DGS~\cite{kerbl20233d} & DeSplat~\cite{wang2025desplat} & SpotlessSplats~\cite{sabour2025spotlesssplats} & Difix3D+~\cite{wu2025difix3d} & Ours & Ours (artifact) \\
    \rowlab{Watermark} &
      \incgqual{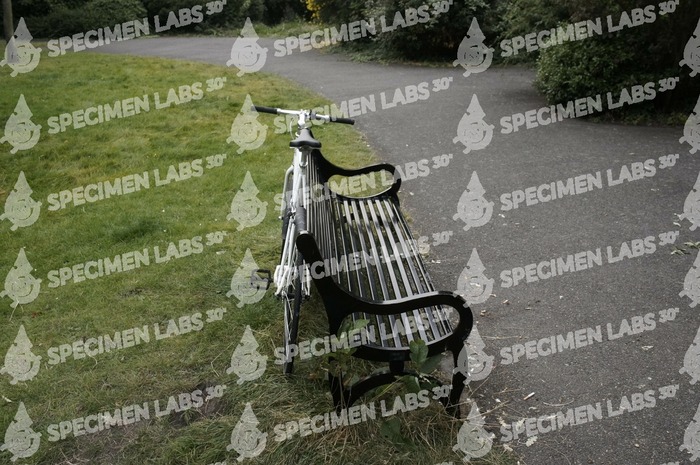} &
      \incgqual{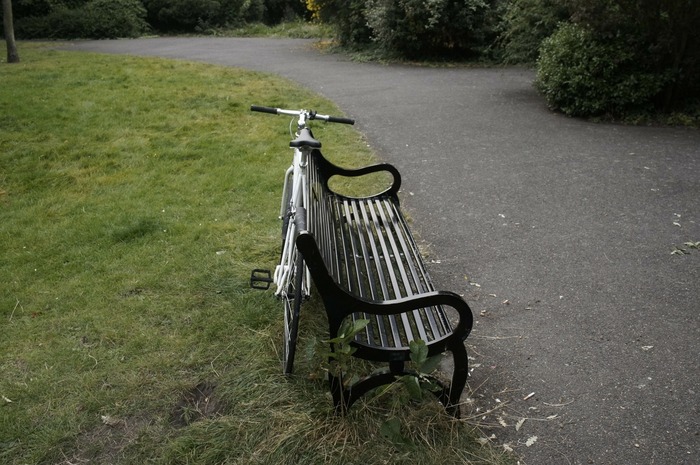} &
      \incgqual{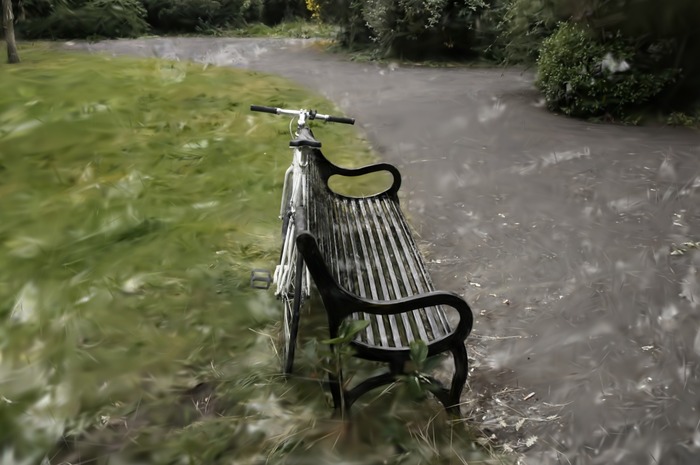} &
      \incgqual{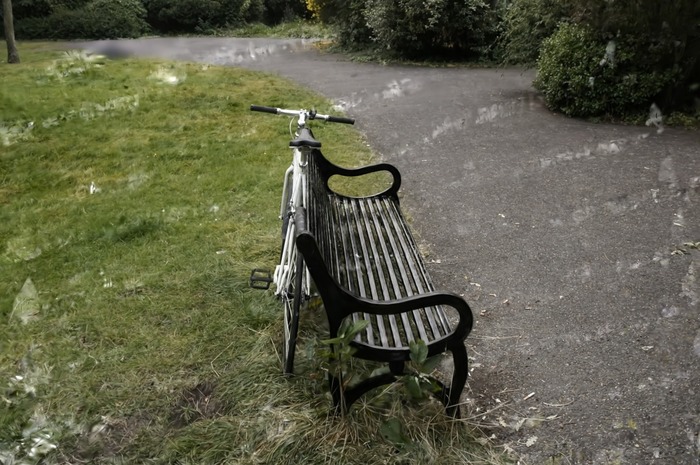} &
      \incgqual{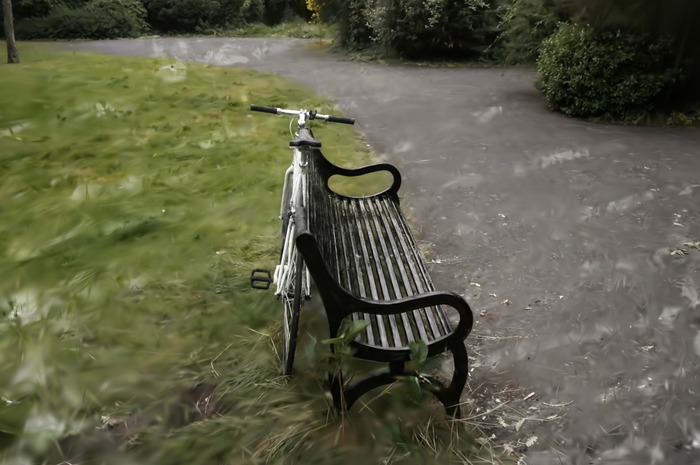} &
      \incgqual{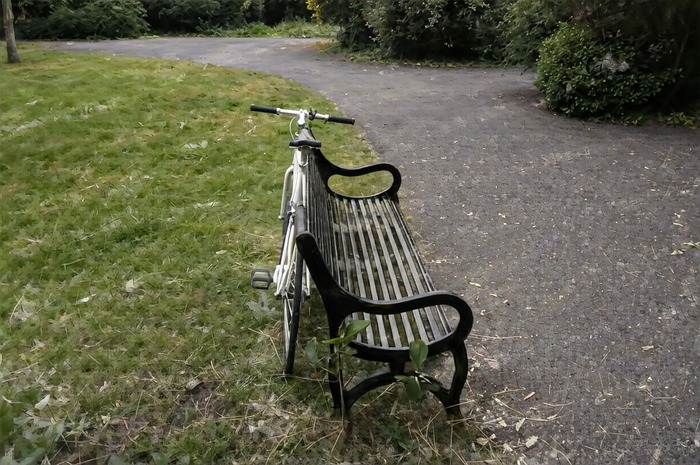} &
      \incgqual{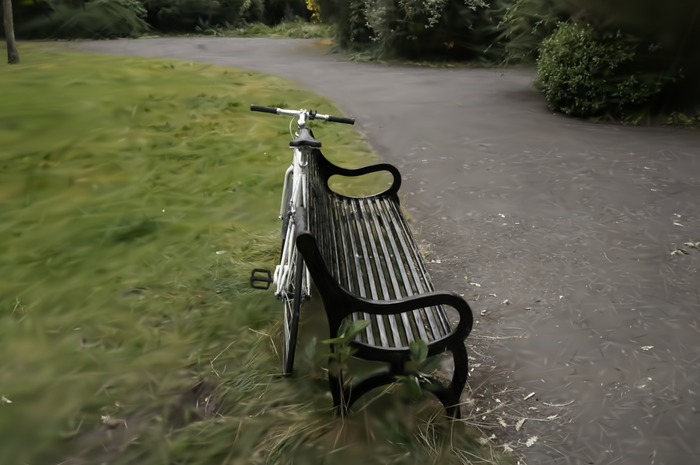} &
      \incgqual{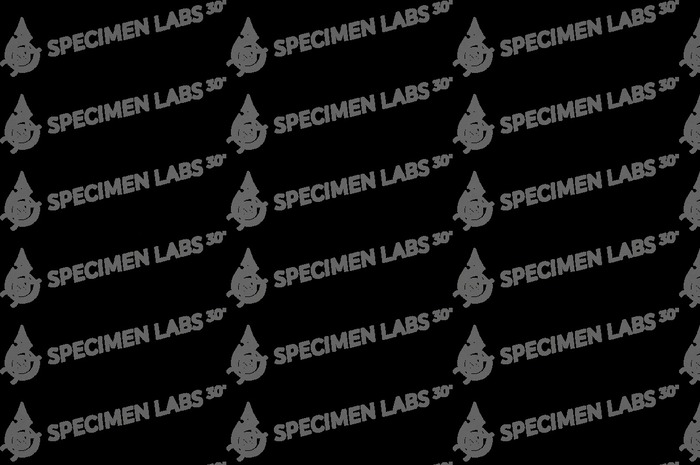} \\  \noalign{\vskip 2\tabcolsep}
    \rowlab{Dashboard} &
      \incgqual{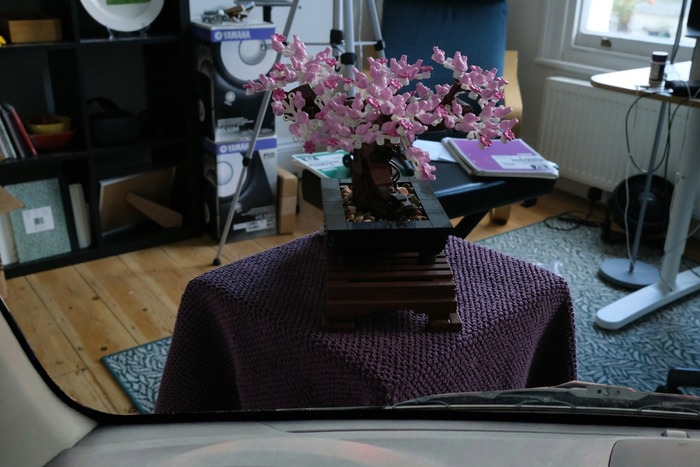} &
      \incgqual{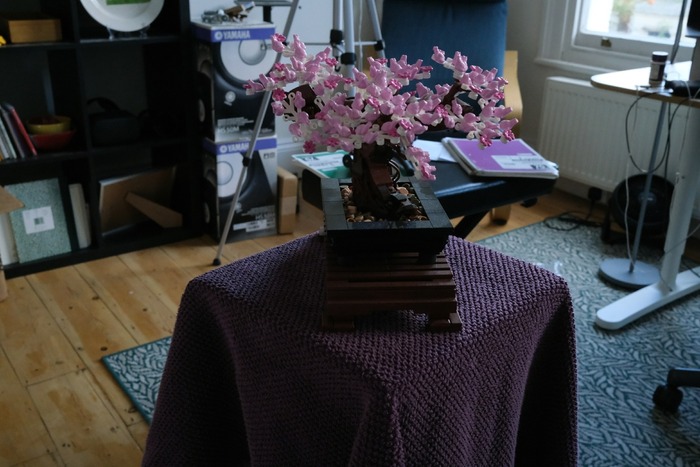} &
      \incgqual{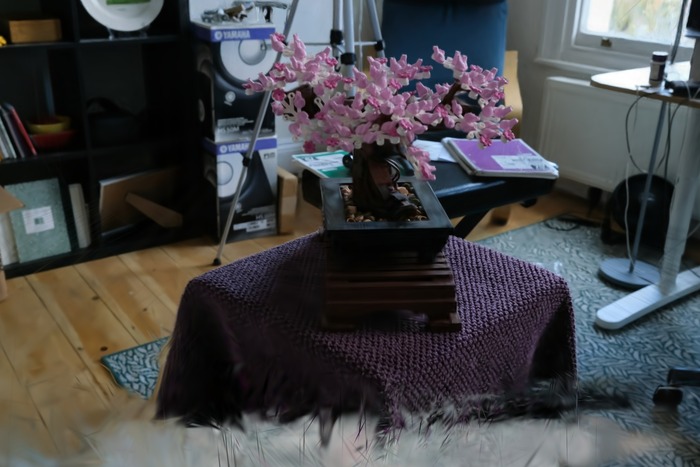} &
      \incgqual{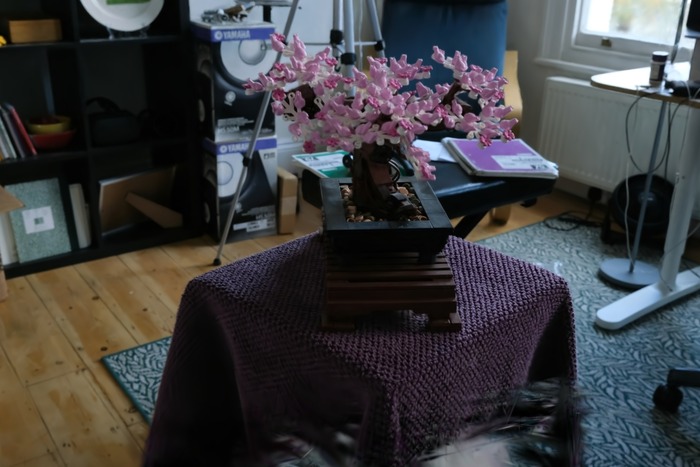} &
      \incgqual{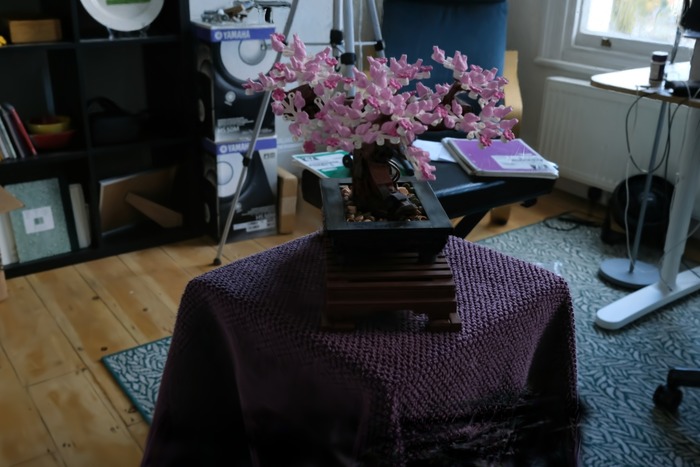} &
      \incgqual{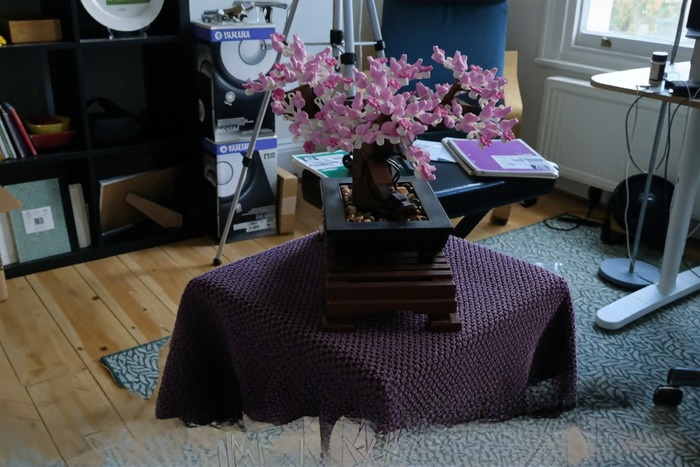} &
      \incgqual{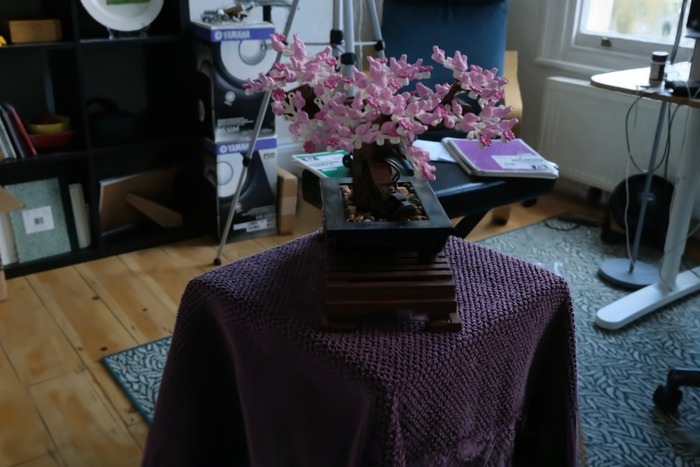} &
      \incgqual{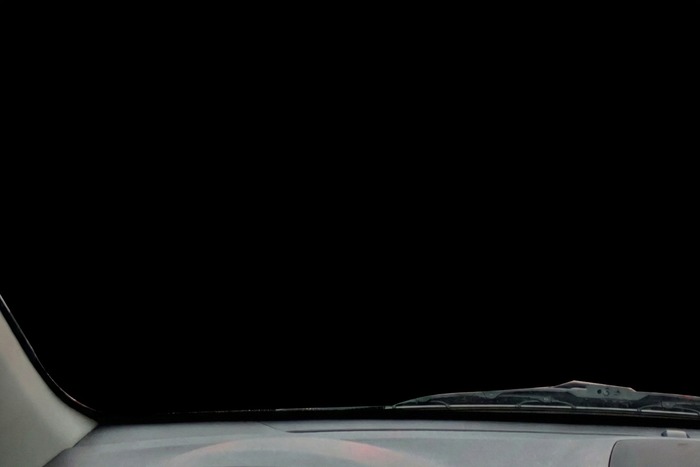} \\  \noalign{\vskip 2\tabcolsep}
    \rowlab{Rain} &
      \incgqual{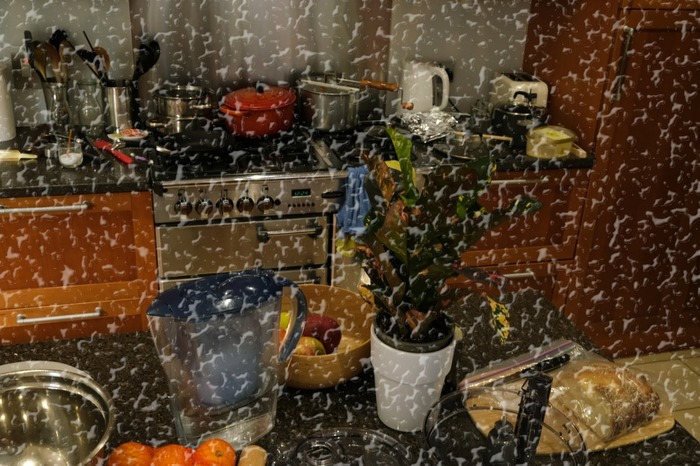} &
      \incgqual{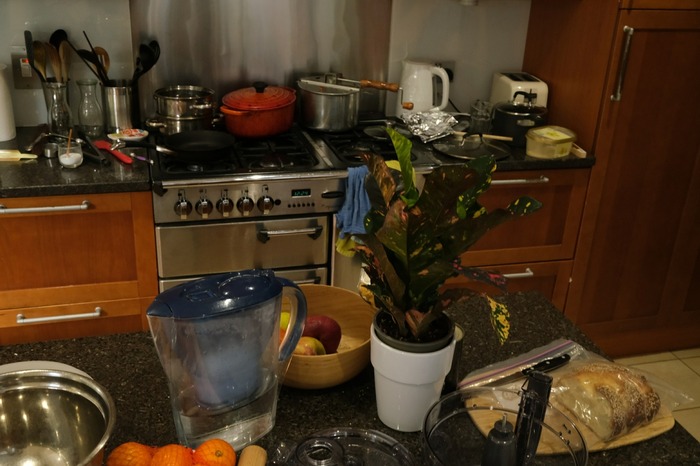} &
      \incgqual{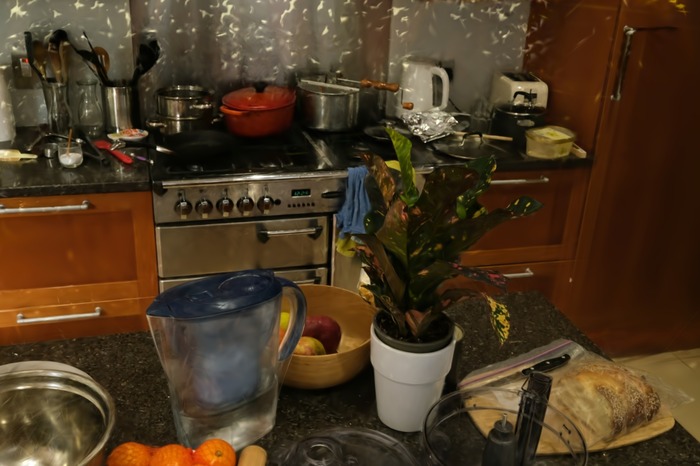} &
      \incgqual{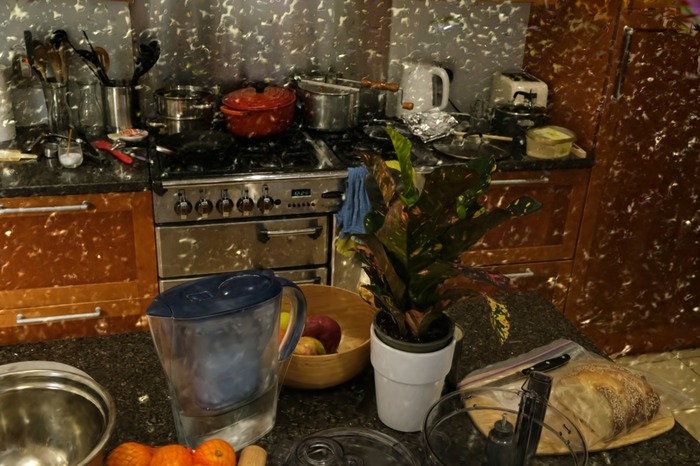} &
      \incgqual{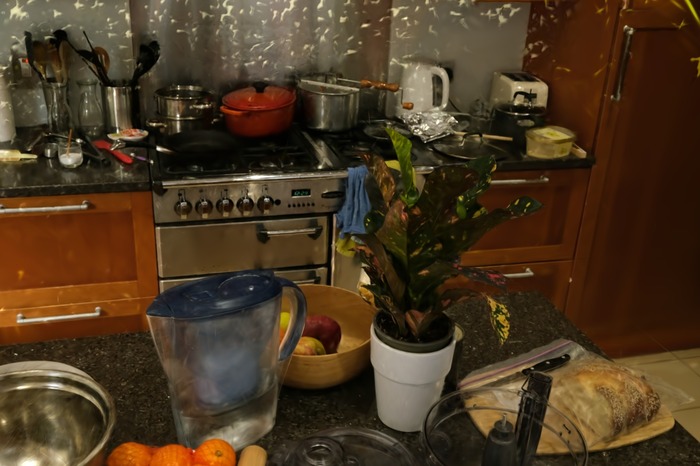} &
      \incgqual{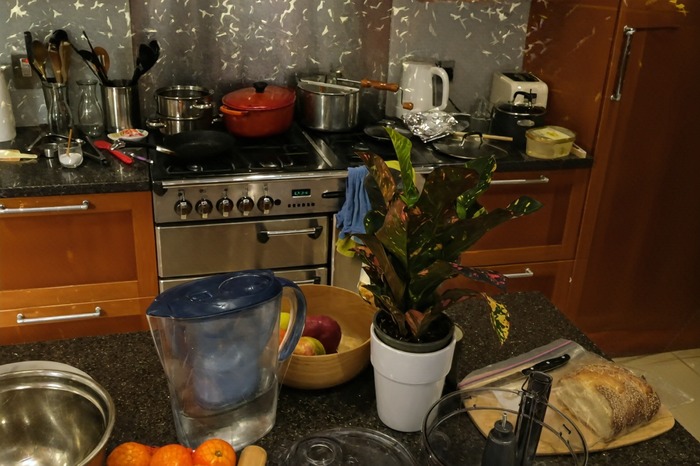} &
      \incgqual{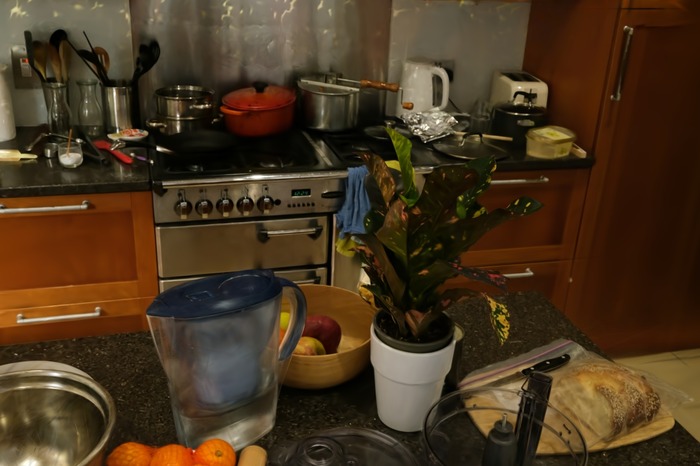} &
      \incgqual{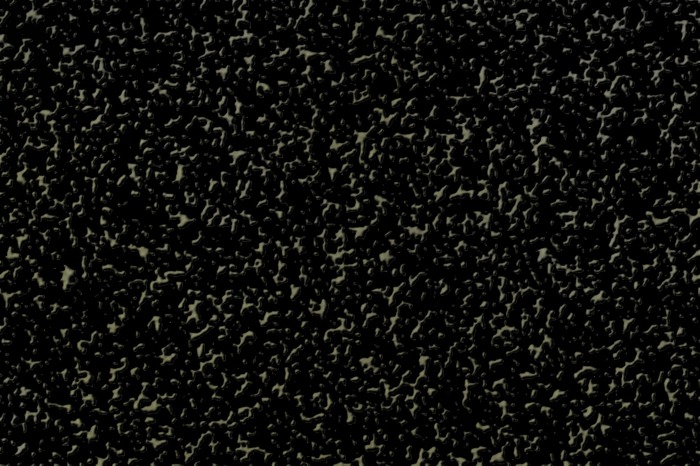} \\  \noalign{\vskip 2\tabcolsep}
    \rowlab{Mud} &
      \incgqual{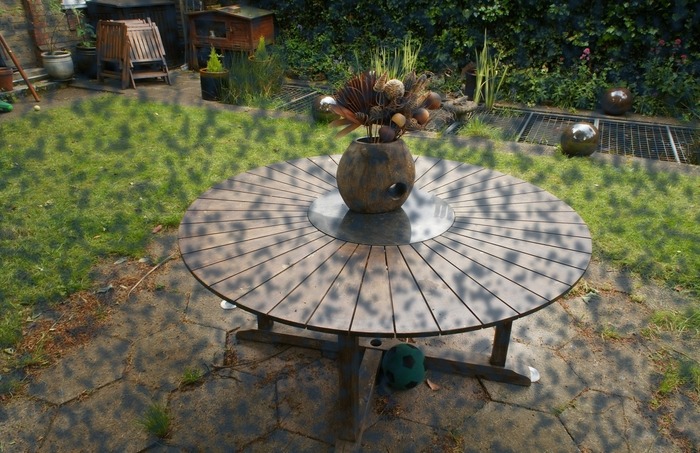} &
      \incgqual{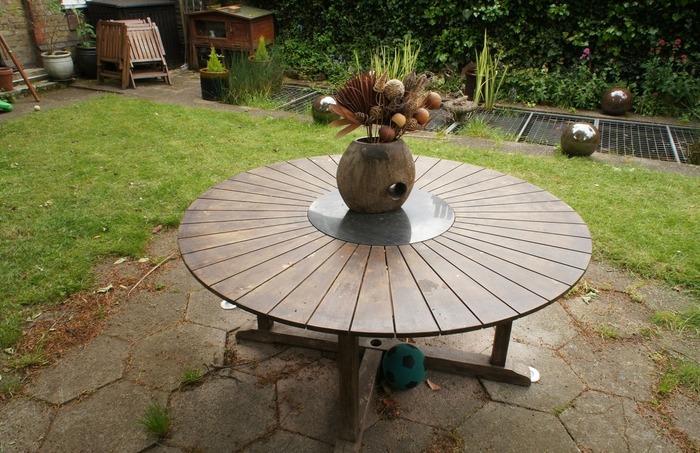} &
      \incgqual{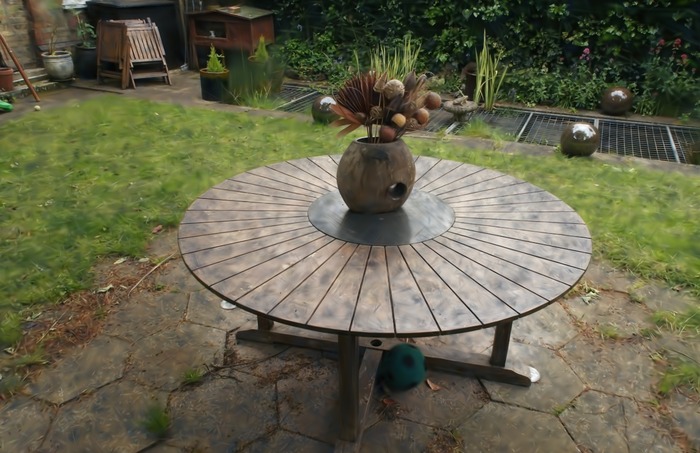} &
      \incgqual{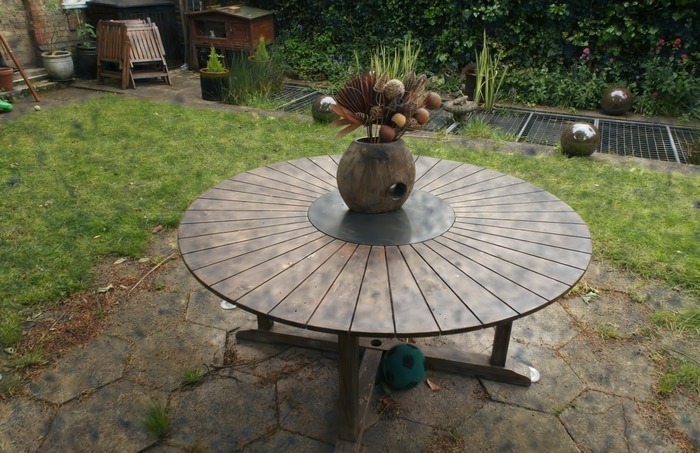} &
      \incgqual{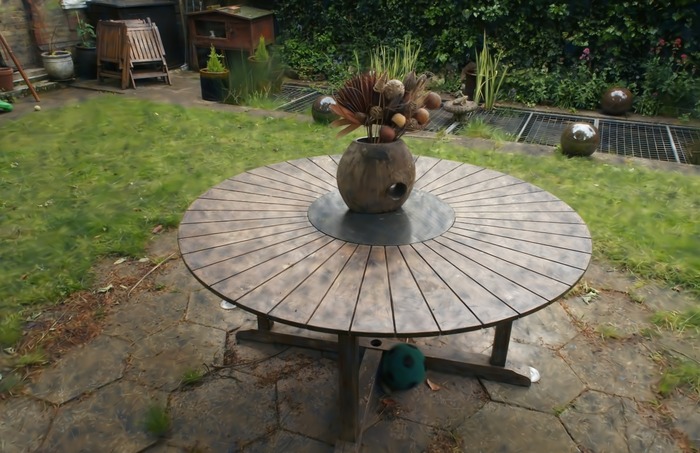} &
      \incgqual{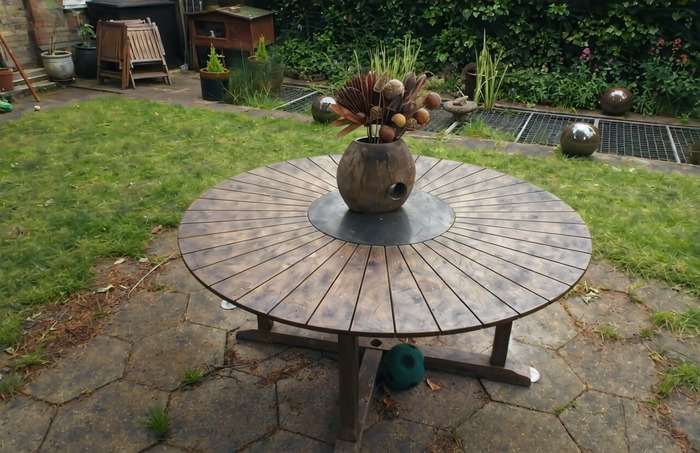} &
      \incgqual{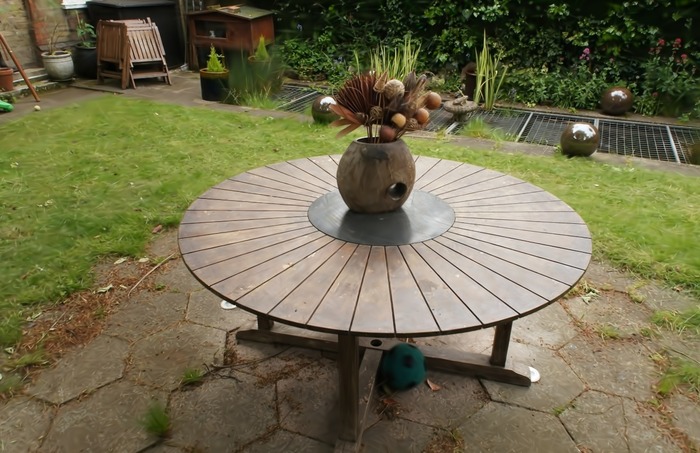} &
      \incgqual{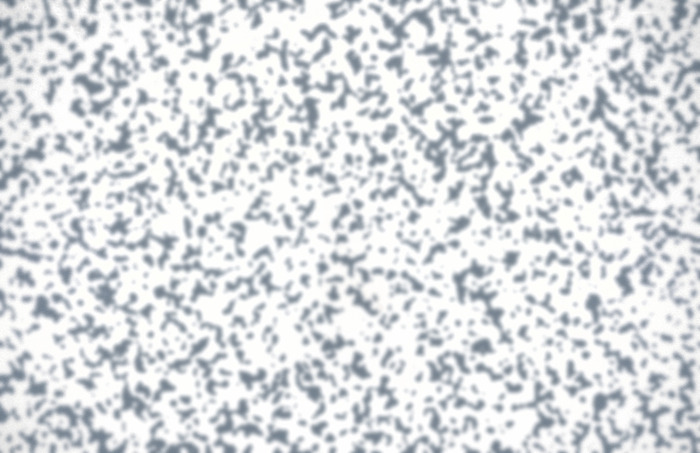} \\
  \end{tabular}%
  }
  \caption{Qualitative results on the synthetic benchmark, one artifact category per row. From left: corrupted training view, clean ground truth, four baselines, our clean reconstruction, and our extracted artifact. Only SSA-3DGS removes the corruption; vanilla 3DGS bakes it in and the other baselines leave it largely in place. Best viewed zoomed in.}
  \label{fig:qualitative}
\end{figure*}

\subsection{Real-world dataset}
\label{sec:realdataset}

\begin{figure}[tbhp]
  \centering
  \setlength{\tabcolsep}{1.5pt}
  \footnotesize
  \begin{tabular}{@{}l@{\,}cc@{\hspace{3pt}}cc@{}}
    & \multicolumn{2}{c}{\textbf{Training} (with artifact)}
    & \multicolumn{2}{c}{\textbf{Clean test} (GT)} \\
    \raisebox{-0.5\height}{\rotatebox{90}{\emph{Finger}}}
      & \dscell{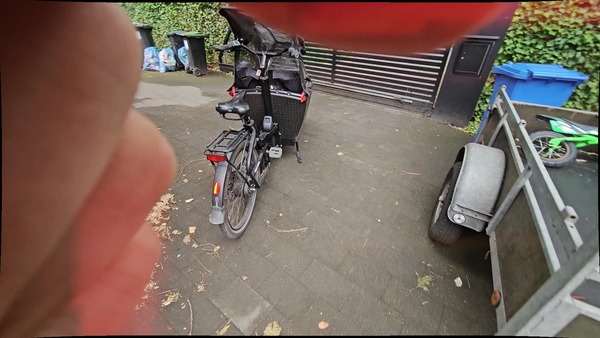} & \dscell{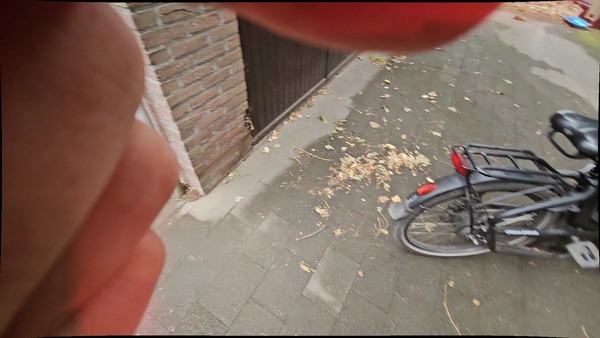}
      & \dscell{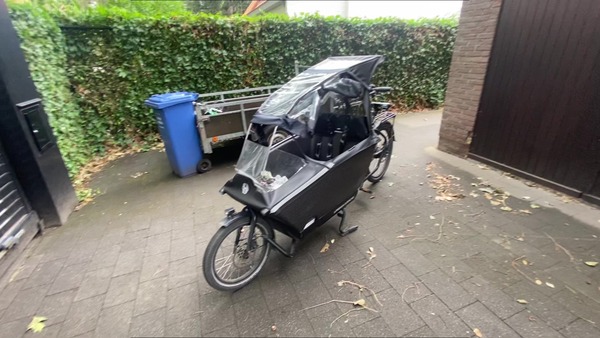}  & \dscell{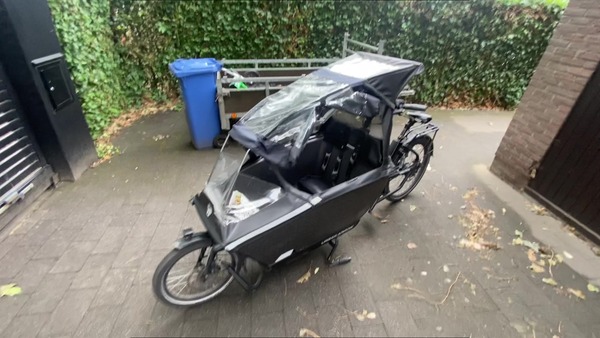}  \\  \noalign{\vskip 2\tabcolsep}
    \raisebox{-0.5\height}{\rotatebox{90}{\emph{Mud}}}
      & \dscell{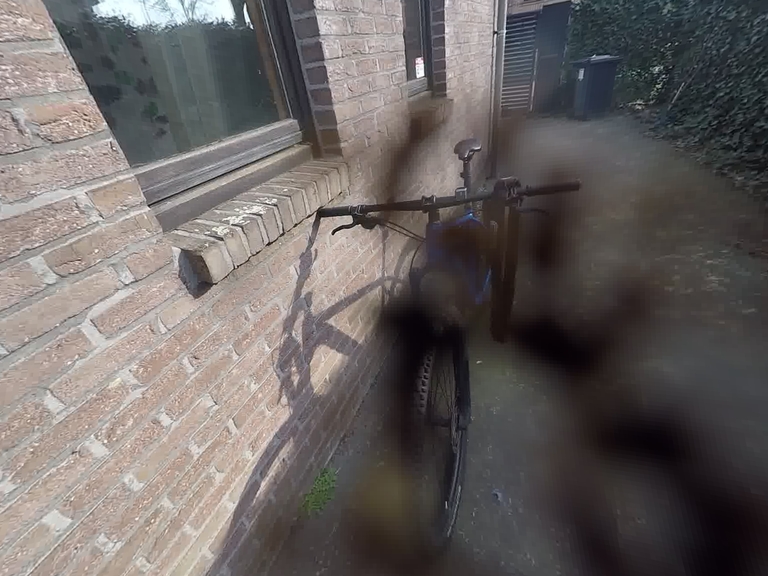} & \dscell{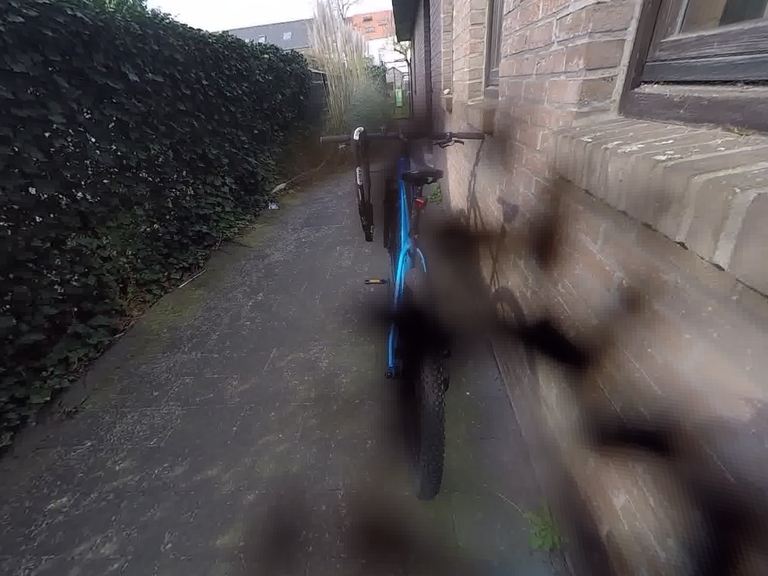}
      & \dscell{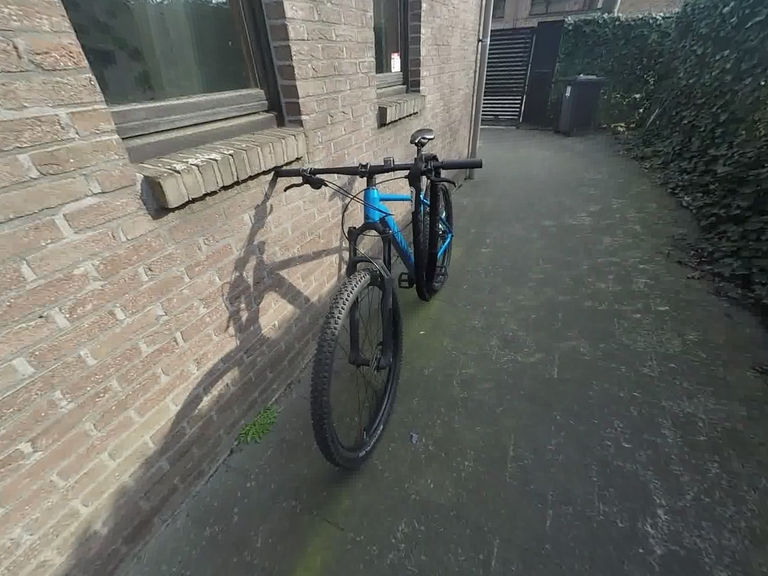}  & \dscell{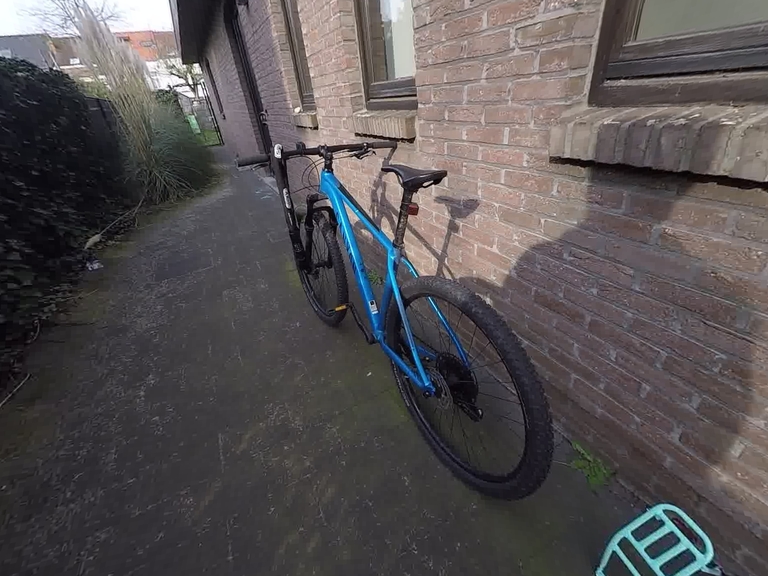}  \\  \noalign{\vskip 2\tabcolsep}
    \raisebox{-0.5\height}{\rotatebox{90}{\emph{Occl.}}}
      & \dscell{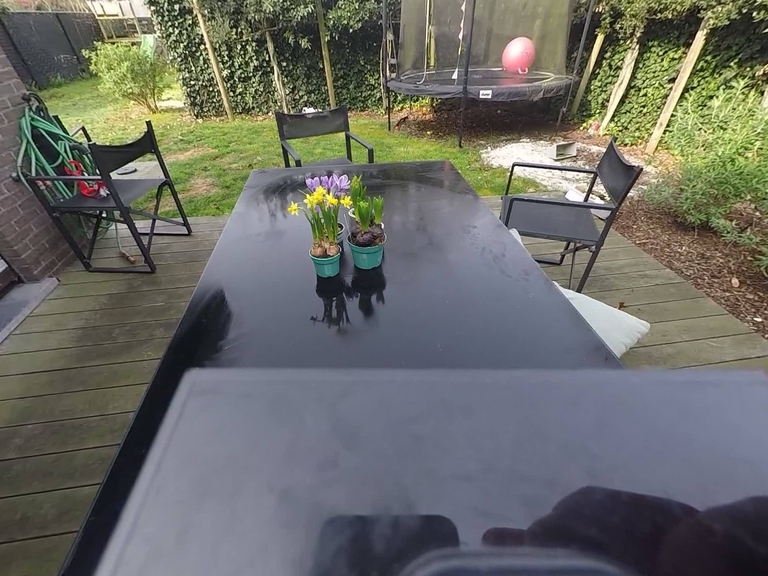} & \dscell{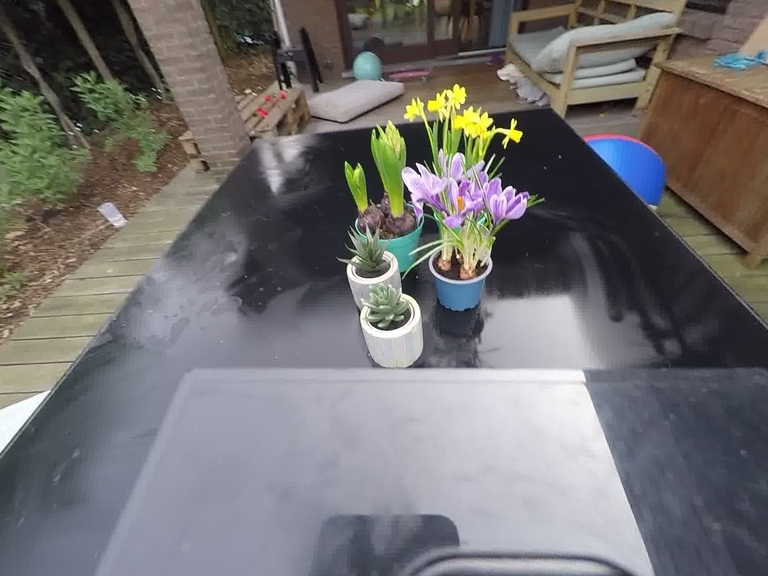}
      & \dscell{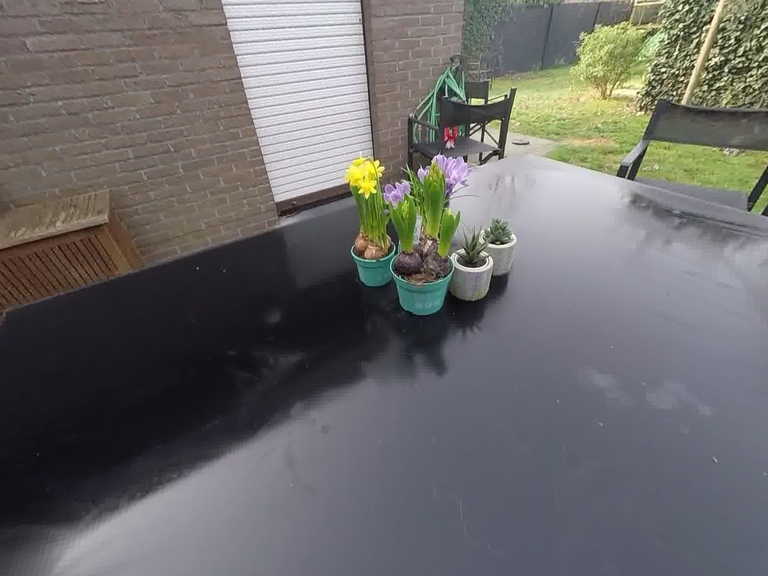}  & \dscell{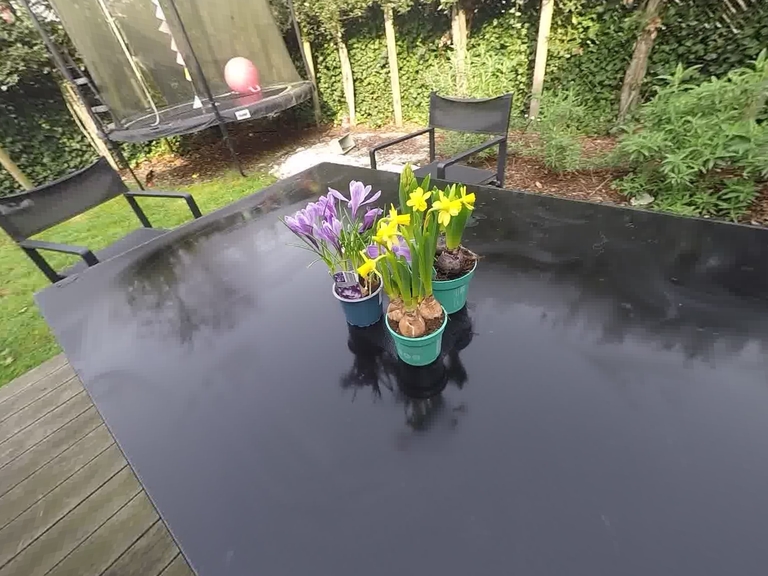}  \\
  \end{tabular}
  \caption{Our real-world dataset: three categories of physically induced screen-space artifacts. Left: corrupted training views; right: separately captured clean test views, registered into the same coordinate frame.}
  \label{fig:dataset}
\end{figure}

To demonstrate that the method transfers to genuine, physically induced artifacts, we additionally captured a real-world dataset of fifteen scenes spanning three categories of screen-space defects: finger obstructions, mud, and occlusions (\cref{fig:dataset}). The \emph{Finger} category is a near-lens obstruction in which the operator's finger appears as a large, heavily defocused, near-opaque blob anchored to the periphery of the sensor frame. \emph{Mud} depicts mud spatters on a camera's transparent enclosure. The \emph{Occlusion} category is a \emph{near}-static artifact where part of the image is occluded by external objects, e.g., a part of the camera rig. Similar to the finger occlusion, such occlusions reflect the surrounding environment and shift slightly from view to view, deliberately evaluating whether our static-overlay model degrades gracefully when its assumptions are not entirely met.

For every scene we record a \emph{training} pass in which the artifact is present, and a separate \emph{test} pass of the same scene with the artifact removed as clean ground truth. The two passes are registered into a single, shared coordinate system using Structure-from-Motion~\cite{colmap}. All corrupted frames are used for training, and all clean views are used for testing. The captures span three acquisition devices: a GoPro~Hero~4, a Samsung~Galaxy~S22, and an iPhone~11~Pro.

\subsection{Baselines}
\label{sec:baselines}
We compare SSA-3DGS against the following baselines:
\begin{itemize}
  \item \textbf{Baseline 3DGS:} vanilla 3DGS \cite{kerbl20233d} trained directly on the corrupted images, representing the behavior of current reconstruction pipelines when fed imperfect data.
  \item \textbf{DeSplat \cite{wang2025desplat}:} a 3DGS variant that separates near-camera ``distractor'' Gaussians from scene Gaussians.
  \item \textbf{SpotlessSplats \cite{sabour2025spotlesssplats}:} masks transient content during optimization using semantic features and a robust loss.
  \item \textbf{RobustNeRF \cite{sabour2023robustnerf}:} a robust loss that down-weighs pixels violating multi-view consistency (i.e., SpotlessSplats without semantic masking).
  \item \textbf{Difix3D+ \cite{wu2025difix3d}:} a single-step diffusion model that removes artifacts from images. Unlike the other baselines, it is a \emph{generative restoration} method that operates as a render-time enhancer for vanilla 3DGS outputs.
\end{itemize}

We additionally report \textbf{Clean (Ref)} on the synthetic dataset---3DGS trained on the artifact-free images---as an upper bound on achievable reconstruction quality. We defer the analysis of the \emph{image-level restoration} model FLUX.1-Kontext to the supplementary material (Sec.~A) due to its extensive inference runtime and failure to remove near-opaque occluders.

\begin{figure*}[htbp]
  \centering
  \setlength{\tabcolsep}{1pt}
  \renewcommand{\arraystretch}{0.5}
  \newcommand{\incg}[1]{\raisebox{-0.5\height}{\includegraphics[width=0.135\textwidth]{media/qualitative_real2/#1}}}   
  \newcommand{\incgp}[1]{\raisebox{-0.5\height}{\includegraphics[width=0.1\textwidth]{media/qualitative_real2/#1}}}  
  \newcommand{\rowlab}[1]{\rotatebox[origin=c]{90}{\footnotesize\emph{#1}}}
  \footnotesize
  \begin{tabular}{@{}c ccccccc@{}}
    & Input & 3DGS~\cite{kerbl20233d} & DeSplat~\cite{wang2025desplat} & SpotlessSplats~\cite{sabour2025spotlesssplats} & Difix3D+~\cite{wu2025difix3d} & Ours & Ours (artifact) \\
    \rowlab{Mud} &
      \incg{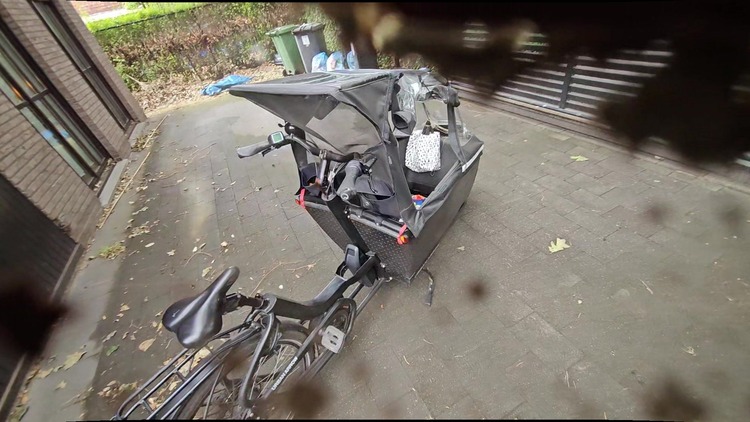} &
      \incg{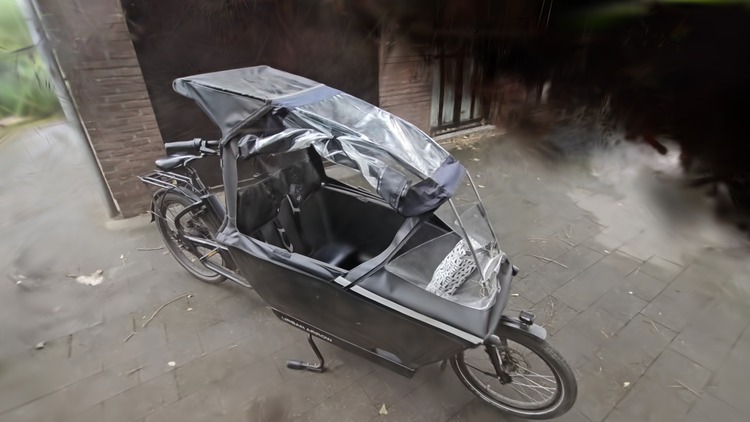} &
      \incg{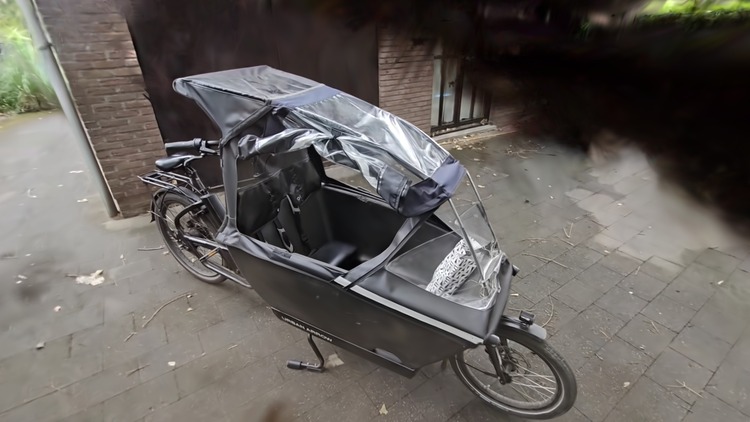} &
      \incg{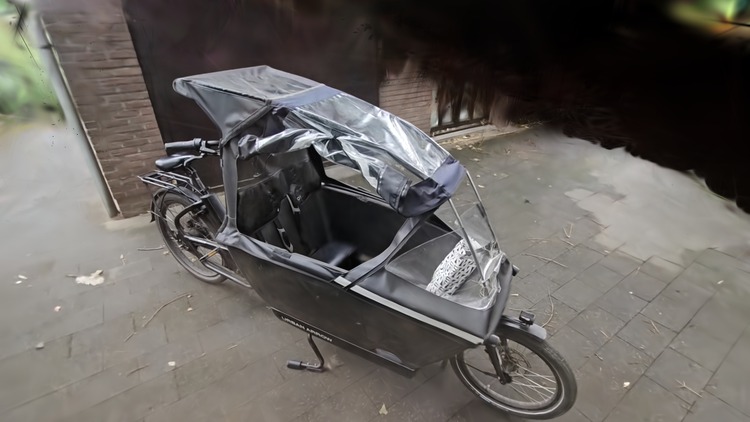} &
      \incg{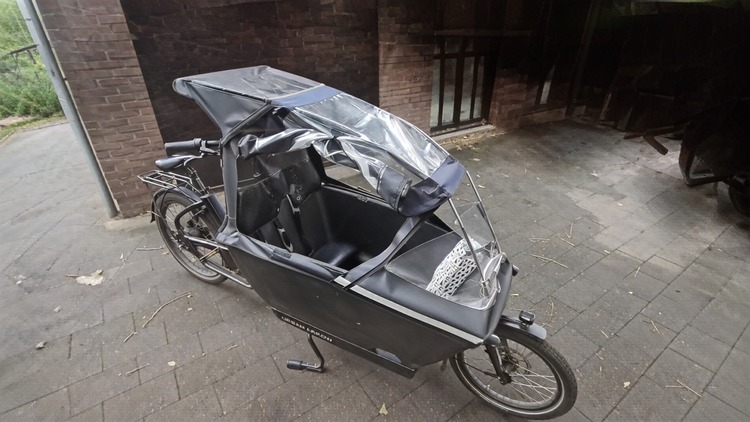} &
      \incg{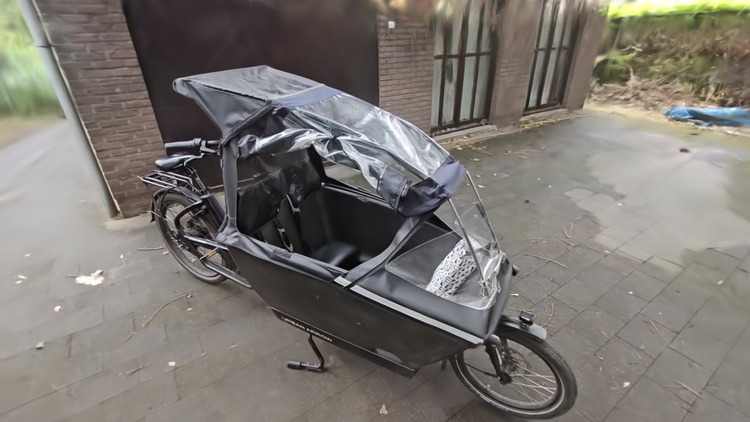} &
      \incg{qreal2_mud_mud_cargobike__val_step29999_0223__artifact.jpg} \\[20pt]
    \rowlab{Mud} &
      \incg{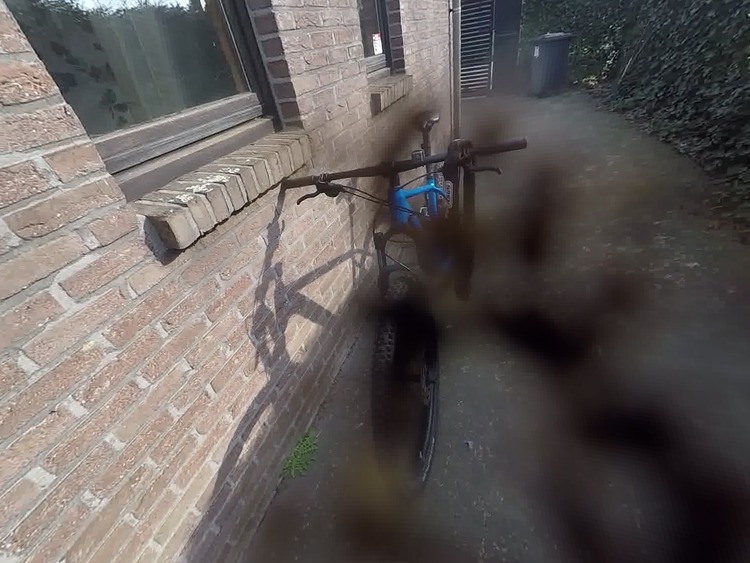} &
      \incg{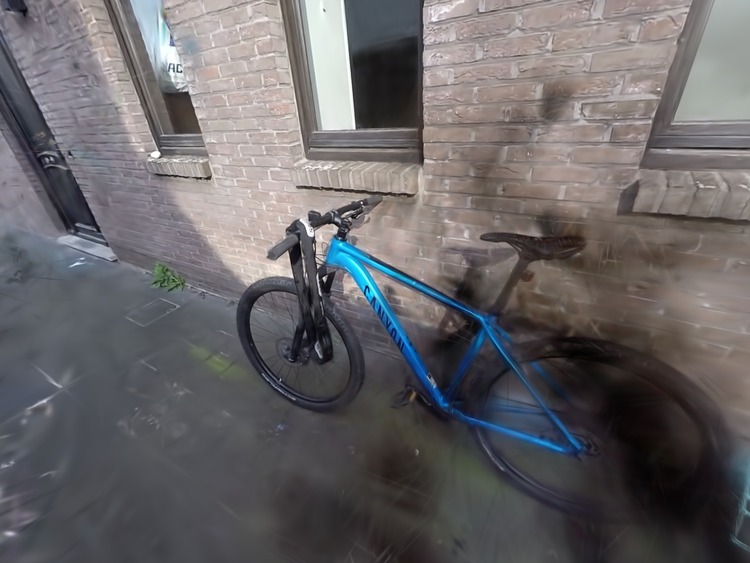} &
      \incg{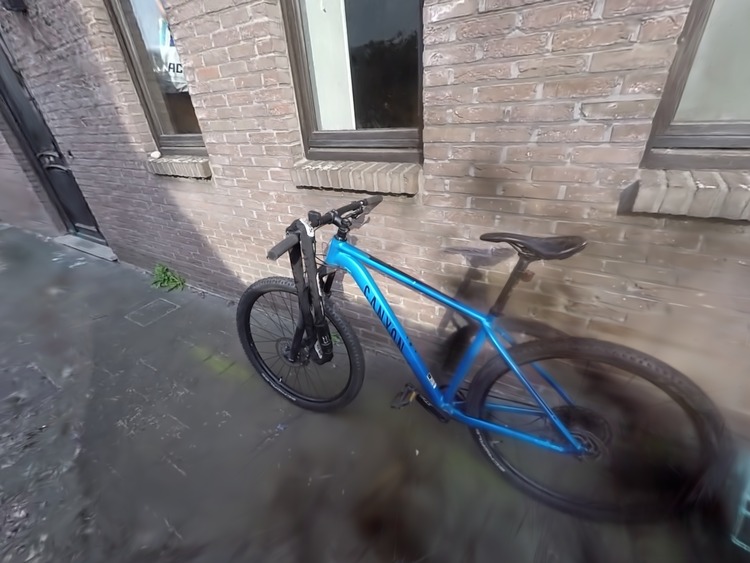} &
      \incg{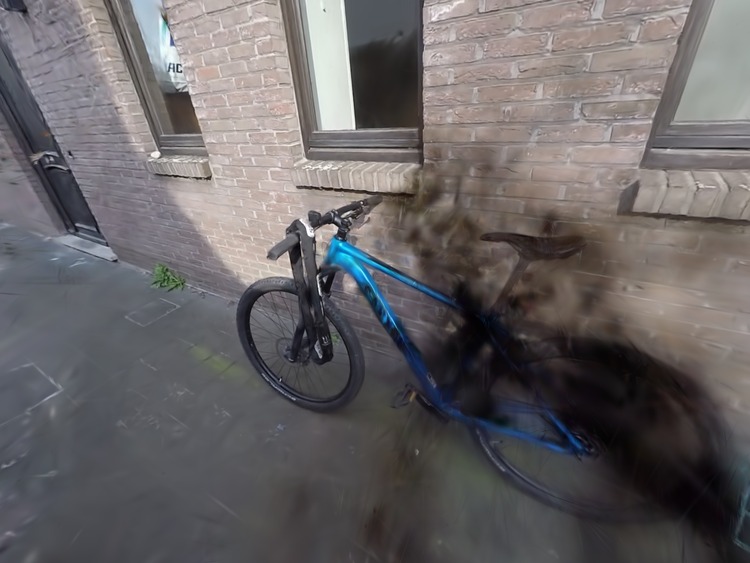} &
      \incg{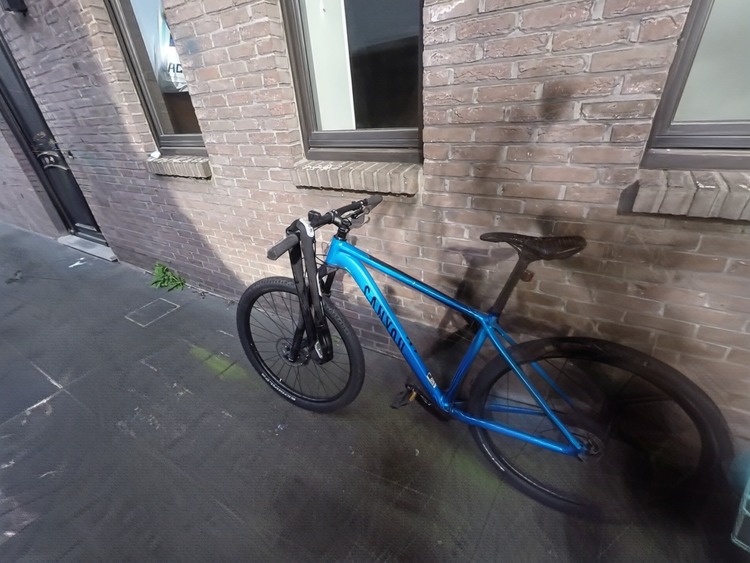} &
      \incg{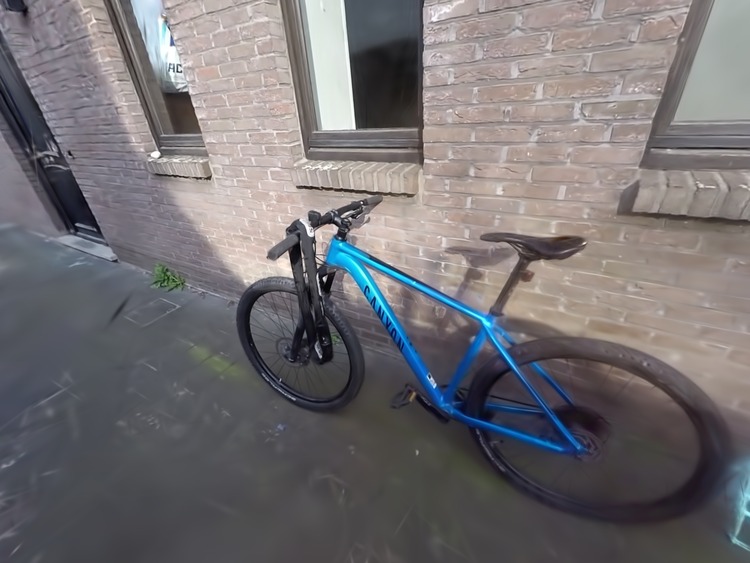} &
      \incg{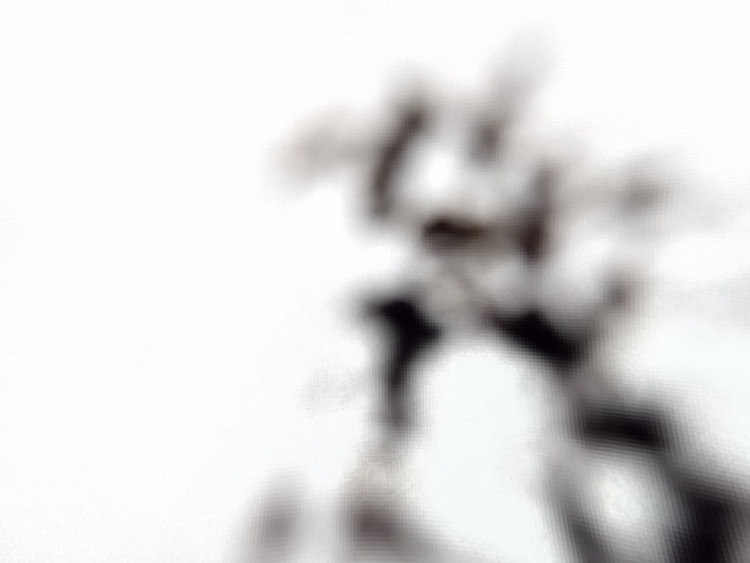} \\[26pt]
    \rowlab{Finger} &
      \incgp{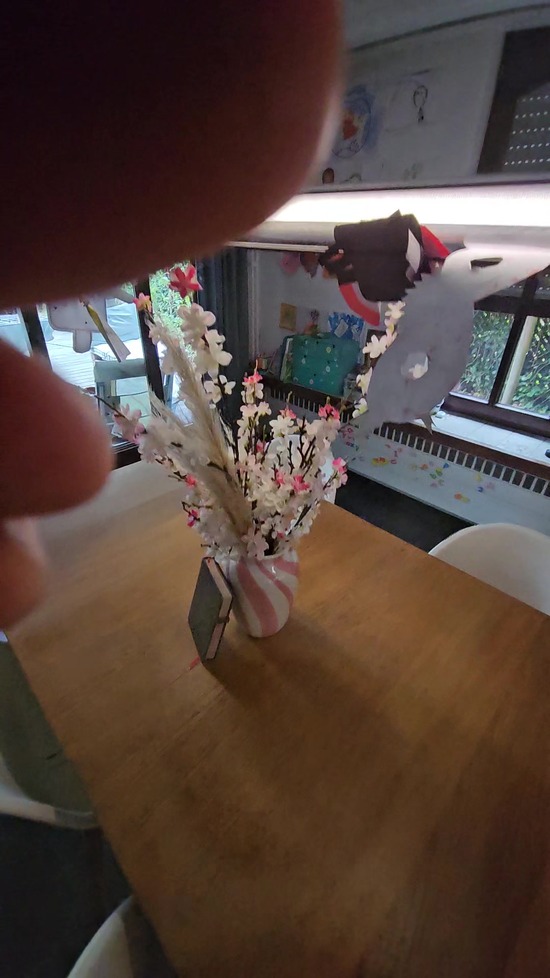} &
      \incgp{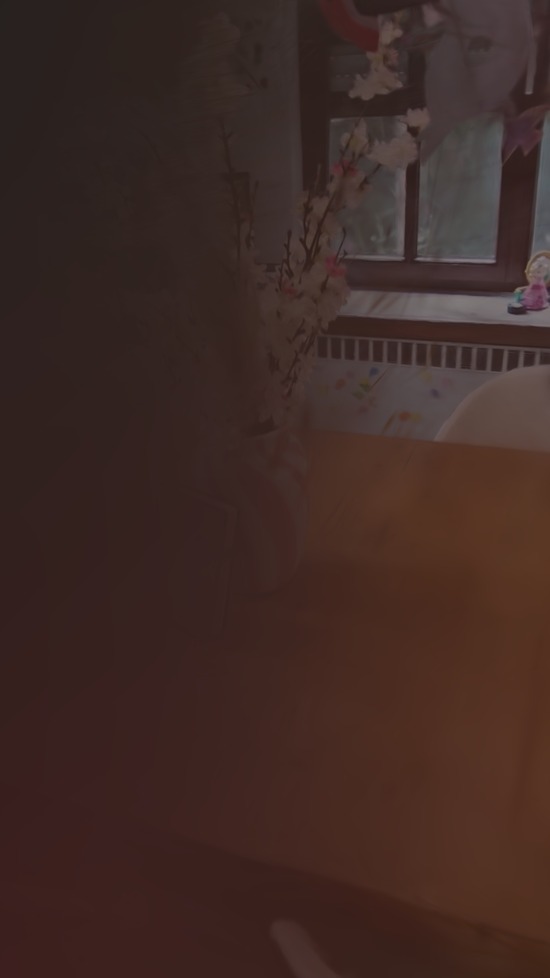} &
      \incgp{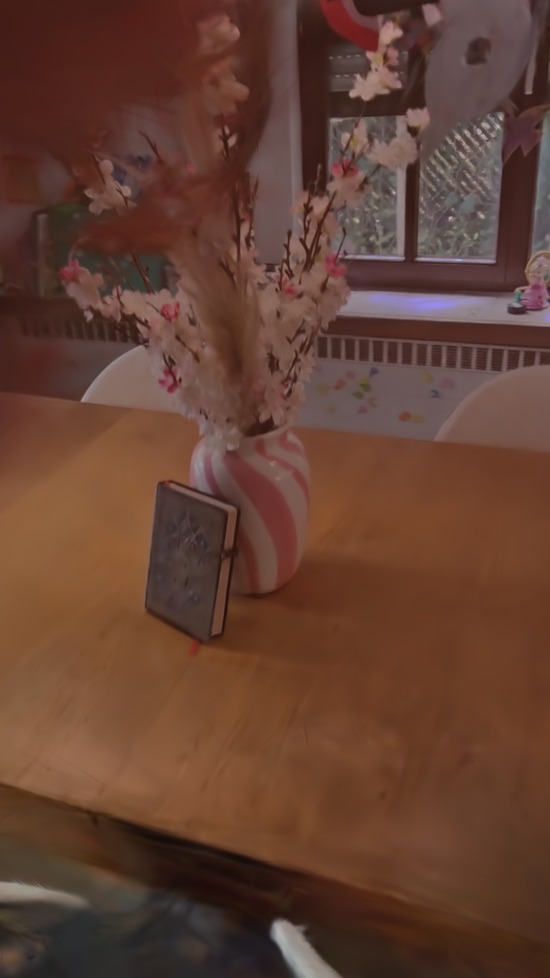} &
      \incgp{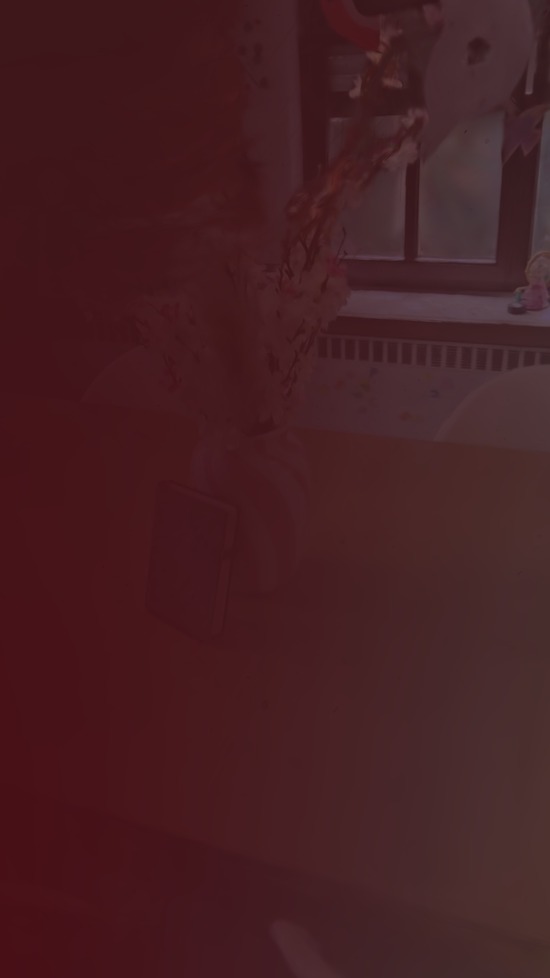} &
      \incgp{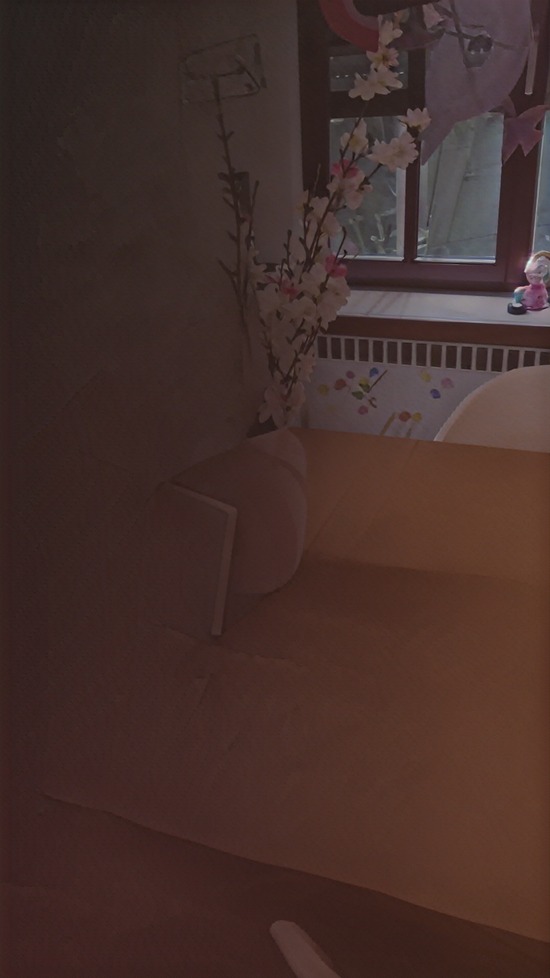} &
      \incgp{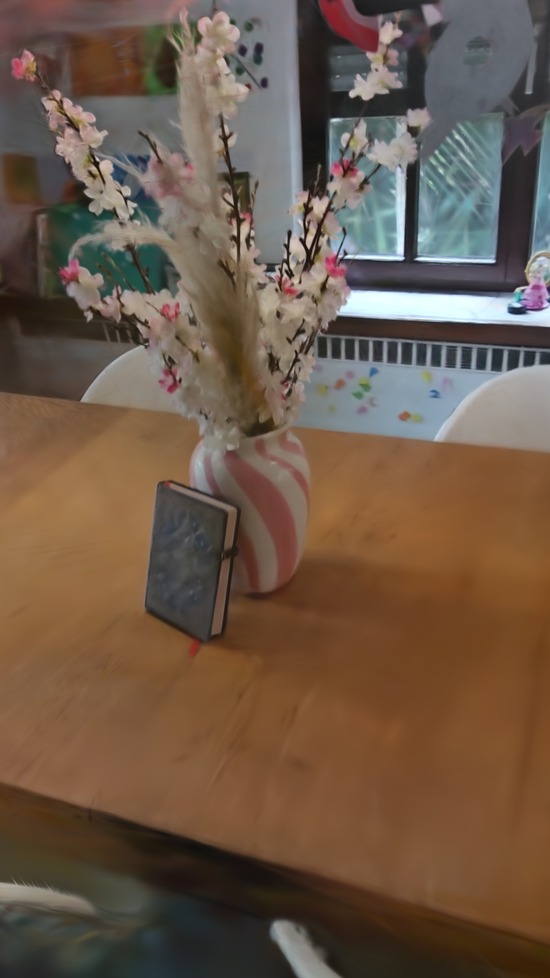} &
      \incgp{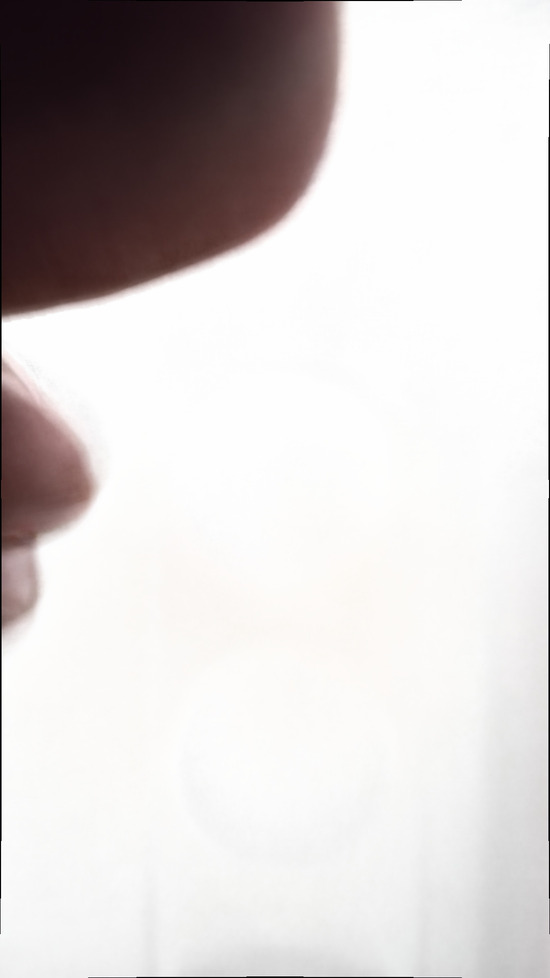} \\[45pt]
    \rowlab{Mud} &
      \incgp{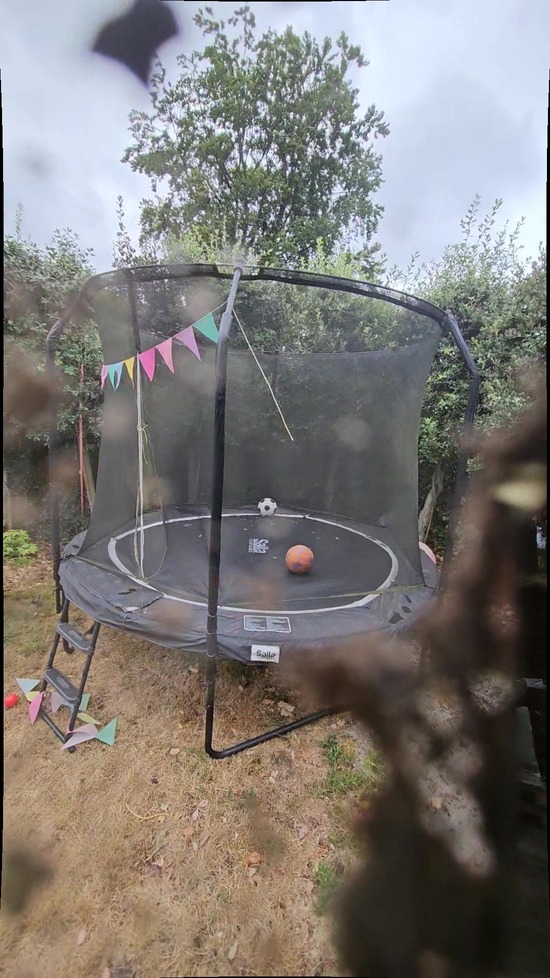} &
      \incgp{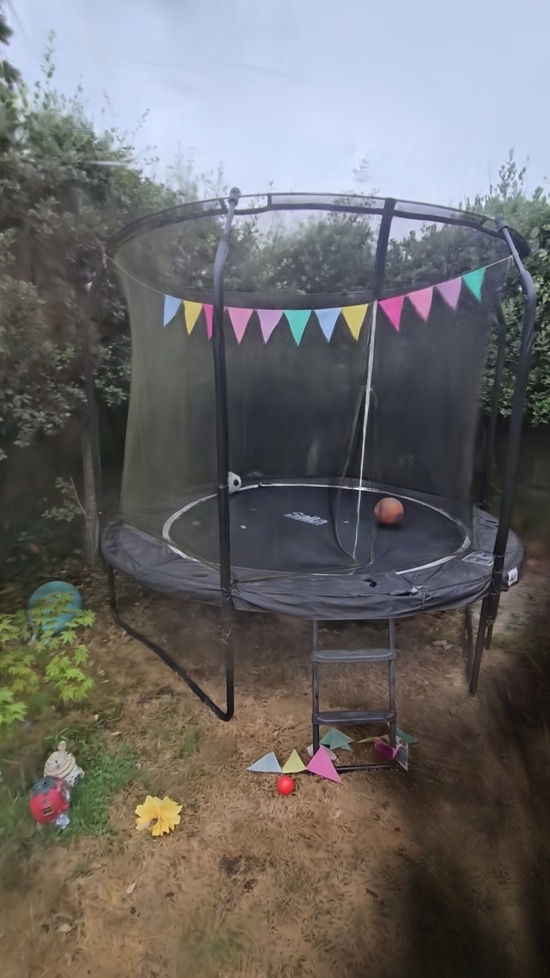} &
      \incgp{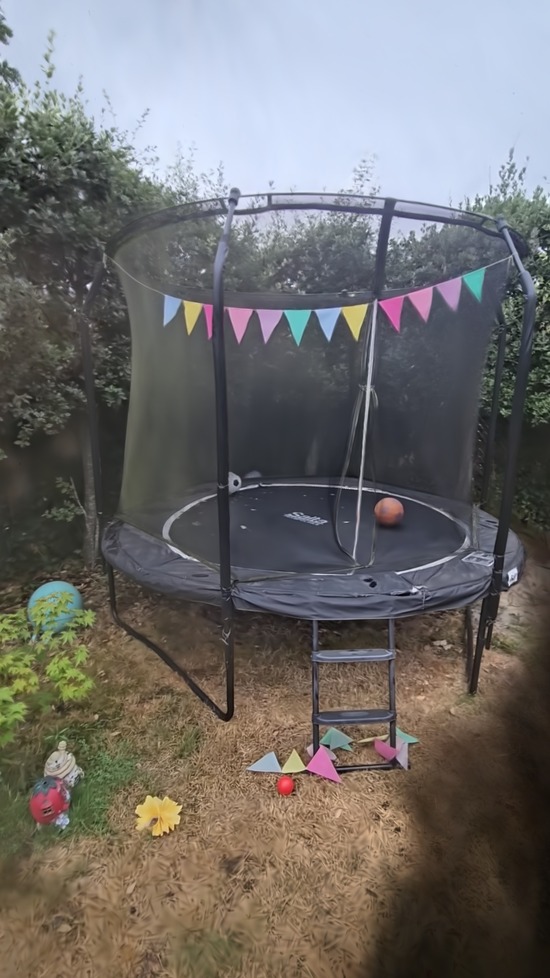} &
      \incgp{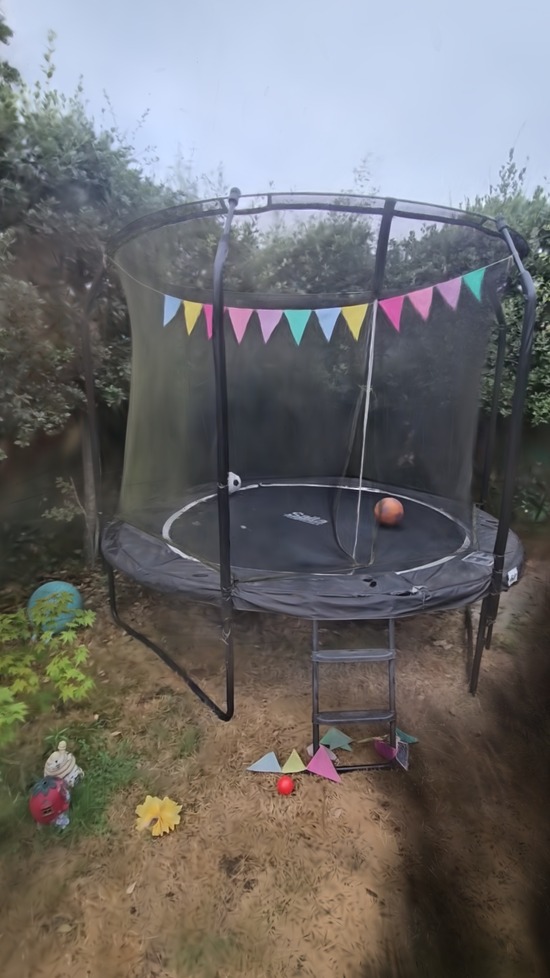} &
      \incgp{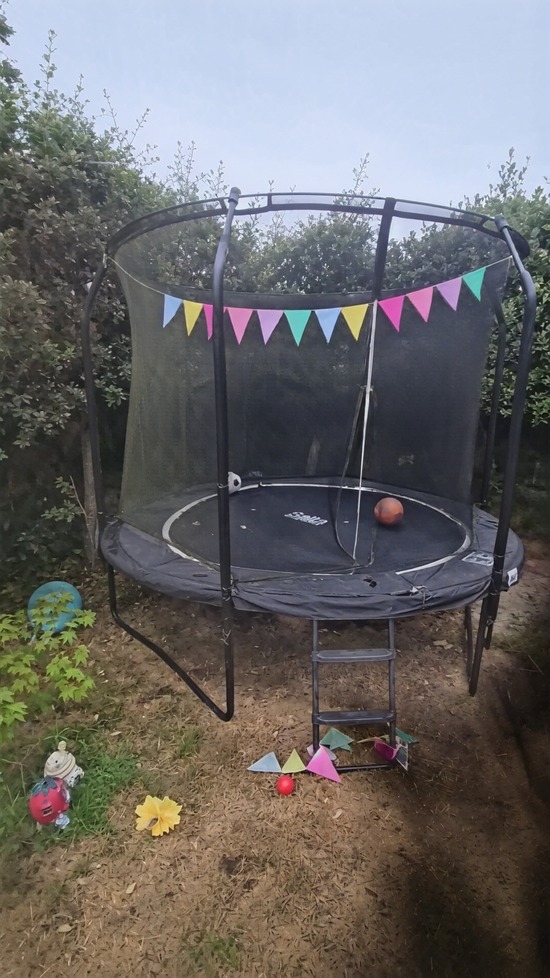} &
      \incgp{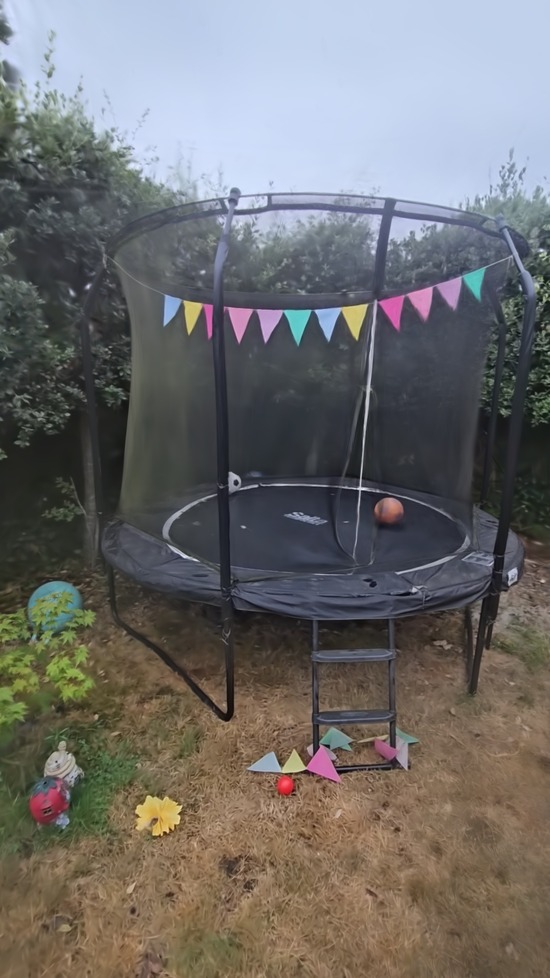} &
      \incgp{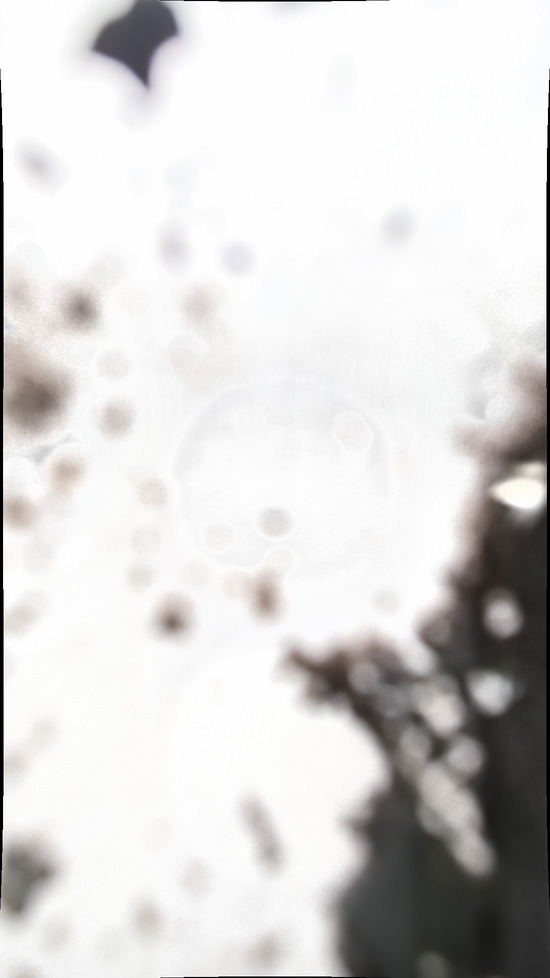} \\[45pt]
    \rowlab{Occl.} &
      \incgp{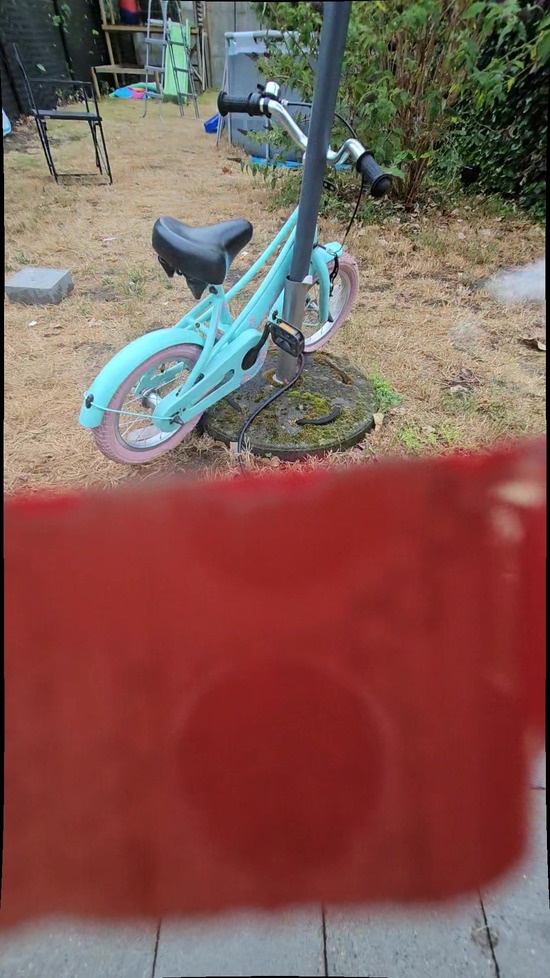} &
      \incgp{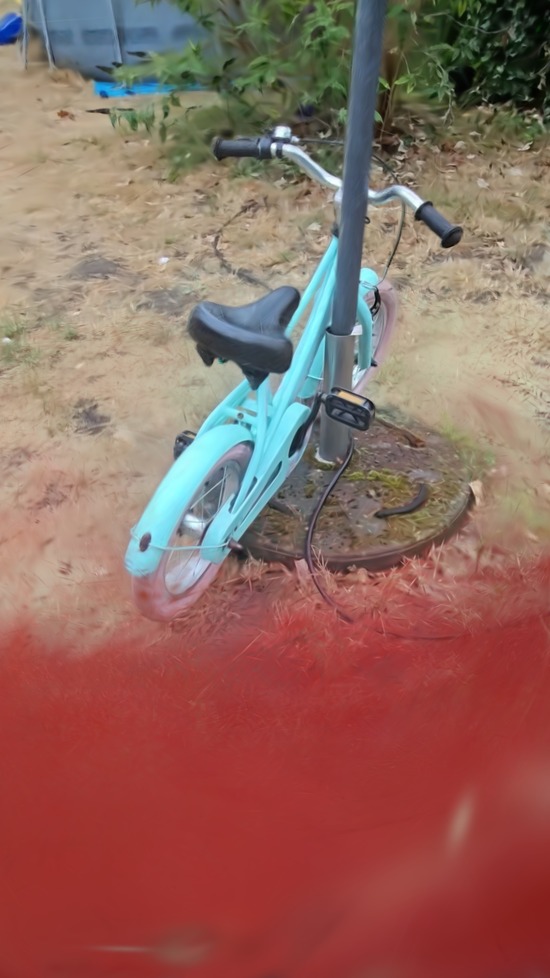} &
      \incgp{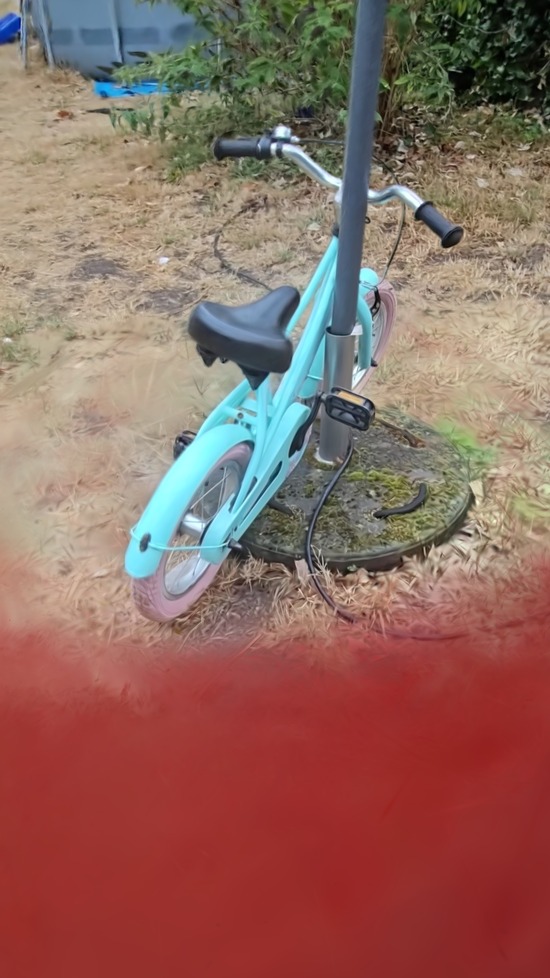} &
      \incgp{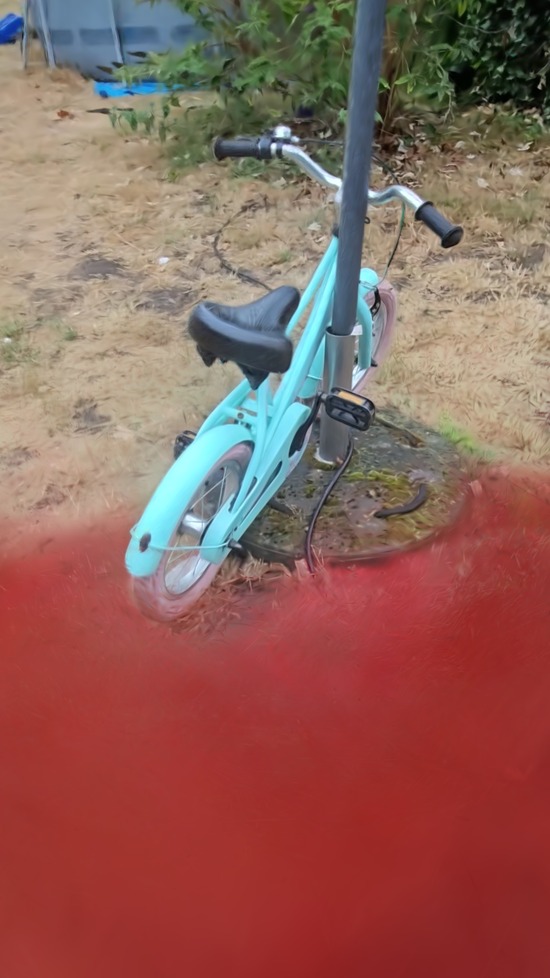} &
      \incgp{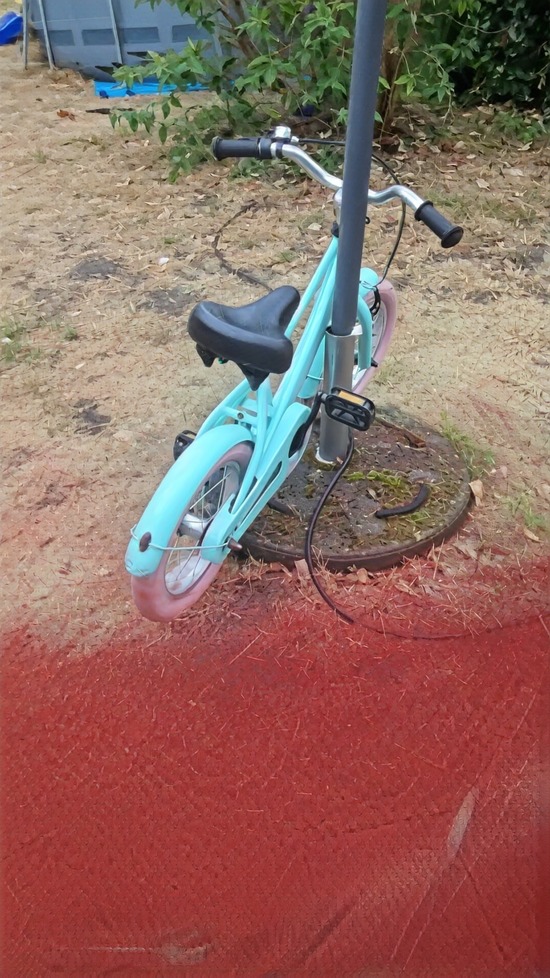} &
      \incgp{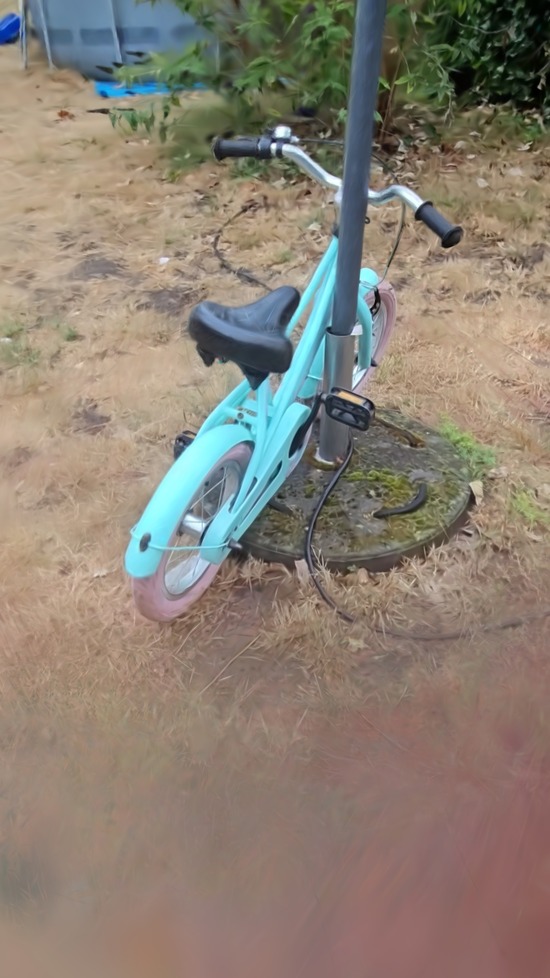} &
      \incgp{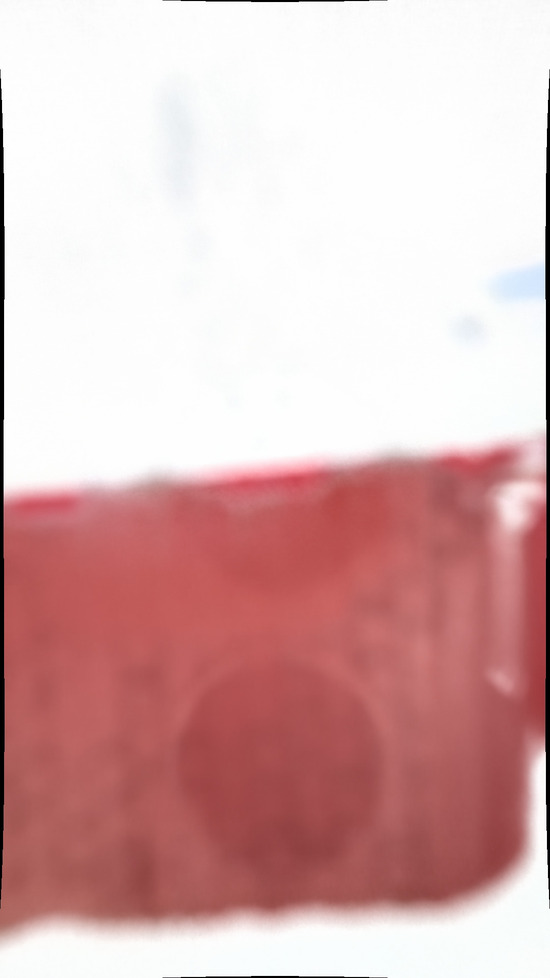} \\
  \end{tabular}
  \caption{Qualitative results on our real-world dataset (\emph{Finger}, \emph{Mud}, \emph{Occlusion}). From left: a corrupted training view, four baselines, our reconstruction, and our extracted artifact. SSA-3DGS largely removes the physical artifact; the baselines leave it in place.}
  \label{fig:qualitative_real}
\end{figure*}

\subsection{Quantitative results}
\label{sec:quantitative}

\begin{table*}[htbp]
\centering
\caption{Results on the synthetic Mip-NeRF~360 benchmark (mean over the seven scenes). 3DGS and Ours report \emph{ADC\,/\,MCMC}; the other baselines use ADC, and Difix3D+ is applied post-hoc to the 3DGS renders. Clean (Ref): 3DGS trained on the uncorrupted images. Best per row and metric in \textbf{bold}. $\dagger$: excludes the \emph{bicycle} scene, on which 3DGS with MCMC diverged.}
\label{tab:results}
\resizebox{\textwidth}{!}{%
\begin{tabular}{l ccc ccc ccc ccc ccc ccc}
\toprule
 & \multicolumn{3}{c}{3DGS~\cite{kerbl20233d}} & \multicolumn{3}{c}{DeSplat~\cite{wang2025desplat}} & \multicolumn{3}{c}{RobustNeRF~\cite{sabour2023robustnerf}} & \multicolumn{3}{c}{SpotlessSplats~\cite{sabour2025spotlesssplats}} & \multicolumn{3}{c}{Difix3D+~\cite{wu2025difix3d}} & \multicolumn{3}{c}{Ours (SSA-3DGS)} \\
\cmidrule(lr){2-4}\cmidrule(lr){5-7}\cmidrule(lr){8-10}\cmidrule(lr){11-13}\cmidrule(lr){14-16}\cmidrule(lr){17-19}
Artifact Type & PSNR$\,\uparrow$ & SSIM$\,\uparrow$ & FLIP$\,\downarrow$ & PSNR$\,\uparrow$ & SSIM$\,\uparrow$ & FLIP$\,\downarrow$ & PSNR$\,\uparrow$ & SSIM$\,\uparrow$ & FLIP$\,\downarrow$ & PSNR$\,\uparrow$ & SSIM$\,\uparrow$ & FLIP$\,\downarrow$ & PSNR$\,\uparrow$ & SSIM$\,\uparrow$ & FLIP$\,\downarrow$ & PSNR$\,\uparrow$ & SSIM$\,\uparrow$ & FLIP$\,\downarrow$ \\
\midrule
Clean (Ref) & 27.760\,/\,29.461 & 0.779\,/\,0.876 & 0.132\,/\,0.113 & -- & -- & -- & -- & -- & -- & -- & -- & -- & -- & -- & -- & -- & -- & -- \\
\midrule
Watermark   & 24.585\,/\,23.578 & 0.728\,/\,0.736 & 0.173\,/\,0.205 & 25.184 & 0.810 & 0.157 & 25.255 & 0.753 & 0.146 & 26.614 & 0.767 & 0.142 & 23.026 & 0.642 & 0.213 & 27.696\,/\,\textbf{29.391} & 0.776\,/\,\textbf{0.874} & 0.131\,/\,\textbf{0.114} \\
UI Elements & 27.735\,/\,28.900 & 0.778\,/\,0.874 & 0.133\,/\,0.117 & 28.541 & 0.859 & 0.120 & 24.298 & 0.730 & 0.156 & 26.828 & 0.789 & 0.131 & 25.043 & 0.682 & 0.177 & 27.791\,/\,\textbf{29.420} & 0.779\,/\,\textbf{0.876} & 0.131\,/\,\textbf{0.114} \\
Dashboard   & 20.756\,/\,20.399$^{\dagger}$ & 0.696\,/\,0.782 & 0.210\,/\,0.203 & 25.751 & 0.819 & 0.145 & 22.166 & 0.711 & 0.184 & 25.515 & 0.763 & 0.153 & 20.309 & 0.616 & 0.241 & 26.889\,/\,\textbf{28.305} & 0.765\,/\,\textbf{0.857} & 0.139\,/\,\textbf{0.124} \\
Rain spots  & 26.077\,/\,26.198 & 0.743\,/\,0.797 & 0.165\,/\,0.174 & 24.557 & 0.767 & 0.186 & 25.694 & 0.750 & 0.144 & 26.437 & 0.757 & 0.157 & 23.818 & 0.650 & 0.208 & 27.323\,/\,\textbf{28.938} & 0.771\,/\,\textbf{0.868} & 0.143\,/\,\textbf{0.126} \\
Mud         & 23.210\,/\,22.845 & 0.708\,/\,0.742 & 0.256\,/\,0.268 & 23.496 & 0.793 & 0.248 & 23.404 & 0.700 & 0.231 & 23.318 & 0.721 & 0.250 & 22.789 & 0.630 & 0.252 & 27.293\,/\,\textbf{28.947} & 0.771\,/\,\textbf{0.871} & 0.147\,/\,\textbf{0.130} \\
\midrule
\textbf{Mean} & 24.473\,/\,24.384 & 0.731\,/\,0.786 & 0.187\,/\,0.193 & 25.506 & 0.810 & 0.171 & 24.163 & 0.729 & 0.172 & 25.742 & 0.759 & 0.167 & 22.997 & 0.644 & 0.218 & 27.398\,/\,\textbf{29.000} & 0.773\,/\,\textbf{0.869} & 0.138\,/\,\textbf{0.122} \\
\bottomrule
\end{tabular}%
}
\end{table*}

\Cref{tab:results} reports per-category results on the synthetic benchmark for both densification strategies. Three observations stand out. First, screen-space corruption severely degrades vanilla 3DGS: with ADC, mean PSNR drops from 27.76~dB (clean training) to 24.47~dB. The distractor-robust baselines recover only part of this loss: DeSplat reaches 25.51~dB and SpotlessSplats 25.74~dB, while the RobustNeRF robust loss (24.16~dB) leaves it essentially unchanged. Second, SSA-3DGS accurately assigns corrupted pixels to the 2D overlay, allowing the 3D Gaussians to converge on the true scene geometry. With ADC, it achieves the best PSNR among ADC-based methods in four of the five categories, as well as the best mean PSNR and FLIP. The exception is the UI element, where the artifact is very compact, so the metrics are dominated by the clean part of the reconstruction. Even though DeSplat retains a higher mean SSIM with ADC (0.810 vs.\ 0.773), both PSNR and FLIP favor our reconstructions, and the qualitative comparison (\cref{fig:qualitative}) shows that its renders remain visibly corrupted. Third, the densification strategies interact differently with corruption. For vanilla 3DGS, MCMC brings no benefit on corrupted inputs: its mean (24.38~dB) is on par with ADC (24.47~dB), and it diverges outright on the \emph{bicycle} scene with the Dashboard occluder. With our overlay absorbing the artifact, MCMC becomes stable and beneficial, adding a consistent 1.4--1.7~dB across all categories and reaching a mean of 29.00~dB---within 0.5~dB of its clean-data reference (29.46~dB) and achieving the best score on nearly every metric. The dashboard remains the most challenging category as it occludes a large, contiguous portion of every frame, so the affected scene regions are observed from far fewer viewpoints, and their geometry is correspondingly harder to recover. Applying Difix3D+~\cite{wu2025difix3d} as a post-hoc enhancer to the vanilla-3DGS (ADC) renders does not recover the clean scene (\cref{tab:results}): mean PSNR \emph{drops} to 23.00~dB, below the un-enhanced 24.47~dB. Difix3D's prior, trained to add detail to \emph{under-observed} regions, leaves the occluder in place while resampling the clean portion away from ground truth.

\begin{table*}[htbp]
\centering
\caption{Results on our real-world dataset, per artifact category and overall (ADC for all methods). Means over the category's scenes (Finger 4, Mud 7, Occlusion 4; Mean: all 15) and all test views, against the separately captured clean pass. Best per column in \textbf{bold}.}
\label{tab:real}
\resizebox{\textwidth}{!}{%
\begin{tabular}{l ccc ccc ccc ccc}
\toprule
 & \multicolumn{3}{c}{\textbf{Finger}} & \multicolumn{3}{c}{\textbf{Mud}} & \multicolumn{3}{c}{\textbf{Occlusion}} & \multicolumn{3}{c}{\textbf{Mean}} \\
\cmidrule(lr){2-4}\cmidrule(lr){5-7}\cmidrule(lr){8-10}\cmidrule(lr){11-13}
\textbf{Method} & PSNR$\,\uparrow$ & SSIM$\,\uparrow$ & FLIP$\,\downarrow$ & PSNR$\,\uparrow$ & SSIM$\,\uparrow$ & FLIP$\,\downarrow$ & PSNR$\,\uparrow$ & SSIM$\,\uparrow$ & FLIP$\,\downarrow$ & PSNR$\,\uparrow$ & SSIM$\,\uparrow$ & FLIP$\,\downarrow$ \\
\midrule
Baseline 3DGS~\cite{kerbl20233d} & 13.45 & 0.575 & 0.506 & 15.94 & 0.574 & 0.402 & 16.16 & 0.564 & 0.436 & 15.34 & 0.572 & 0.439 \\
DeSplat~\cite{wang2025desplat} & 13.75 & 0.605 & 0.481 & 16.15 & 0.616 & 0.382 & 16.45 & \textbf{0.592} & 0.424 & 15.59 & 0.606 & 0.420 \\
RobustNeRF~\cite{sabour2023robustnerf} & 13.89 & 0.601 & 0.462 & 15.88 & 0.579 & 0.385 & 15.77 & 0.552 & 0.443 & 15.32 & 0.578 & 0.421 \\
SpotlessSplats~\cite{sabour2025spotlesssplats} & 13.66 & 0.597 & 0.471 & 15.95 & 0.586 & 0.393 & 15.97 & 0.569 & 0.438 & 15.34 & 0.585 & 0.426 \\
Difix3D+~\cite{wu2025difix3d} & 13.68 & 0.540 & 0.499 & 15.95 & 0.536 & 0.401 & 16.35 & 0.535 & 0.436 & 15.45 & 0.537 & 0.437 \\
Ours (SSA-3DGS) & \textbf{16.67} & \textbf{0.678} & \textbf{0.389} & \textbf{19.25} & \textbf{0.645} & \textbf{0.316} & \textbf{17.66} & 0.585 & \textbf{0.388} & \textbf{18.14} & \textbf{0.638} & \textbf{0.355} \\
\bottomrule
\end{tabular}%
}
\end{table*}

Finally, \cref{tab:real} reports results on our self-captured real-world dataset, aggregated per artifact category over all scenes and test views. All methods use ADC because vanilla 3DGS with MCMC densification diverges outright on the real captures. The pattern from the synthetic benchmark carries over: SSA-3DGS attains the best PSNR and FLIP on every artifact category, improving on the strongest baseline by $2.8$~dB on \emph{Finger} ($16.67$ vs.\ $13.89$), $3.1$~dB on \emph{Mud} ($19.25$ vs.\ $16.15$), and $1.2$~dB on \emph{Occlusion} ($17.66$ vs.\ $16.45$). The distractor-robust (DeSplat, SpotlessSplats, RobustNeRF) and generative (Difix3D+) baselines remain within a few tenths of a dB of vanilla 3DGS, confirming that a corruption shared by every training view is not the multi-view inconsistency they are built to reject. \Cref{fig:qualitative_real} shows representative renders.

\subsection{Qualitative results}
\label{sec:qualitative}

Visual inspection of \cref{fig:qualitative} confirms the quantitative results on the synthetic benchmark. Vanilla 3DGS bakes the screen-space artifacts into the scene as semi-transparent, near-camera clouds that shadow the true geometry, and the distractor-robust baselines (DeSplat, SpotlessSplats) and the generative Difix3D+ enhancer leave the corruption largely in place. SSA-3DGS instead produces clean renders that closely match the underlying scene. \Cref{fig:qualitative_real} shows the qualitative comparison on our real-world dataset, where our method largely removes the physically induced finger, mud, and rig occlusions while the baselines do not.

\subsection{Ablation study}
\label{sec:ablation}
To validate our design choices, we ablate the sparsity regularizer and the overlay initialization. \Cref{tab:ablation} reports the ablation on both the synthetic benchmark (mean over the artifact categories, under MCMC densification) and the real scenes (mean over all scenes, under ADC). On the synthetic benchmark, adding sparsity raises mean PSNR by $1.93$~dB ($27.08\!\to\!29.00$) and lowers FLIP from $0.167$ to $0.122$. In real scenes, its impact on these metrics is less pronounced. The sparsity regularizer's goal, apart from preventing scene content from being absorbed, is to prevent a low-opacity overlay from absorbing a \emph{global} color/exposure shift from the clean reconstruction. To quantify this, we measure, per frame, the deviation of the render's mean color from ground truth, $\Delta\bar{c}=\lVert\bar{c}_{\mathrm{render}}-\bar{c}_{\mathrm{GT}}\rVert_2$ on the $0$--$255$ scale, averaged over all test views. Adding $\mathcal{L}_{sparse}$ roughly halves the mean color deviation on the synthetic benchmark ($\Delta\bar{c}=\mathbf{3.19}$ vs.\ $6.43$ without it). Because this bias is a near-uniform tint, it barely moves per-pixel PSNR/SSIM even when it visibly shifts the reconstruction's color.

Replacing the statistics initialization by a constant overlay costs $1.31$~dB on the synthetic benchmark ($29.00\!\to\!27.69$), almost entirely on the Dashboard category ($28.31\!\to\!23.12$), and $1.56$~dB on the real scenes ($18.14\!\to\!16.57$). Conversely, the statistics alone do not suffice: applying the statistics-based overlay as a fixed per-frame correction and training vanilla 3DGS on the corrected frames reaches only $26.04$~dB on the synthetic benchmark and $15.67$~dB on the real scenes, i.e., $2.96$ and $2.47$~dB below the full method. Per-pixel statistics conflate stable scene content (sky, textureless regions) with artifacts and misestimate the opacity of semi-transparent occluders; only joint optimization corrects both.

\begin{table}[htbp]
\centering
\caption{Ablation of the sparsity regularizer, the overlay initialization, and the joint optimization. Synthetic: mean over the five artifact categories (MCMC). Real: mean over the fifteen scenes (ADC).}
\label{tab:ablation}
\resizebox{\columnwidth}{!}{%
\begin{tabular}{l ccc ccc}
\toprule
 & \multicolumn{3}{c}{\textbf{Synthetic}} & \multicolumn{3}{c}{\textbf{Real}} \\
\cmidrule(lr){2-4}\cmidrule(lr){5-7}
\textbf{Configuration} & PSNR$\,\uparrow$ & SSIM$\,\uparrow$ & FLIP$\,\downarrow$ & PSNR$\,\uparrow$ & SSIM$\,\uparrow$ & FLIP$\,\downarrow$ \\
\midrule
No regularization & 27.075 & 0.859 & 0.167 & 18.006 & 0.640 & 0.355 \\
$\mathcal{L}_{sparse}$ + statistics init & 29.000 & 0.869 & 0.122 & 18.137 & 0.638 & 0.355 \\
\midrule
$\mathcal{L}_{sparse}$ + const.\ init & 27.691 & 0.841 & 0.147 & 16.573 & 0.601 & 0.396 \\
Statistics pre-cleanup + 3DGS & 26.038 & 0.803 & 0.182 & 15.665 & 0.585 & 0.429 \\
\bottomrule
\end{tabular}%
}
\end{table}
\section{Limitations}
Our method relies on multi-view consensus to separate the static overlay from the scene. A sufficiently dense capture is required to prevent 3DGS from explaining artifacts through view-dependent effects that are not penalized by overlapping views. Our overlay model also assumes that the artifact is fixed in screen space. Artifacts that are only \emph{approximately} static are only partially captured by the overlay. While SSA-3DGS still improves substantially over the baselines, fully view-dependent occluders remain an open problem. 

\section{Conclusion}

We have presented \textit{SSA-3DGS}, a robust framework that extends 3D Gaussian Splatting to handle input imagery corrupted by static screen-space artifacts. By explicitly modeling the image formation process as a composition of a world-space 3D scene and a view-independent image-space 2D overlay, we cast reconstruction as a dual-objective joint optimization that simultaneously disentangles sensor-plane obstructions and reconstructs clean 3D geometry without supervision. Our approach restores high-fidelity 3D scenes from heavily corrupted inputs, including synthetic watermarks, natural environmental effects, and genuine, self-captured real-world artifacts. These results suggest that explicitly modeling capture imperfections is a viable path toward reliable 3D reconstruction on unconstrained, ``in-the-wild'' captures, significantly reducing the reliance on laborious manual cleaning or ground-truth masks.

\section*{Acknowledgments}
This project was funded by the Flanders Make ADDIL project. We gratefully acknowledge imec IDLab for computational resources via the iLab.t infrastructure.
{
    \small
    \bibliographystyle{ieeenat_fullname}
    \bibliography{main}

@String(CVPR= {IEEE Conf. Comput. Vis. Pattern Recog.})

@String(ICCV= {Int. Conf. Comput. Vis.})

@String(ECCV= {Eur. Conf. Comput. Vis.})

@String(NIPS= {Adv. Neural Inform. Process. Syst.})

@String(TOG= {ACM Trans. Graph.})

@String(AAAI = {AAAI})

@String(CSVT = {IEEE Trans. Circuit Syst. Video Technol.})

@String(CVPR  = {CVPR})

@String(ICCV  = {ICCV})

@String(ECCV  = {ECCV})

@String(NIPS  = {NeurIPS})

@String(TOG   = {ACM TOG})

@String(CSVT = {IEEE TCSVT})

@String(WACV  = {WACV})

@String(ICRA  = {ICRA})

@String(GCPR  = {GCPR})

@String(ICCVW = {ICCVW})

@String(JMLR  = {JMLR})

@String(CACM  = {Commun. ACM})

@String(CG    = {Comput. Graph.})

@String(IEICE = {IEICE Trans. Inf. Syst.})

@String(PACMCGIT = {Proc. ACM Comput. Graph. Interact. Tech.})

@String(ICML  = {ICML})

@article{mildenhall2021nerf,
  title={{NeRF}: Representing scenes as neural radiance fields for view synthesis},
  author={Mildenhall, Ben and Srinivasan, Pratul P. and Tancik, Matthew and Barron, Jonathan T. and Ramamoorthi, Ravi and Ng, Ren},
  journal = CACM,
  volume={65},
  number={1},
  pages={99--106},
  year={2021}
}

@article{kerbl20233d,
  title={{3D} {Gaussian} splatting for real-time radiance field rendering},
  author={Kerbl, Bernhard and Kopanas, Georgios and Leimk{\"u}hler, Thomas and Drettakis, George},
  journal = TOG,
  volume={42},
  number={4},
  pages={139:1--139:14},
  year={2023}
}

@inproceedings{suvorov2022resolution,
  title={Resolution-robust large mask inpainting with {Fourier} convolutions},
  author={Suvorov, Roman and Logacheva, Elizaveta and Mashikhin, Anton and Remizova, Anastasia and Ashukha, Arsenii and Silvestrov, Aleksei and Kong, Naejin and Goka, Harshith and Park, Kiwoong and Lempitsky, Victor},
  booktitle = WACV,
  pages={2149--2159},
  year={2022}
}

@inproceedings{fridovich2022plenoxels,
  title={Plenoxels: Radiance fields without neural networks},
  author={Fridovich-Keil, Sara and Yu, Alex and Tancik, Matthew and Chen, Qinhong and Recht, Benjamin and Kanazawa, Angjoo},
  booktitle = CVPR,
  pages={5501--5510},
  year={2022}
}

@article{muller2022instant,
  title={Instant neural graphics primitives with a multiresolution hash encoding},
  author={M{\"u}ller, Thomas and Evans, Alex and Schied, Christoph and Keller, Alexander},
  journal = TOG,
  volume={41},
  number={4},
  pages={102:1--102:15},
  year={2022}
}

@inproceedings{martin2021nerf,
  title={{NeRF} in the wild: Neural radiance fields for unconstrained photo collections},
  author={Martin-Brualla, Ricardo and Radwan, Noha and Sajjadi, Mehdi S. M. and Barron, Jonathan T. and Dosovitskiy, Alexey and Duckworth, Daniel},
  booktitle = CVPR,
  pages={7210--7219},
  year={2021}
}

@inproceedings{sabour2023robustnerf,
  title={{RobustNeRF}: Ignoring distractors with robust losses},
  author={Sabour, Sara and Vora, Suhani and Duckworth, Daniel and Krasin, Ivan and Fleet, David J. and Tagliasacchi, Andrea},
  booktitle = CVPR,
  pages={20626--20636},
  year={2023}
}

@article{sabour2025spotlesssplats,
  title={{SpotLessSplats}: Ignoring distractors in {3D} {Gaussian} splatting},
  author={Sabour, Sara and Goli, Lily and Kopanas, George and Matthews, Mark and Lagun, Dmitry and Guibas, Leonidas and Jacobson, Alec and Fleet, David and Tagliasacchi, Andrea},
  journal = TOG,
  volume={44},
  number={2},
  pages={1--11},
  year={2025}
}

@inproceedings{xu2024wild,
  title={{Wild-GS}: Real-time novel view synthesis from unconstrained photo collections},
  author={Xu, Jiacong and Mei, Yiqun and Patel, Vishal},
  booktitle = NIPS,
  pages={103334--103355},
  year={2024}
}

@InProceedings{wang2025desplat,
    title={{DeSplat}: Decomposed {Gaussian} Splatting for Distractor-Free Rendering}, 
    author={Yihao Wang and Marcus Klasson and Matias Turkulainen and Shuzhe Wang and Juho Kannala and Arno Solin},
    year={2025},
    booktitle = CVPR
}

@inproceedings{seiskari2024gaussian,
  title={{Gaussian} Splatting on the Move: Blur and Rolling Shutter Compensation for Natural Camera Motion},
  author={Otto Seiskari and Jerry Ylilammi and Valtteri Kaatrasalo and Pekka Rantalankila and Matias Turkulainen and Juho Kannala and Arno Solin},
  booktitle = ECCV,
  pages={160--177},
  year={2024}
}

@inproceedings{hertz2019blind,
  title={Blind visual motif removal from a single image},
  author={Hertz, Amir and Fogel, Sharon and Hanocka, Rana and Giryes, Raja and Cohen-Or, Daniel},
  booktitle = CVPR,
  pages={6858--6867},
  year={2019}
}

@article{su2025deep,
  title={Deep learning for visible watermark removal: A survey},
  author={Su, Peixian and Zhang, Yong},
  journal={Comput. Intell.},
  volume={41},
  number={3},
  pages={e70072},
  year={2025}
}

@inproceedings{barron2022mip,
  title={{Mip-NeRF} 360: Unbounded anti-aliased neural radiance fields},
  author={Barron, Jonathan T. and Mildenhall, Ben and Verbin, Dor and Srinivasan, Pratul P. and Hedman, Peter},
  booktitle = CVPR,
  pages={5470--5479},
  year={2022}
}

@inproceedings{zhang2018unreasonable,
  title={The unreasonable effectiveness of deep features as a perceptual metric},
  author={Zhang, Richard and Isola, Phillip and Efros, Alexei A. and Shechtman, Eli and Wang, Oliver},
  booktitle = CVPR,
  pages={586--595},
  year={2018}
}

@article{ye2025gsplat,
  title={{gsplat}: An open-source library for {Gaussian} splatting},
  author={Ye, Vickie and Li, Ruilong and Kerr, Justin and Turkulainen, Matias and Yi, Brent and Pan, Zhuoyang and Seiskari, Otto and Ye, Jianbo and Hu, Jeffrey and Tancik, Matthew and Kanazawa, Angjoo},
  journal = JMLR,
  volume={26},
  number={34},
  pages={1--17},
  year={2025}
}

@inproceedings{qiao2025restorgs,
  title={{RestorGS}: Depth-aware {Gaussian} Splatting for Efficient {3D} Scene Restoration},
  author={Qiao, Yuanjian and Shao, Mingwen and Meng, Lingzhuang and Xu, Kai},
  booktitle = CVPR,
  pages={11177--11186},
  year={2025}
}

@inproceedings{zhu2024revisiting,
  title={Revisiting single image reflection removal in the wild},
  author={Zhu, Yurui and Fu, Xueyang and Jiang, Peng-Tao and Zhang, Hao and Sun, Qibin and Chen, Jinwei and Zha, Zheng-Jun and Li, Bo},
  booktitle = CVPR,
  pages={25468--25478},
  year={2024}
}

@inproceedings{guo2014robust,
  title={Robust separation of reflection from multiple images},
  author={Guo, Xiaojie and Cao, Xiaochun and Ma, Yi},
  booktitle = CVPR,
  pages={2187--2194},
  year={2014}
}

@inproceedings{eigen2013restoring,
  title={Restoring an image taken through a window covered with dirt or rain},
  author={Eigen, David and Krishnan, Dilip and Fergus, Rob},
  booktitle = ICCV,
  pages={633--640},
  year={2013}
}

@article{tanaka2006removal,
  title={Removal of adherent waterdrops from images acquired with a stereo camera system},
  author={Tanaka, Yuu and Yamashita, Atsushi and Kaneko, Toru and Miura, Kenjiro T.},
  journal = IEICE,
  volume={89},
  number={7},
  pages={2021--2027},
  year={2006}
}

@article{luo2020weakly,
  title={Weakly supervised learning for raindrop removal on a single image},
  author={Luo, Wenjie and Lai, Jianhuang and Xie, Xiaohua},
  journal = CSVT,
  volume={31},
  number={5},
  pages={1673--1683},
  year={2020}
}

@inproceedings{wan2020reflection,
  title={Reflection scene separation from a single image},
  author={Wan, Renjie and Shi, Boxin and Li, Haoliang and Duan, Ling-Yu and Kot, Alex C.},
  booktitle = CVPR,
  pages={2398--2406},
  year={2020}
}

@inproceedings{hao2019learning,
  title={Learning from synthetic photorealistic raindrop for single image raindrop removal},
  author={Hao, Zhixiang and You, Shaodi and Li, Yu and Li, Kunming and Lu, Feng},
  booktitle = ICCVW,
  year={2019}
}

@inproceedings{huang20253d,
  title={{3D} {Gaussian} inpainting with depth-guided cross-view consistency},
  author={Huang, Sheng-Yu and Chou, Zi-Ting and Wang, Yu-Chiang Frank},
  booktitle = CVPR,
  pages={26704--26713},
  year={2025}
}

@article{zhou2025high,
  title={High-fidelity {3D} {Gaussian} inpainting: Preserving multi-view consistency and photorealistic details},
  author={Zhou, Jun and Li, Dinghao and Li, Nannan and Wang, Mingjie},
  journal = CG,
  pages={104362},
  year={2025}
}

@article{wang2025low,
  title={Low-frequency first: Eliminating floating artifacts in {3D} {Gaussian} splatting},
  author={Wang, Jianchao and Zhou, Peng and Li, Cen and Quan, Rong and Qin, Jie},
  journal={arXiv preprint arXiv:2508.02493},
  year={2025}
}

@inproceedings{ungermann2024robust,
  title={Robust {3D} {Gaussian} splatting for novel view synthesis in presence of distractors},
  author={Ungermann, Paul and Ettenhofer, Armin and Nie{\ss}ner, Matthias and Roessle, Barbara},
  booktitle = GCPR,
  pages={153--167},
  year={2024}
}

@article{Andersson2020,
  author    = {Pontus Andersson and
               Jim Nilsson and
               Tomas Akenine{-}M{\"{o}}ller and
               Magnus Oskarsson and
               Kalle {\AA}str{\"{o}}m and
               Mark D. Fairchild},
  title     = "{{FLIP:} {A} Difference Evaluator for Alternating Images}",
  journal   = PACMCGIT,
  volume    = {3},
  number    = {2},
  pages     = {15:1--15:23},
  year      = {2020},
  doi={10.1145/3406183}
}

@inproceedings{kheradmand20243d,
  title={{3D} {Gaussian} Splatting as {Markov} Chain {Monte} {Carlo}},
  author={Kheradmand, Shakiba and Rebain, Daniel and Sharma, Gopal and Sun, Weiwei and Tseng, Yang-Che and Isack, Hossam and Kar, Abhishek and Tagliasacchi, Andrea and Yi, Kwang Moo},
  booktitle = NIPS,
  pages={80965--80986},
  year={2024}
}

@inproceedings{colmap,
    author={Sch\"{o}nberger, Johannes Lutz and Frahm, Jan-Michael},
    title={Structure-from-{Motion} Revisited},
    booktitle = CVPR,
    year={2016},
}

@inproceedings{wu2025difix3d,
    title={{Difix3D+}: Improving {3D} Reconstructions with Single-Step Diffusion Models},
    author={Wu, Jay Zhangjie and Zhang, Yuxuan and Turki, Haithem and Ren, Xuanchi and Gao, Jun and Shou, Mike Zheng and Fidler, Sanja and Gojcic, Zan and Ling, Huan},
    booktitle = CVPR,
    year={2025},
}

@inproceedings{liu2024enhancer,
    title={{3DGS-Enhancer}: Enhancing Unbounded {3D} {Gaussian} Splatting with View-consistent {2D} Diffusion Priors},
    author={Liu, Xi and Zhou, Chaoyi and Huang, Siyu},
    booktitle = NIPS,
    year={2024},
}

@InProceedings{yin2025gsfixer,
  title = 	 {{GSF}ixer: Improving 3{D} {G}aussian Splatting with Reference-Guided Video Diffusion Priors},
  author =       {Yin, Xingyilang and Zhang, Qi and Chang, Jiahao and Feng, Ying and Fan, Qingnan and Yang, Xi and Pun, Chi-Man and Zhang, Huaqi and Cun, Xiaodong},
  booktitle = ICML,
  pages = 	 {149259--149284},
  year = 	 {2026},
  editor = 	 {Zhang, Tong and Dudik, Miroslav and Jaggi, Martin and Agarwal, Alekh and Li, Sharon and Schuurmans, Dale and Zhu, Jerry and Berkenkamp, Felix and Dong, Hanze and Bietti, Alberto},
  volume = 	 {306},
  series = 	 {Proceedings of Machine Learning Research},
  month = 	 {06--11 Jul},
  publisher =    {PMLR},
  url = 	 {https://proceedings.mlr.press/v306/yin26m.html}
}

@misc{flux2025kontext,
      title={{FLUX.1 Kontext}: Flow Matching for In-Context Image Generation and Editing in Latent Space}, 
      author={Black Forest Labs and Stephen Batifol and Andreas Blattmann and Frederic Boesel and Saksham Consul and Cyril Diagne and Tim Dockhorn and Jack English and Zion English and Patrick Esser and Sumith Kulal and Kyle Lacey and Yam Levi and Cheng Li and Dominik Lorenz and Jonas Müller and Dustin Podell and Robin Rombach and Harry Saini and Axel Sauer and Luke Smith},
      year={2025},
      eprint={2506.15742},
      archivePrefix={arXiv},
      primaryClass={cs.GR},
      url={https://arxiv.org/abs/2506.15742},
}

@inproceedings{zamfir2025moceir,
    title={Complexity Experts are Task-Discriminative Learners for Any Image Restoration},
    author={Zamfir, Eduard and Wu, Zongwei and Mehta, Nancy and Tan, Yuedong and Paudel, Danda Pani and Zhang, Yulun and Timofte, Radu},
    booktitle = CVPR,
    year={2025},
}

@inproceedings{cui2025bioir,
    title={Bio-Inspired Image Restoration},
    author={Cui, Yuning and Ren, Wenqi and Knoll, Alois},
    booktitle = NIPS,
    year={2025},
}

@inproceedings{qian2025weathergs,
  title={{WeatherGS}: {3D} Scene Reconstruction in Adverse Weather Conditions via {Gaussian} Splatting},
  author={Qian, Chenghao and Guo, Yuhu and Li, Wenjing and Markkula, Gustav},
  booktitle = ICRA,
  pages={185--191},
  year={2025}
}

@inproceedings{liu2025deraings,
  title={{DeRainGS}: {Gaussian} Splatting for Enhanced Scene Reconstruction in Rainy Environments},
  author={Liu, Shuhong and Chen, Xiang and Chen, Hongming and Xu, Quanfeng and Li, Mingrui},
  booktitle = AAAI,
  volume={39},
  number={5},
  pages={5558--5566},
  year={2025}
}

@inproceedings{li2024derainnerf,
  title={{DerainNeRF}: {3D} Scene Estimation with Adhesive Waterdrop Removal},
  author={Li, Yunhao and Wu, Jing and Zhao, Lingzhe and Liu, Peidong},
  booktitle = ICRA,
  pages={2787--2793},
  year={2024}
}

@article{gu2009removing,
  title={Removing Image Artifacts Due to Dirty Camera Lenses and Thin Occluders},
  author={Gu, Jinwei and Ramamoorthi, Ravi and Belhumeur, Peter N. and Nayar, Shree K.},
  journal = TOG,
  volume={28},
  number={5},
  year={2009}
}

@misc{overdulve2026,
  author       = {Overdulve, Kristof},
  title        = {{SSA3DGS}},
  howpublished = {Hugging Face dataset of synthetic and real-world screen-space artifacts, \url{https://doi.org/10.57967/hf/10495}},
  year         = {2026},
  doi          = {10.57967/hf/10495},
  url          = {https://huggingface.co/datasets/KeejOow/SSA3DGS}
}
}

\clearpage
\appendix

\section{Comparison to image-level restoration}
\label{sec:imagelevel}

An alternative to reconstruction-integrated removal is to first restore each corrupted frame using an off-the-shelf 2D editor, then reconstruct the scene with vanilla 3DGS. We do this using a state-of-the-art model, the instruction-tuned \textbf{FLUX.1-Kontext}~\cite{flux2025kontext} (a 12B-parameter rectified-flow image editor), in its all-in-one editing mode (guidance scale $2.5$, $28$ denoising steps, fixed seed), prompting it per artifact category to remove the corruption---for example \emph{``remove the car dashboard occluding the view''} or \emph{``remove the mud splatter on the camera lens''}. We run it on the corrupted training views of the \emph{bicycle} scene, for which an aligned clean original is available for every restored frame. However, we considered this workflow ill-suited for our goals for the following reasons.

\begin{figure}[htbp]
  \centering
  \setlength{\tabcolsep}{1pt}
  \renewcommand{\arraystretch}{0.9}
  \newcommand{\ilincg}[1]{\includegraphics[width=0.485\columnwidth]{media/restoration/#1}}
  \begin{tabular}{@{}cc@{}}
    \footnotesize Corrupted input & \footnotesize FLUX.1-Kontext cleanup \\
    \ilincg{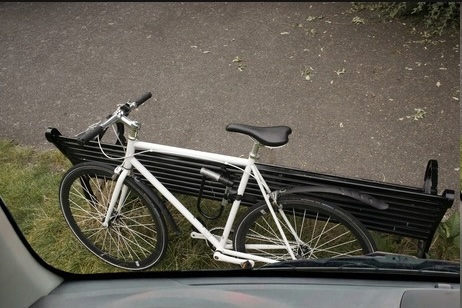} & \ilincg{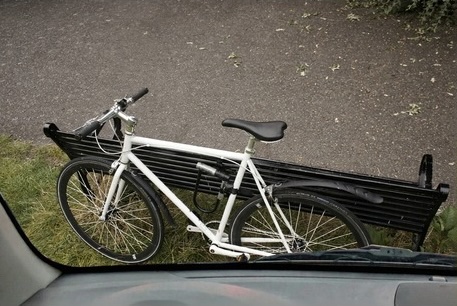} \\
    \ilincg{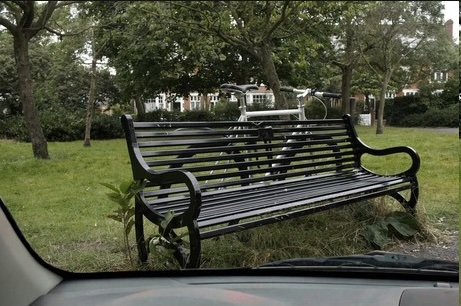} & \ilincg{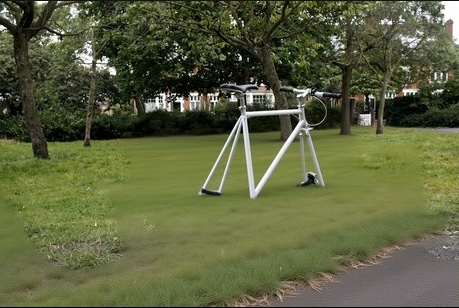} \\
  \end{tabular}
  \caption{\textbf{Image-level restoration fails on opaque occluders.} Two \emph{Dashboard} views of \emph{bicycle}: corrupted input (left) and per-frame FLUX.1-Kontext~\cite{flux2025kontext} cleanup (right). Top: the occluder is left intact. Bottom: the real bench is erased and a bicycle hallucinated in its place.}
  \label{fig:imagelevel}
\end{figure}

\paragraph{It fails on opaque occluders.} \Cref{fig:imagelevel} shows two representative \emph{Dashboard} views: the editor either leaves the occluder essentially intact, or removes it while erasing real content (the bench) and hallucinating an incorrect replacement. Highly transparent, thin overlays, such as some watermarks, are, by contrast, cleaned well. Quantifying this with the per-frame RMS discrepancy from the clean original ($169$ frames per category), the physically induced contaminations are recovered far less faithfully than the compositing overlays---mean PSNR of only $20.6$~dB for the dashboard, $23.0$~dB for rain, and $24.3$~dB for mud, versus $29.0$ and $26.7$~dB for UI and watermark---so a downstream reconstruction would start from inputs that are still corrupted.

\paragraph{It is multi-view inconsistent.} Because the editor operates on one frame at a time, with no access to the multi-view geometry, its residual occluders and hallucinated completions disagree from view to view. A subsequent 3DGS optimization, which assumes photometric agreement across views, cannot fuse such mutually inconsistent frames into a clean scene and instead averages them into blur and floaters---precisely the regime our reconstruction-integrated overlay avoids by modeling a single sensor-frame overlay shared across all views.

\paragraph{It is prohibitively slow.} FLUX.1-Kontext requires $28$ denoising iterations per frame, which on a Tesla V100-SXM3-32GB take on average $22.18$~s each, i.e.\ $\approx\!10.4$~min per frame; a typical capture of $\approx\!400$ views therefore incurs $\approx\!69$~hours of per-frame cleanup \emph{before} 3DGS reconstruction can even begin, whereas our method adds no such preprocessing. For these reasons, we do not train a full FLUX$\rightarrow$3DGS reconstruction, but we show the failure directly at the image level, where it originates.

\section{Per-scene real-world results}
\label{sec:supp_perscene}
\Cref{tab:real_perscene} gives the complete per-scene breakdown behind the per-category means of Tab.~2 of the main paper: PSNR, SSIM, and FLIP for every method on all fifteen real scenes, each computed over all test views. Alongside the baselines and the full SSA-3DGS, we include the two ablations of our method---constant overlay initialization (Ours$_{\text{ci}}$) and no sparsity regularization (Ours$_{\text{nr}}$). A variant of our method attains the best PSNR on every scene: the full method or its no-regularization variant on 13 of the 15 scenes, and the constant-initialization variant on \emph{Finger/Hallway} and \emph{Occlusion/Selfie}. The largest margins occur on the pure screen-space \emph{Mud} and near-lens \emph{Finger} artifacts, consistent with the per-category summary.

\begin{sidewaystable*}[htbp]
\centering
\caption{Per-scene results on the real-world dataset (ADC, all test views). Ours$_{\text{ci}}$/Ours$_{\text{nr}}$: constant-initialization and no-regularization ablations; Ours: full SSA-3DGS. Best per scene and metric in \textbf{bold}.}
\label{tab:real_perscene}
\resizebox{\textheight}{!}{%
\begin{tabular}{ll ccc ccc ccc ccc ccc ccc ccc ccc}
\toprule
\textbf{Artifact} & \textbf{Scene} & \multicolumn{3}{c}{3DGS} & \multicolumn{3}{c}{DeSplat} & \multicolumn{3}{c}{RobustNeRF} & \multicolumn{3}{c}{Spotless} & \multicolumn{3}{c}{Difix3D+} & \multicolumn{3}{c}{Ours$_{\text{ci}}$} & \multicolumn{3}{c}{Ours$_{\text{nr}}$} & \multicolumn{3}{c}{Ours} \\
\cmidrule(lr){1-1}\cmidrule(lr){2-2}\cmidrule(lr){3-5}\cmidrule(lr){6-8}\cmidrule(lr){9-11}\cmidrule(lr){12-14}\cmidrule(lr){15-17}\cmidrule(lr){18-20}\cmidrule(lr){21-23}\cmidrule(lr){24-26}
 & & PSNR$\,\uparrow$ & SSIM$\,\uparrow$ & FLIP$\,\downarrow$ & PSNR$\,\uparrow$ & SSIM$\,\uparrow$ & FLIP$\,\downarrow$ & PSNR$\,\uparrow$ & SSIM$\,\uparrow$ & FLIP$\,\downarrow$ & PSNR$\,\uparrow$ & SSIM$\,\uparrow$ & FLIP$\,\downarrow$ & PSNR$\,\uparrow$ & SSIM$\,\uparrow$ & FLIP$\,\downarrow$ & PSNR$\,\uparrow$ & SSIM$\,\uparrow$ & FLIP$\,\downarrow$ & PSNR$\,\uparrow$ & SSIM$\,\uparrow$ & FLIP$\,\downarrow$ & PSNR$\,\uparrow$ & SSIM$\,\uparrow$ & FLIP$\,\downarrow$ \\
\midrule
Finger & Cargobike & 13.92 & 0.629 & 0.495 & 14.14 & 0.634 & 0.477 & 14.88 & 0.650 & 0.436 & 14.96 & 0.670 & 0.440 & 14.39 & 0.551 & 0.482 & 16.52 & 0.698 & 0.395 & 16.47 & 0.698 & \textbf{0.390} & \textbf{16.58} & \textbf{0.700} & 0.395 \\
 & Diningtable & 14.78 & 0.702 & 0.450 & 14.92 & 0.729 & 0.431 & 14.09 & 0.703 & 0.466 & 13.90 & 0.674 & 0.471 & 14.79 & 0.690 & 0.452 & 16.25 & 0.735 & 0.391 & \textbf{17.18} & \textbf{0.790} & \textbf{0.384} & 16.95 & 0.785 & 0.398 \\
 & Gardenchair & 12.49 & 0.307 & 0.565 & 12.76 & 0.328 & 0.567 & 13.54 & 0.351 & 0.492 & 13.35 & 0.362 & 0.486 & 12.49 & 0.282 & 0.566 & 12.60 & 0.310 & 0.545 & 14.07 & \textbf{0.411} & \textbf{0.427} & \textbf{14.47} & 0.407 & \textbf{0.427} \\
 & Hallway & 12.63 & 0.661 & 0.515 & 13.18 & 0.728 & 0.448 & 13.05 & 0.701 & 0.456 & 12.42 & 0.682 & 0.486 & 13.03 & 0.639 & 0.498 & \textbf{19.04} & 0.816 & \textbf{0.319} & 18.64 & \textbf{0.824} & 0.335 & 18.67 & 0.820 & 0.339 \\
Mud & Cargobike & 12.30 & 0.564 & 0.529 & 12.96 & 0.602 & 0.487 & 13.18 & 0.582 & 0.474 & 12.09 & 0.546 & 0.531 & 12.57 & 0.518 & 0.518 & 13.16 & 0.566 & 0.503 & 18.53 & 0.723 & 0.344 & \textbf{18.75} & \textbf{0.729} & \textbf{0.340} \\
 & Gardenchair & 13.29 & 0.326 & 0.474 & 13.13 & 0.375 & 0.458 & 13.05 & 0.329 & 0.456 & 13.36 & 0.349 & 0.449 & 13.16 & 0.295 & 0.476 & 13.69 & 0.352 & 0.446 & 16.67 & \textbf{0.392} & 0.371 & \textbf{16.78} & 0.389 & \textbf{0.365} \\
 & Kidsbike & 15.24 & 0.431 & 0.406 & 15.32 & 0.529 & 0.383 & 14.45 & 0.430 & 0.408 & 15.83 & 0.481 & 0.377 & 14.99 & 0.386 & 0.409 & 15.86 & 0.476 & 0.379 & 19.83 & \textbf{0.537} & 0.294 & \textbf{19.94} & 0.532 & \textbf{0.291} \\
 & Kitchen & 18.30 & 0.750 & 0.336 & 18.29 & 0.774 & 0.320 & 18.43 & 0.769 & 0.315 & 19.32 & 0.776 & 0.307 & 18.41 & 0.729 & 0.331 & 19.50 & 0.783 & 0.302 & 19.47 & 0.789 & 0.304 & \textbf{19.70} & \textbf{0.791} & \textbf{0.299} \\
 & Kitchencounter & 19.15 & 0.739 & 0.337 & 17.62 & 0.745 & 0.357 & 18.09 & 0.733 & 0.345 & 18.84 & 0.750 & 0.337 & 19.06 & 0.720 & 0.340 & 19.66 & 0.753 & 0.326 & 20.47 & 0.776 & 0.308 & \textbf{20.84} & \textbf{0.777} & \textbf{0.299} \\
 & Mountainbike & 18.17 & 0.721 & 0.319 & 20.08 & 0.755 & 0.288 & 19.13 & 0.724 & 0.289 & 17.30 & 0.714 & 0.325 & 18.30 & 0.656 & 0.322 & 19.85 & 0.733 & 0.287 & \textbf{22.80} & \textbf{0.781} & \textbf{0.230} & 22.16 & 0.775 & 0.247 \\
 & Trampoline & 15.17 & 0.490 & 0.415 & 15.66 & 0.529 & 0.383 & 14.83 & 0.485 & 0.408 & 14.90 & 0.488 & 0.423 & 15.11 & 0.449 & 0.411 & 15.97 & 0.513 & 0.376 & 16.54 & \textbf{0.531} & \textbf{0.370} & \textbf{16.58} & 0.524 & 0.371 \\
Occl. & Diningtable & 18.25 & 0.787 & 0.370 & 18.47 & \textbf{0.810} & 0.356 & 17.27 & 0.772 & 0.385 & 17.46 & 0.779 & 0.382 & 18.17 & 0.770 & 0.378 & 18.62 & 0.774 & 0.358 & \textbf{19.18} & 0.803 & \textbf{0.348} & 19.15 & 0.803 & 0.351 \\
 & Gardenchair & 13.36 & 0.291 & 0.567 & 13.68 & \textbf{0.323} & 0.553 & 13.52 & 0.292 & 0.563 & 13.63 & 0.303 & 0.561 & 13.60 & 0.276 & 0.565 & 13.42 & 0.299 & 0.551 & 14.71 & 0.317 & 0.514 & \textbf{15.13} & 0.313 & \textbf{0.508} \\
 & Kidsbike & 14.49 & 0.365 & 0.528 & 14.16 & 0.400 & 0.525 & 14.24 & 0.350 & 0.537 & 14.13 & 0.367 & 0.540 & 14.75 & 0.340 & 0.524 & 14.86 & 0.381 & 0.495 & 16.44 & \textbf{0.406} & 0.435 & \textbf{16.90} & 0.400 & \textbf{0.432} \\
 & Selfie & 18.53 & 0.813 & 0.280 & 19.48 & \textbf{0.834} & 0.263 & 18.05 & 0.796 & 0.285 & 18.67 & 0.827 & 0.268 & 18.88 & 0.754 & 0.279 & \textbf{19.60} & 0.825 & \textbf{0.261} & 19.09 & 0.824 & 0.271 & 19.45 & 0.825 & 0.261 \\
\bottomrule
\end{tabular}%
}
\end{sidewaystable*}

\section{Sensitivity to the sparsity weight}
\label{sec:supp_lambda}
\Cref{tab:lambda} reports SSA-3DGS under a uniform sparsity weight $\lambda_{sparse}\in\{0, 0.01, 0.05\}$, on the synthetic benchmark (MCMC densification, mean over the seven scenes per category) and on the real dataset (ADC, mean over the scenes per category). Watermark, UI, and Dashboard are insensitive to the choice. The two ends of the artifact spectrum, however, pull in opposite directions: the small, semi-transparent rain spots require the strong prior ($0.01$ costs $3.3$~dB, and $\lambda_{sparse}=0$ collapses), whereas the large, opaque mud spatters and all real artifacts prefer the weak one ($0.05$ costs $1.3$~dB on synthetic mud and $0.9$~dB on the real scenes). This motivates the rule of Sec.~4.1 of the main paper: $0.05$ by default and $0.01$ for artifacts that occlude large, contiguous regions.

\begin{table}[htbp]
\centering
\caption{Sensitivity to a \emph{uniform} sparsity weight $\lambda_{sparse}$ (PSNR$\,\uparrow$/SSIM$\,\uparrow$/FLIP$\,\downarrow$). Synthetic: MCMC, mean over the seven scenes per category; real: ADC, mean per category, and the real \emph{Mean} row is the mean of the three category means (Tab.~2 of the main paper averages over all fifteen scenes). Last synthetic row: the per-category rule of the paper ($0.05$, mud $0.01$). Best per row in \textbf{bold}.}
\label{tab:lambda}
\resizebox{\columnwidth}{!}{%
\begin{tabular}{ll ccc}
\toprule
 & \textbf{Category} & $\lambda_{sparse}=0$ & $\lambda_{sparse}=0.01$ & $\lambda_{sparse}=0.05$ \\
\midrule
\multirow{7}{*}{\rotatebox[origin=c]{90}{Synthetic}}
 & Watermark & 28.29\,/\,0.868\,/\,0.137 & 29.09\,/\,\textbf{0.874}\,/\,0.120 & \textbf{29.39}\,/\,\textbf{0.874}\,/\,\textbf{0.114} \\
 & UI Elements & 28.65\,/\,0.872\,/\,0.130 & 29.24\,/\,\textbf{0.876}\,/\,0.117 & \textbf{29.42}\,/\,\textbf{0.876}\,/\,\textbf{0.114} \\
 & Dashboard & 27.67\,/\,0.854\,/\,0.139 & 28.24\,/\,\textbf{0.858}\,/\,0.127 & \textbf{28.30}\,/\,0.857\,/\,\textbf{0.124} \\
 & Rain spots & 22.30\,/\,0.832\,/\,0.288 & 25.66\,/\,0.856\,/\,0.206 & \textbf{28.94}\,/\,\textbf{0.868}\,/\,\textbf{0.126} \\
 & Mud & 28.47\,/\,0.867\,/\,0.139 & \textbf{28.95}\,/\,\textbf{0.871}\,/\,\textbf{0.130} & 27.63\,/\,0.848\,/\,0.165 \\
 & Mean & 27.07\,/\,0.859\,/\,0.167 & 28.24\,/\,\textbf{0.867}\,/\,0.140 & \textbf{28.74}\,/\,0.864\,/\,\textbf{0.129} \\
 & Mean (paper rule: 0.05, mud 0.01) & \multicolumn{3}{c}{\textbf{29.00}\,/\,\textbf{0.869}\,/\,\textbf{0.122}} \\
\midrule
\multirow{4}{*}{\rotatebox[origin=c]{90}{Real}}
 & Finger & 16.59\,/\,\textbf{0.681}\,/\,\textbf{0.384} & \textbf{16.67}\,/\,0.678\,/\,0.389 & 15.90\,/\,0.648\,/\,0.422 \\
 & Mud & 19.19\,/\,\textbf{0.647}\,/\,0.317 & \textbf{19.25}\,/\,0.645\,/\,\textbf{0.316} & 18.53\,/\,0.634\,/\,0.338 \\
 & Occlusion & 17.35\,/\,\textbf{0.587}\,/\,0.392 & \textbf{17.66}\,/\,0.585\,/\,\textbf{0.388} & 16.44\,/\,0.570\,/\,0.430 \\
 & Mean & 17.71\,/\,\textbf{0.638}\,/\,\textbf{0.364} & \textbf{17.86}\,/\,0.636\,/\,0.365 & 16.95\,/\,0.617\,/\,0.397 \\
\bottomrule
\end{tabular}%
}
\end{table}

\section{Qualitative initialization ablation}
\label{sec:supp_init}
\begin{figure}[htbp]
  \centering
  \setlength{\tabcolsep}{1pt}
  \renewcommand{\arraystretch}{0.9}
  \newcommand{\aincg}[1]{\includegraphics[width=0.485\columnwidth]{media/ablation/#1}}
  \footnotesize
  \begin{tabular}{@{}cc@{}}
    Corrupted & GT (clean) \\
    \aincg{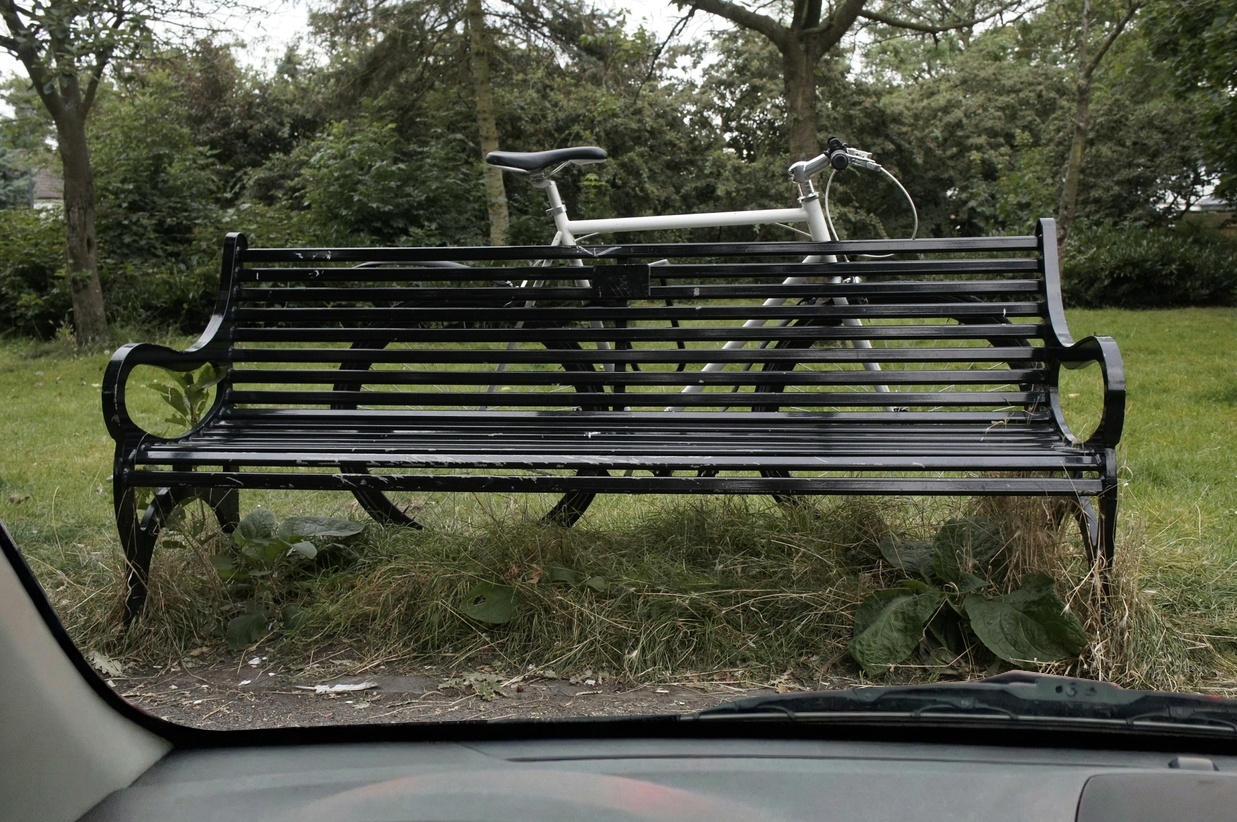} & \aincg{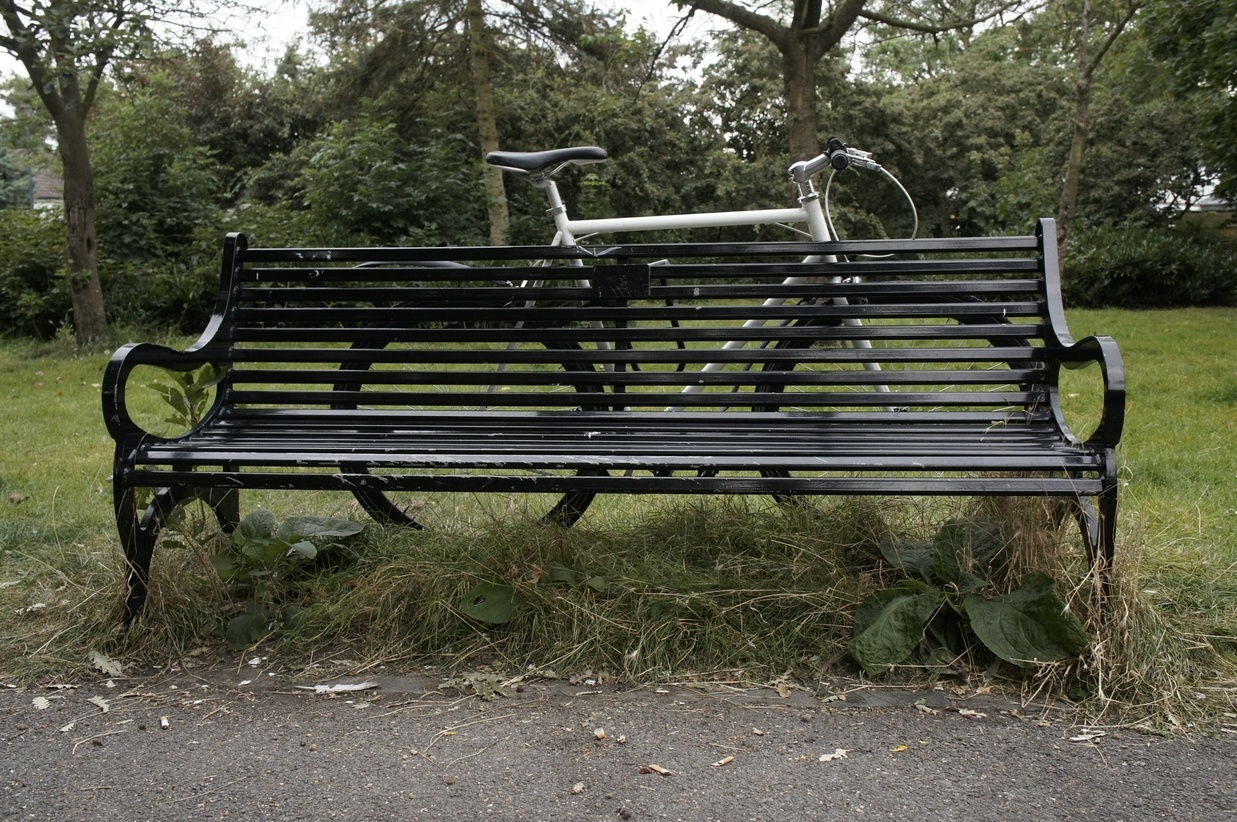} \\
    Ours & Ours (const.\ init) \\
    \aincg{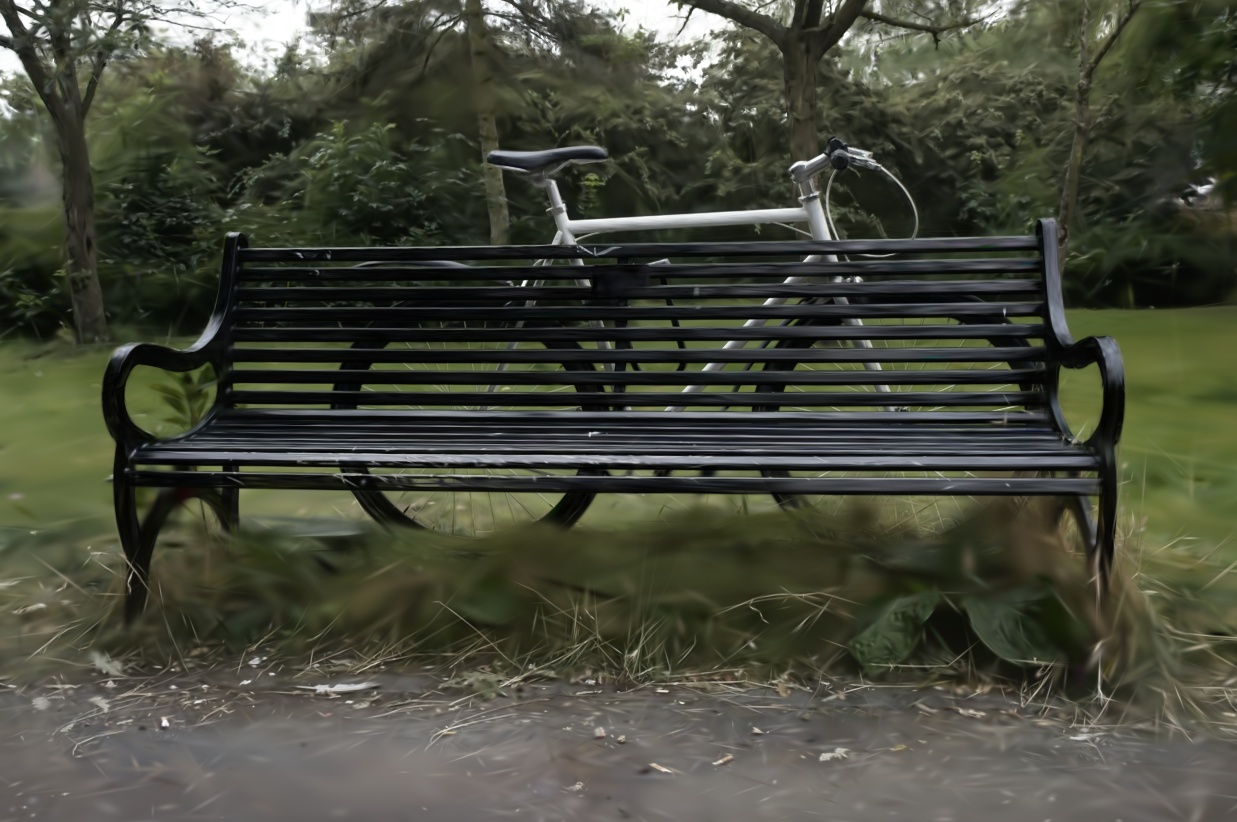} & \aincg{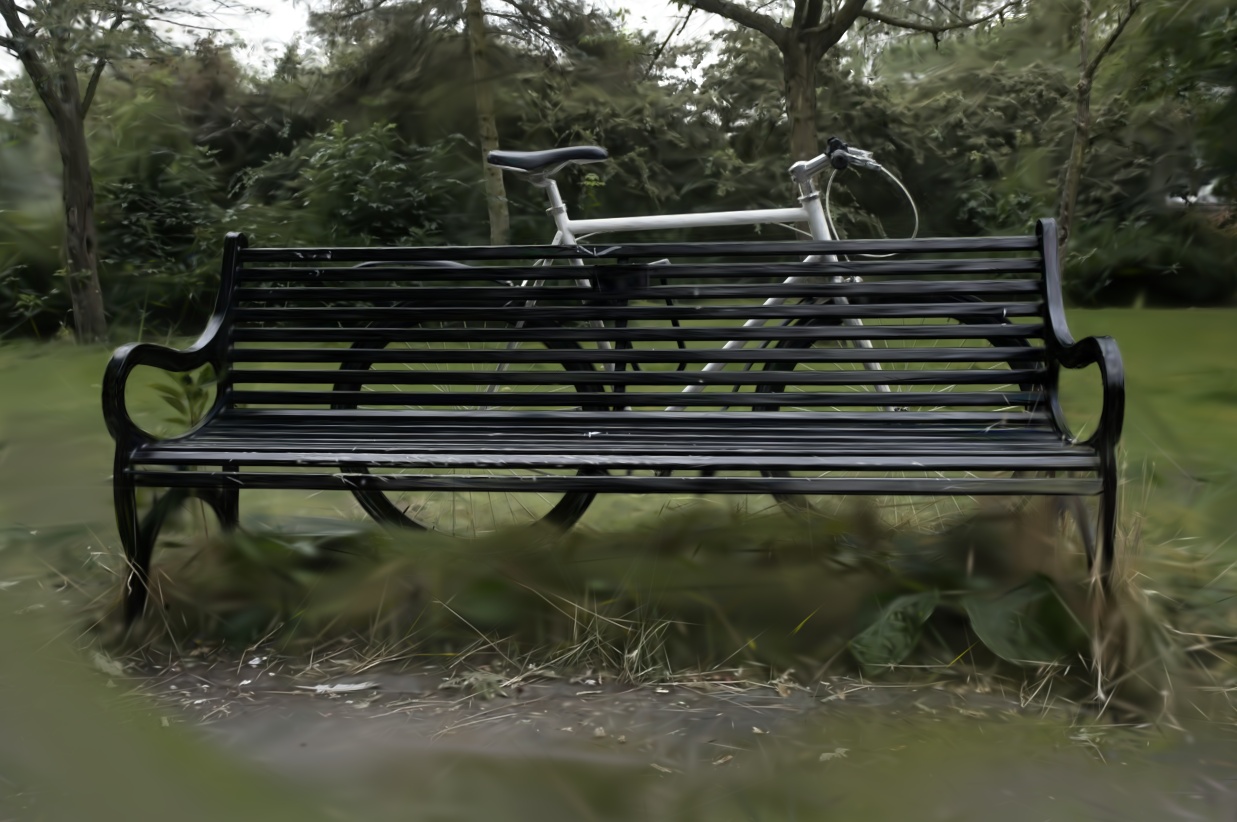} \\
  \end{tabular}
  \caption{Overlay initialization ablation (\emph{Dashboard}, \emph{bicycle}). Top: corrupted input and clean ground truth. Bottom: statistics init (ours) vs.\ constant init. Both remove the occluder; the constant init recovers the occluded region less faithfully.}
  \label{fig:ablation_qual}
\end{figure}

\Cref{fig:ablation_qual} complements the initialization ablation of Sec.~4.8 of the main paper with a \emph{Dashboard} view of the \emph{bicycle} scene: with the statistics prior, our clean render matches the ground truth, whereas with a constant initialization the 3D scene starts explaining the occluded region before the overlay can claim it, and that region is recovered less faithfully.

\end{document}